pdfoutput=1

\documentclass[%
 superscriptaddress,
reprint,
amsmath,amssymb,
pra,
floatfix,
]{revtex4-1}

\newcommand{\Rsd}{R_\mathrm{sd}}
\newcommand{\Rse}{R_\mathrm{se}}
\newcommand{\Rop}{R_\mathrm{op}}

\newcommand{\TBW}{T_{\text{bw}}}

\usepackage{graphicx}
\usepackage{dcolumn}
\usepackage{bm}
\usepackage{xparse}
\usepackage{mathrsfs}  
\usepackage{upgreek}
\usepackage{siunitx}
\usepackage{bbold}
\usepackage{stmaryrd}
\usepackage{bbm}
\usepackage{ulem}
\usepackage{color}
\usepackage{braket,bigints}
\DeclareSIUnit{\amagat}{amg}
\DeclareSIUnit{\sample}{Sa}

\newcommand{\supin}{^{(\mathrm{in})}}
\newcommand{\supout}{^{(\mathrm{out})}}
\usepackage{mathtools}

\definecolor{mygreen}{rgb}{0,0.5,0}
\definecolor{mygrey}{rgb}{0.5,0.5,0.5}
\definecolor{myred}{rgb}{0.75,0,0}
\definecolor{myblue}{rgb}{0,0,0.75}
\definecolor{mymagenta}{cmyk}{0,1,0,0.12}
\definecolor{mycyan}{cmyk}{1,0,0,0.12}
\definecolor{myorange}{rgb}{1,0.5,0}
\definecolor{myviolet}{rgb}{0.5,0.0,0.75}
\definecolor{mybrown}{cmyk}{0,0.50,1,0.41}

\newcommand{\BE}{\begin{equation}}
\newcommand{\EE}{\end{equation}}
\newcommand{\BEw}{\begin{widetext}\begin{equation}}
\newcommand{\EEw}{\end{equation}\end{widetext}}

\newcommand{\BEA}{\begin{eqnarray}}
\newcommand{\EEA}{\end{eqnarray}}
\newcommand{\NAtoms}{N_\mathrm{at}}

\renewcommand\bra[1]{{\langle{#1}|}}
\makeatletter
\renewcommand\ket[1]{%
\@ifnextchar\bra{\k@t{#1}\!}{\k@t{#1}}%
}
\newcommand\k@t[1]{{|{#1}\rangle}}
\makeatother

\begin{document}

\preprint{APS/123-QED}

\newcommand{\thetitle}{
Spin projection noise and the magnetic sensitivity of optically pumped magnetometers \\
}
\title{\thetitle}

\newcommand{\myaffiliation}{\affiliation}

\newcommand{\ITE}{\myaffiliation{Institute of Electronic Structure and Laser, Foundation for Research and Technology, 71110 Heraklion, Greece}}

\newcommand{\ICFO}
{\myaffiliation{ICFO - Institut de Ci\`encies Fot\`oniques, The Barcelona Institute of Science and Technology, 08860 Castelldefels (Barcelona), Spain}}

\newcommand{\ICREA}{\myaffiliation{ICREA - Instituci\'{o} Catalana de Recerca i Estudis Avan{\c{c}}ats, 08010 Barcelona, Spain}}

\newcommand{\UOC}{\myaffiliation{Department of Physics, University of Crete, Heraklion 71003, Greece}}

\author{K. Mouloudakis}
\email[Corresponding author: ]{kostas.mouloudakis@icfo.eu}
\ICFO

\author{V. Koutrouli}
\ITE

\author{I. K. Kominis}
\UOC

\author{M. W. Mitchell}
\ICFO
\ICREA

\author{G. Vasilakis}
\email[Corresponding author: ]{gvasilak@iesl.forth.gr}
\ITE

\date{\today}

\begin{abstract}
Present protocols for obtaining the ultimate magnetic sensitivity of optically pumped magnetometers (OPMs) utilizing alkali-metal ensembles rely on uncorrelated atoms in stretched states. A new approach for calculating the spin projection noise (SPN)-limited signal to noise ratio (SNR) and  the magnetic sensitivity of OPMs is proposed. Our model is based solely on the mean-field density matrix dynamics and in contrast to previous models, it applies to both low and high field regimes, it takes into account the degree of spin polarization, the intra- and interhyperfine correlations, the decoherence processes, the atom-light coupling and the effects of the spin dynamics on the spin-noise spectra. Fine tuning of the probe frequency allow us to explore different hyperfine states and ground-state correlations. Especially in the spin-exchange-relaxation-free (SERF) regime, alongside the magnetic resonance narrowing and the increased number density, hallmarks of SERF magnetometers, we report on a new SERF feature; the reduction of spin-projection noise at the spin precession frequency as a consequence of strongly-correlated hyperfine spins that attenuate and redistribute SPN when properly probed. 

\end{abstract}

\maketitle


\newcommand{\acronym}{\rm SISNI}
\newcommand{\boldacronym}{\textbf{SISNI}}

\section{Introduction} 
Continuous quantum measurement of many-body systems is central in quantum sensing, where one or more parameters of a statistical distribution (i.e., a magnetic field) are estimated through non-destructive monitoring of suitable observables, while the  system to be measured  is usually prepared in an ideal quantum state \cite{helstrom1976quantum,Colangelo2017}. Aside technical limitations, the quality of the estimation is characterized by the uncertainty in estimating the unknown parameter, in turn depending on i) the \textit{signal to noise ratio} of the observable \cite{PhysRevLett.95.063004,PhysRevA.77.033408} and ii) the efficiency of the measurement \cite{Deutsch2010,PhysRevA.107.012611}. 

Optically pumped magnetometers (OPMs) utilizing ensembles of alkali-metal atoms constitute a great example of quantum sensing \cite{RevModPhys.92.021001,RevModPhys.89.035002}. For example, it is shown that by exploiting quantum-correlated probes the quality of estimation is improved at specific spectral regions \cite{PhysRevLett.127.193601,PhysRevLett.131.133602,Jia2023}. In addition, it is well-known that quantum nondemolition (QND) measurements using classical probes are utilized for generating and detecting non-classical correlations between the particles in such atomic media, when back-action noise is tactfully avoided \cite{PhysRevLett.113.093601,PhysRevLett.111.103601,PhysRevLett.106.143601,Vasilakis2015,MitchellNatureCom,PhysRevX.2.031016}. 

In an ensemble of $\NAtoms$ classically correlated and non-interacting atoms the standard quantum limit (SQL) in the estimation of the magnitude of the magnetic field is set by spin projection noise (SPN) and yields \cite{BudkerRomalis2007}
\begin{equation}
  \delta B = \frac{\hbar}{g_F \mu_B \sqrt{2F}}\Big(\frac{\Gamma}{\NAtoms T}\Big)^{1/2}, 
\label{eq:Equation One}  
\end{equation}
where $\delta B$ is the uncertainty of the estimation, $g_F$ is the L\'ande factor of the particular spin used, $\mu_B $ is the Bohr magneton, $F$ is the total angular momentum of the spin system, $\Gamma$ is the spin-relaxation (decoherence) rate and $T$ is the duration of the continuous quantum measurement of the spins, precessing in the magnetic field. Multiple quantum information protocols leveraging quantum resources, like entanglement between the particles, have shown that it is possible to achieve or even surpass the Heisenberg scaling $\delta B \propto 1/\NAtoms$, both under ideal \cite{PhysRevLett.85.1594}, or under dissipative and noisy environments \cite{Napolitano2011,PhysRevX.5.031010, PhysRevLett.109.253605}.  

SPN that sets the SQL limit in Eq.\eqref{eq:Equation One} is rooted in the discreteness of the atom and the quantization of atomic observables and arises through the uncertainty of the atomic spin in the particular quantum state of the ensemble subjected to continuous measurement \cite{PhysRevA.47.3554,RomalisQuantumNoiseChapterInBudkerBook2013}. Often, the experimentally measured  uncertainty is compared against the SQL, derived assuming spin-noise variance in the ``stretched state'', where all atoms are in the fully polarized state $\ket{FF}$ (or $\ket{F-F}$) with $F=I+1/2$ and $I$ being the nuclear quantum number \cite{budker2020sensing}. Nevertheless, OPMs that operate continuously under steady state conditions of pump and probe light-fields, rely on partially polarized atoms. These atoms exhibit spin-variance levels that fall between those of stretched states and fully-mixed (thermal) states. Furthermore, in the most interesting case of high alkali-metal density, spin-exchange collisions play a significant role in the dynamics; they induce non-linear effects and give rise to non-trivial correlations among ground-state observables, impacting the spectral distribution of noise \cite{PhysRevA.106.023112}. It is therefore apparent that the application of Eq.\eqref{eq:Equation One} is not guaranteed.

Of particular interest are OPMs operating in the spin-exchange-relaxation-free (SERF) regime \cite{Happer-Tang,Happer-Tam,SavukovRomalis}. In this regime, high atomic density and line-narrowing of the magnetic resonances led to a demonstrated enhancement in the magnetic sensitivity \cite{Allred,Kominis2003}, enabling a range of applications including magnetoencephalography in shielded \cite{boto} or ambient environments \cite{PhysRevApplied.14.011002, Zhang2020} and tests of fundamental physics \cite{Bloch2023,VasilakisPRL2009}. Although SERF has revolutionized hot-vapor magnetometry, the obtained sensitivities are based on signal improvement arguments and are still away from the SQL, mostly due to technical limitations \cite{Kominis2003}.

Based on these considerations it is emergent that proper characterization of OPMs requires a quantitative understanding of of how quantum noise impacts the measurement and ultimately the magnetic sensitivity, and involves examining the noise spectra of the experimental observables. Although the protocols discussed above provide fundamental bounds, a more reliable strategy would be to investigate application-specific models, taking into account the whole alkali structure, the different sources of noise and the atomic physics involved. In general, three sources of quantum noise could potentially limit the sensitivity of the OPMs: spin projection noise, photon shot noise (PSN) and AC-Stark shifts.

Strategies to account for spin projection noise so far were mostly based on two-level Bloch models. These are simplified representations of the actual dynamics; they assume that spin-correlation functions decay with a single rate, and ignore correlations between the hyperfine manifolds of the ground electronic state \cite{RomalisQuantumNoiseChapterInBudkerBook2013}.
Recent studies with unpolarized ensembles, including both single- and dual-species vapors \cite{PhysRevA.106.023112, mouloudakis2023anomalous, PhysRevA.108.052822}, found qualitative and quantitative agreement between experimental spin-noise spectra of atomic ensembles and quantum noise models derived from the master equation and noise-balance considerations. Such comparisons were made at near-zero spin polarization. 

In this paper, we extend the theory of spin-noise dynamics to spin-polarized ensembles \cite{PhysRevLett.104.013601,PhysRevLett.106.143601} of high alkali-metal densities. By considering the master equation for spin-polarized atomic ensembles, we study the effects of SPN and PSN on the magnetic sensitivity of an AC OPM and obtain general analytic results that go beyond the standard quantum limit given by Eq.\eqref{eq:Equation One}. Our approach takes into account the hyperfine structure of the atoms and properly addresses the correlations that spontaneously build up between the two ground-state hyperfine manifolds. The model predicts that under certain probing conditions in the SERF regime, the measured SPN is attenuated at frequencies around the magnetic resonance. Such a spectral reshaping of measured spin-noise was suggested before for unpolarized vapors \cite{PhysRevA.106.023112}, however it was not clear from this work whether the predicted effect will be apparent also in spin-polarized ensembles.

The structure of the paper is as follows: In \ref{sec:fromalism} we develop the theoretical framework for calculating spin dynamics and quantum-noise spectra in spin-polarized ensembles. In \ref{sec:Optical readout} we describe the atom-light coupling of the Faraday probe, the rf modulation, the demodulated signal and the signal to noise ratio of the rf OPM. Finally, in \ref{sec:conclusion} we summarize our results.



\section{Formalism: Spin polarized ensembles}
\label{sec:fromalism}

\subsection{Spin dynamics}

\begin{figure}[t]
	\centering
\includegraphics[width=1.0\columnwidth]{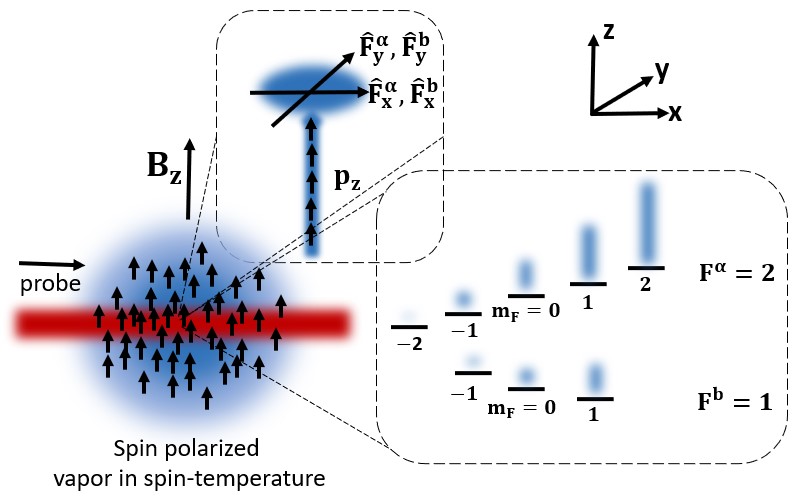}	
\caption{Experimental  configuration. A spin-polarized atomic ensemble (here $^{87}$Rb in a spin temperature distribution along $\mathbf{\hat{z}}$ is prepared by continuous optical pumping along $\mathbf{\hat{z}}$, parallel to a DC magnetic field. for a  \textsuperscript{87}Rb vapor. An off-resonant linearly-polarized probe-light along the $\mathbf{\hat{x}}$-axis records the transverse collective-spin fluctuations through a quantum non-demolition measurement.} 
\label{fig:fig3}
\end{figure}

We consider a continuously pumped, high-density alkali-metal atomic ensemble, spin-polarized parallel to a DC magnetic field $B$ along the $\mathbf{\hat{z}}$-direction. 
A continuous, off-resonant and linearly polarized monochromatic probe light monitors through the paramagnetic Faraday effect \cite{RevModPhys.74.1153}  the ensemble-collective transverse spin component along the direction of propagation, taken here to be the $x$-axis (see Fig.\ref{fig:fig3}). This magnetometer configuration has realized the highest sensitivities among the OPMs \cite{RevModPhys.92.021001}.

Neglecting atomic motion and Hamiltonian dynamics due to probe-light (i.e. dynamics induced from light-shifts), the master equation describing ground-state spin dynamics is written as \cite{HapperBook2010}:
\begin{eqnarray}
 \frac{d  \rho}{d t} &= & A_{\rm{hfs}}\frac{[ \mathbf{I} \cdot \mathbf{S},\rho]}{i\hbar}  +g_s \mu_{_B} \frac{[\mathbf{S} \cdot \mathbf{B} (t), \rho]}{i \hbar}+R_{\mathrm{se}} (\rho_{\langle \mathbf{S} \rangle} - \rho)  \nonumber \\
& & +R_{\text{op}} \mathcal{L}_{\rm{op}}(\rho)  + R_{\rm{sd}} \mathcal{L}_{\rm{sd}}(\rho),
  \label{eq:mastereq}
 \end{eqnarray} 
where $\mathbf{S} $ and $\mathbf{I}$ are the electron and nuclear spin operators,
$A_{\rm{hfs}}$ is the hyperfine coupling constant of the alkali-metal medium, $g_s \approx 2$ is the g-factor of the electron, $\mu_{_B} \approx 9.27 \times 10^{-24}$ \si{\joule\per\tesla} the Bohr magneton, $R_{\rm{se}}$, $R_{\rm{op}}$ and  $R_{\rm{sd}}$ are the spin-exchange, optical pumping and spin-destruction rates, respectively, and $\rho_{\langle \mathbf{S} \rangle}\equiv \left( \mathbbm{1}/{2} + 2 \langle \mathbf{S} \rangle  \cdot \mathbf{S}\right) \otimes \mathrm{Tr}_\mathrm{S}[\rho]$ is the state formed by processes (here spin-exchange collisions between alkali-metal atoms) that drive the spin polarization to have mean (single-atom) electronic spin $\langle \mathbf{S} \rangle$ and leave the nuclear state unchanged \cite{RevModPhys.44.169}. The operators $\mathcal{L}_{\rm{op}}$ and $\mathcal{L}_{\rm{sd}}$ describe optical pumping and spin destruction dynamics, respectively. The exact form of these operators depends on the specifics of the ensemble (e.g., presence of buffer gas, wavelength of probe and pump light), but in general for high alkali-metal densities that are of interest in this work the rate of spin-exchange collisions is much larger than $R_{\text{op}}$ and $R_{\text{sd}}$.

In order to study spin noise, we adopt the methodology outlined in \cite{PhysRevA.108.052822} and leverage the quantum regression theorem (QRT). The QRT posits that if the equations of motion for the expectation values of specific operators exhibit linearity, then the corresponding time-correlation functions conform to the same equations \cite{gardiner2009stochastic,PhysRevA.106.023112}. Due to the spin-exchange dynamics (third term in the right hand side of Eq.\eqref{eq:mastereq}), the master equation is non-linear with respect to $\rho$. When examining quantum spin-noise in atomic ensembles, there are only small fluctuations from the equilibrium steady state, as the fractional spin fluctuations are on the order of $1/\sqrt{N_{\text{at}}}$, where $N_{\text{at}}$ is the number of probed atoms in the ensemble. Given the small magnitude of the fluctuations, we can linearize the equations of motion around the steady state and apply the QRT to find spin correlations and noise.

It is important to note that the single-atom mean spin dynamics derived from the (mean-field) density matrix equation correspond to the dynamics for the means of the collective spin variables measured in the experiment (see discussion in \cite{PhysRevA.108.052822}). As long as there are no equal-time correlations between distinct atoms, the noise spectra for the measured ensemble-collective spin-observables can be derived directly from the single atom quantities by simply scaling the noise with the effective number of atoms contributing to the measurement. 


In the following, we outline the derivation of equations of motion for the mean spin values, clarifying the linearization process, and present the spin-noise spectrum.
Analytic calculations become significantly simpler when performing the analysis (expressing the density matrix and all spin operators) in terms of spherical tensors in the coupled basis defined by \cite{SuppInfo}:
\begin{align}
T_{LM}(FF') = \sum_{m} |Fm\rangle & \langle F' m-M|(-1)^{m-M-F'} \nonumber \\
&\times C^{L M}_{Fm;F'(M-m)}, \label{eq:SphericalTensors}
\end{align}
where $C$ denotes Clebsch-Gordan coefficient. This is because the dynamical processes considered in Eq.\eqref{eq:mastereq} do not couple operators with different projections $M$'s, thus enabling an analysis within a lower-dimensional space as compared to employing spin tensors in the Cartesian basis. Furthermore, in the absence of resonant microwave fields, hyperfine coherences (represented by tensors with $F\neq F'$) remain small, induced only by noise processes. As a result, they have a negligible effect on the dynamic evolution of Zeeman coherences ($F=F'$) and are ignored in the analysis. Within this approximation, the density matrix is written as: 
\begin{equation}
\rho = \sum_{LMF}\langle T_{LM}(FF) \rangle T^{\dag}_{LM}(FF).
\end{equation}

From the properties of the Hermitian conjugate we also obtain: $T_{LM}^{\dag}(FF)=(-1)^M T_{L-M}(FF)$. By multiplying both sides of Eq.\eqref{eq:mastereq} by $T_{LM}(FF)$ and subsequently taking the trace, we derive equations describing the dynamic evolution of $\langle T_{LM}(FF) \rangle$. The resulting differential equations incorporate non-linear terms of the form: $\langle T_{LM}(FF) \rangle \langle T_{L'M'}(F'F')\rangle$, arising from the spin-exchange dynamics. In the context of noise analysis where only small deviations from the steady state are considered, Zeeman coherences (expressed by the operators $T_{L M\neq0 }(FF)$) are small, and the product of two coherences can be ignored in the dynamics. On the other hand, population terms (expressed by the operators $T_{L M=0 }(FF)$) are not negligible for polarized spin ensembles, and the associated non-linear terms should be taken into account. Nonetheless, for noise considerations, the departure of the population terms from their equilibrium value is small, on the order of transverse coherences, and the non-linear terms involving populations can be linearized by making the approximation: $\langle T_{LM}(FF) \rangle \langle T_{L'M'=0}(F'F') \rangle \approx \langle T_{LM}(FF) \rangle   \text{Tr}\left[T_{L'M'=0}(F'F') \rho_{0}\right]$, where $\rho_0$ is the equilibrium density matrix, i.e., the steady state solution of Eq.\eqref{eq:mastereq}. A comprehensive analysis of the linearization process is enclosed in \cite{SuppInfo}.

Using this linearization, the differential equations for the mean values of the coherences can be cast in a matrix form equation:
\begin{equation}
\frac{d \langle \vec{T}_M (t) \rangle}{dt}= \mathcal{A}_M \langle \vec{T}_M (t) \rangle,
\label{eq:meanT}
\end{equation}
where the drift matrix $\mathcal{A}_M$ encapsulates the impact of the processes affecting the dynamics \cite{SuppInfo}, and can be summarized as:
\begin{equation}
\mathcal{A}_M=\mathcal{A}_{\text{MG},M}+\mathcal{A}_{\text{SE}} +\mathcal{A}_{\text{SD},M} +\mathcal{A}_{\text{OP}} 
\end{equation}
where $\mathcal{A}_{\text{MG},M}$ is the drift matrix associated with the influence of the external magnetic field on the atomic spin and $\mathcal{A}_{\text{SE}}$, $\mathcal{A}_{\text{SD},M}$ and $\mathcal{A}_{\text{OP}} $  are the drift matrices associated to spin-exchange, spin destruction and optical pumping processes, respectively. Exact, analytic expressions for the preceding drift matrices are derived in \cite{SuppInfo}. The state vector:
\begin{equation}
\begin{split}
\vec{T}_M &= [  T_{1M}(aa),  T_{1M}(bb), T_{2M}(aa), T_{2M}(bb)\\
& ...T_{LM}(aa), T_{LM}(bb),...]^T,
\label{eq:Tvector}
\end{split}
\end{equation}
is a column vector in $4I$-dimensional space with $ L=1,2,...2I$, and $a=I+1/2$ and $b=I-1/2$ denoting the ground hyperfine manifolds. 
For polarized atomic ensembles, spins with different multipolarities become coupled to each other \cite{RevModPhys.44.169}. Although the experimentally observable quantity depends on the vector spin multipole ($L=1$), and for buffer gas-free cells also on the second-rank spin-multipole \cite{HapperMathur}, for a comprehensive treatment of polarized atomic ensembles, the complete multipole spectrum should be considered \cite{kozbial2023spin}.

\subsection{Noise spectra}

The noise properties can be described by the transverse covariance matrix or equivalently by the power spectral density matrix. In the spherical basis, the covariance matrix $\tilde{\mathcal{R}}$ for arbitrary lag time $\tau$ has matrix elements $\tilde{\mathcal{R}}_{ij} = [\langle x_i(\tau) x_j(0)+x_j(0) x_i(\tau) \rangle]/2$, where $x_i$ is an element of the phase-space vector $\displaystyle \bigoplus_{M} \vec{T}_M$. According to QRT, the covariance matrix evolves with the lag time following a dynamic equation identical to that governing the evolution of mean values (see Eq.~\ref{eq:meanT}). This gives: $\tilde{\mathcal{R}}(\tau) = e^{\mathcal{A} \tau} \tilde{\mathcal{R}}(0)$ for $\tau>0$, with $\tilde{\mathcal{R}}(0)$ determined from the equilibrium state $\rho_0$, and $\mathcal{A} = \bigoplus_{M} \mathcal{A}_M$. The power spectral density matrix $\tilde{S}$ at frequency $\omega$ is related to the covariance matrix through a Fourier transform: $\tilde{S}(\nu) = \int_{-\infty}^{\infty} \tilde{\mathcal{R}}(\tau) e^{i 2 \pi \nu \tau} d\tau$.


Longitudinal pumping creates steady state spin-polarization rotational symmetric around the $z$-axis. Consequently, the equal-time correlation $\langle T_{LM}(FF) T_{L'M'}(F'F')\rangle$ is non-zero only for $M=-M'$. Since the linear dynamics considered here do not couple spherical tensors with different values of $M$, the complete covariance matrix, and similarly the complete spectrum matrix, can be partitioned into distinct non-mixing blocks, each characterized by the absolute value of the azimuthal quantum number $M$. The power spectral density matrix reads \cite{gardiner2009stochastic, PhysRevA.106.023112,PhysRevA.108.052822,SuppInfo}:
\begin{align}
&\tilde{S}_{|M|} (\nu) = - \left( -\mathcal{A}_{|M|}+i 2 \pi \nu \right)^{-1} \nonumber \\
& \times \left( \mathcal{A}_{|M|} \tilde{\mathcal{R}}_{|M|}(0)+\tilde{\mathcal{R}}_{|M|} (0)\mathcal{A}_{|M|}^{\top}  \right)    \left( -\mathcal{A}_{|M|}^{\top}-i 2 \pi \nu \right)^{-1}, \label{eq:SpectrumSpherical0}
\end{align}
where: $\mathcal{A}_{|M|} = \mathcal{A}_{M} \bigoplus \mathcal{A}_{-M}$ is the drift matrix for the combined $\vec{T}_{|M|}=[\vec{T}_{M}, \vec{T}_{-M} ]^{\top}$ vector, and the equal-time transverse covariance matrix $\tilde{\mathcal{R}}_{|M|}(0)$ for $\pm M$ has the form of a symmetric block anti-diagonal matrix \cite{SuppInfo}.

\begin{figure}[htp]
\centering
\includegraphics[width=8.0cm]{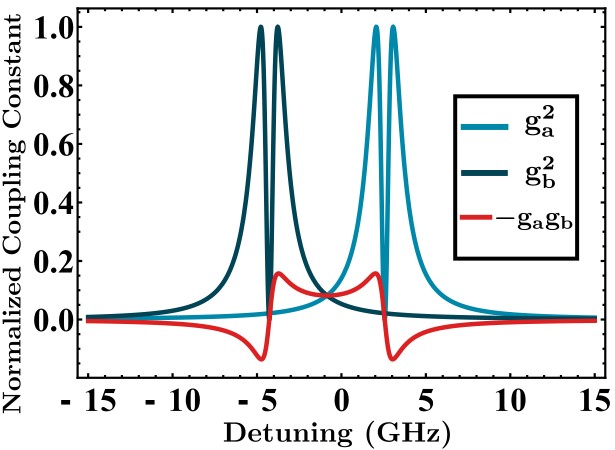} 
\caption{Coupling constants $g_a^2$, $g_b^2$ and $-g_a g_b$ appearing in Eq.\eqref{Eq:auto-corr} as a function of the optical detuning from the $D_1$ transition of $^{87}$Rb. The spectrum corresponds to an optical homogeneous linewidth of $\Delta \upnu=\SI{1}{\giga\hertz}$ and the lineshapes are normalized to the maximum value of $g_a^2$.}
\label{fig:detuning}
\end{figure} 

Although the spherical basis is convenient for calculations,  for a direct comparison with the experiment (see below) it is useful to have an expression for the spectrum matrix $S(\omega)$ of the transverse Cartesian spin components. These are related only to the vector spherical tensors $L=1, M=\pm 1$. Therefore, $S(\omega)$ can be found from the transformation: $S(\omega) = \mathcal{M} \tilde{S}_{|1|}(\omega) \mathcal{M}^{T}$, where $\mathcal{M}$ represents the change of basis matrix from the spherical basis to the Cartesian spin components \cite{SuppInfo}. A similar transformation also holds for the covariance matrix.

\subsection{Variances}

For relaxation processes sudden with respect to the nuclear spin dynamics and for continuous optical pumping, the equilibrium state of the ensemble is the spin-temperature density matrix \cite{appelt, RevModPhys.44.169}: 
\begin{equation}
\rho_0=\rho_{\text{ST}} = \frac{e^{\beta S_z}}{Z(\frac{1}{2},\beta)} \frac{e^{\beta I_z}}{Z(I,\beta)}, Z(K,\beta) = \frac{\sinh \left[ \beta (K+\frac{1}{2})\right]}{\sinh\left[ \beta/2\right]}.
\label{eq:SpinTemperature}
\end{equation}
The spin-temperature parameter $\beta$ is related to the mean spin through:
\begin{equation}
\begin{split}
\langle K_z \rangle & =\frac{1}{2} (2K+1) \coth (\frac{\beta}{2}) \coth \left[ \beta (K+1/2) \right] \tanh{( \frac{\beta}{2})}\\
&-\frac{1}{2}\coth^2(\frac{\beta}{2})\tanh{( \frac{\beta}{2} )},
\end{split}
\end{equation}
with the degree of spin polarization defined as $p=2 | \langle S_z\rangle |$. For this steady state, the transverse spin variances of the two hyperfine spins comprising the Cartesian equal-time covariance matrix take the form:
\begin{widetext} 
\begin{equation}
\text{Var} \left[ F_{a,x}\right] = \text{Var} \left[ F_{a,y}\right] = \frac{[2 (I+1) p+1] (p-1)^{2 I+2}+(p+1)^{2 I+2} [1-2 (I+1) p]}{8 p^2 \left[(p-1)^{2 I+1}-(p+1)^{2 I+1}\right]}
\end{equation}
\begin{equation}
\text{Var} \left[ F_{b,x}\right] = \text{Var} \left[ F_{b,y}\right] =\frac{\left(p^2-1\right) \left[(p+1)^{2 I} (2 I p-1)-(p-1)^{2 I} (2 I p+1)\right]}{8 p^2 \left[(p-1)^{2 I+1}-(p+1)^{2 I+1}\right]}
\end{equation}
\end{widetext} 
The above formulas apply to any alkali-metal atom with  arbitrary spin polarization $p$. 

\section{Optical readout}
\label{sec:Optical readout}
\subsection{Faraday probing}
The transverse spin in the $x$ direction is measured by detecting the optical rotation induced in the probe light upon interaction with the atomic ensemble (see Fig.~\ref{fig:fig3}). In the small angle approximation (relevant when studying noise and fundamental sensitivity), the detected light-observable can be written in the form \cite{PhysRevA.106.023112}:
\begin{align}
\mathcal{S}_2\supout (t) \approx \mathcal{S}_2\supin (t) + & \frac{(-1)^{j+1/2}}{2}  \Big[g_a \sum_i F_{a,x}^{(i)}(t) \nonumber \\  
&-g_{b} \sum_i F_{b,x}^{(i)} (t) \Big] \Phi, \label{eq:polarimeter}
\end{align}
where the summation is performed over all the atoms interacting with the probe beam. For simplification, we assume homogeneous coupling of light to atoms in the ensemble
The superscripts out (in) denote the light observable after (before) interaction with the ensemble, respectively, $\mathcal{S}_2$ is the Stokes polarization component quantifying the difference in fluxes of linearly polarized photons at angles of $\pm 45^\circ$ from the input polarization axis, and $\Phi$ corresponds to the photon flux measured at the ensemble's output. For conditions pertinent to high-sensitivity magnetometers, where the homogeneous broadening (e.g., from collisions) dominates the optical linewidth or the probe detuning is significantly larger than the Doppler broadening, the dispersive atom-light coupling constants $g_\alpha$, $\alpha  \in \left \{a= I + 1=2; b = I - 1/2 \right \}$, are given by:
\begin{equation}
g_{\alpha}  = \frac{4}{(2j+1)(2I+1)} \frac{c r_e f_{\mathrm{osc}}}{A_\mathrm{{eff}}} D_{\alpha} (\upnu)= G D_{\alpha}(\upnu),
\label{eq:gjDef}
\end{equation}
\begin{equation}
     D_\alpha (\upnu) = \frac{(\upnu-\upnu_\alpha)/\Gamma}{[(\nu-\nu_\alpha)/\Gamma]^2+1}, \label{eq:DetuningFactor}
\end{equation}
where $r_e=2.83 \times 10^{-15}$ \si{\meter} is the classical electron radius, $f_{\mathrm{osc}}$ is the oscillator strength associated with the particular optical transition, $A_{\mathrm{eff}}$ is the effective beam area \cite{mouloudakis2023anomalous}, $\Gamma$ is the half width at half-maximum optical linewidth, and $ \upnu-\upnu_{\alpha}$ is the detuning of probe light (with frequency $\upnu$) from the optical frequency $\upnu_{\alpha}$ associated with the transition from manifold $\alpha$ in the ground electronic state to the excited state. In Eq.~\ref{eq:gjDef} 
the coupling strength $G$ is related to the optical cross section at resonance $\sigma_0$:
\begin{equation}
    G =  \frac{4}{(2j+1)(2I+1)} \frac{\sigma_0}{A_\mathrm{{eff}}}.
\end{equation}
Typically, for high-sensitivity magnetometers, the optical detuning or the optical linewidth significantly exceeds the hyperfine splitting in the excited electronic state; in this case, the hyperfine structure in the excited state gives only a small correction to Eq.~\ref{eq:gjDef} \cite{HapperMathur}, while optical rotation induced from the ensemble's circular dichroism can be neglected \cite{HapperMathur}.  

The correlation function and correspondingly the noise spectrum for the measured light-variable yield:
\begin{widetext}
\begin{equation}
\mathcal{R}_{\mathcal{S}_2\supout,\mathcal{S}_2\supout}(\tau) \approx \frac{ \Phi}{2} \delta(\tau)  + \frac{\Phi^2}{4} N_{\text{at}} G^2\left\{
D_a^2 \mathcal{R}_{F_{a,x};F_{a,x}}(\tau)  + D_b^2  \mathcal{R}_{F_{b,x};F_{b,x}}(\tau) - D_a D_b [\mathcal{R}_{F_{a,x};F_{b,x}}(\tau) + \mathcal{R}_{F_{b,x};F_{a,x}}(\tau) ] \right\}, \label{Eq:auto-corr}
\end{equation}
\begin{align}
S_{\mathcal{S}_2\supout,\mathcal{S}_2\supout}(\nu) & = \frac{\Phi }{2}+\frac{\Phi^2}{4} N_{\text{at}} G^2\left\{ D_a^2 S_{F_{a,x};F_{a,x}}(\nu)  + D_b^2 S_{F_{b,x};F_{b,x}}(\nu) - D_a D_b [S_{F_{a,x};F_{b,x}}(\nu) + S_{F_{b,x};F_{a,x}}(\nu) ] \right\}, \label{Eq:MeasuredSpectrum} \\
&= \frac{\Phi }{2}+\frac{\Phi^2}{4} N_{\text{at}} G^2 \mathrm{S}(\nu) \label{Eq:MeasuredSpectrum2}
\end{align}
\end{widetext}
where the first term describes photon shot noise (coherent light was assumed in Eq.~\ref{Eq:MeasuredSpectrum}) and the rest spin-noise correlations. $\mathrm{S}(\nu)$ expresses the noise spectrum from the effective spin that is probed in the measurement. Note that collective spin correlations (spectra) have been expressed in relation to single-atom correlations (spectra) scaled by the total number of atoms that contribute to the measurement (see \cite{PhysRevA.108.052822} for details). The different atom-light coupling constants as a function of probe wavelength are plotted in Fig.\ref{fig:detuning} for $^{87}$Rb and conditions stated in the caption. 

\subsection{rf modulation}

In order to evaluate the sensitivity of magnetometer, the measured noise should be compared against the measured response to a known magnetic field stimulus. In the following, we analyze the response of the spin-ensemble to a coherent, sinusoidally-driven, transverse magnetic field: $\mathbf{B}_{\perp}= B_{0\perp} \left[ \cos(b) \cos(\omega t) \mathbf{\hat{x}} +\sin(b) \cos(\omega t +\varphi) \mathbf{\hat{y}} \right]$, with $\varphi$ denoting the relative phase between the two field components, while $b$ parametrizes the relative amplitude in the two transverse directions. The presence of the transverse magnetic field introduces an additional term in the right hand side of Eq.~\eqref{eq:meanT}, which for the $\langle T_{LM}(FF) \rangle$ element js given by: $\langle i \left[ g_s \mu_B \mathbf{B}_{\perp} \cdot \mathbf{S}, T_{LM}(FF) \right]\rangle/\hbar = \pm \gamma_F \mathbf{B}_{\perp} \cdot \langle \left[ \mathbf{F}, T_{LM}(FF) \right]\rangle $, where $\gamma_F = g_s \mu_B/(2I+1)\hbar$ is the atomic gyromagnetic ratio, and the $+$ ($-$) applies to the $F=a$ ($F=b$) case. We assume a small magnetic field excitation (Rabi frequency $\Omega_{0}=\gamma_F B_{0\perp}$ much smaller than the precession frequency and the effective relaxation rate) oscillating at a frequency $\omega$ close to the spin precession frequency. We neglect coherences in the harmonics of $\omega$, setting for the coherent response: $\langle T_{LM}(FF) \rangle =0$ for $|M|>1$. We also take that $\langle T_{L0}(FF) \rangle \approx \text{Tr}[ T_{L0}(FF) \rho_0 ]$, i.e. the transverse magnetic field is small enough to have a negligible effect on the longitudinal polarization.  Under these approximations we find:
\begin{equation}
\begin{split}
\frac{d}{dt} \langle \vec{T}_{M=1} \rangle &= \mathcal{A}_{1} \langle \vec{T}_{1} \rangle+  B_{0\perp}  \Big \{i \cos(b)\left( e^{i \omega t}+e^{-i \omega t} \right)/2 \\
&- \sin(b) \left( e^{i \omega t +i \varphi}+e^{-i \omega t -i \varphi} \right)/2 \Big \} \mathcal{B}, \label{eq:MatrixEquationForCoherentEvolution}
\end{split}
\end{equation}
where the column vector $\mathcal{B}$ is given by:
\begin{equation}
\mathcal{B}^{\top} = \bigoplus_{L}\sqrt{L(L+1)/2} [  \langle T_{L0}(aa), -\langle T_{L0}(bb) \rangle ].
\end{equation}
The (steady state) long-time limit -- i.e. $t$ much larger than the slowest relaxation time scale -- can be found by considering a steady state solution of the form $ \langle \vec{T}_1 \rangle_{\infty}^+ e^{i \omega t}$ (or $ \langle \vec{T}_1 \rangle_{\infty}^- e^{i \omega t}$) and equating the terms proportional to $e^{i \omega t}$ (or $e^{-i \omega t}$) in the two sides of Eq.\eqref{eq:MatrixEquationForCoherentEvolution}. This gives:
\begin{equation}
\langle \vec{T}_1 \rangle_{\infty}(t) = \langle \vec{T}_1 \rangle_{\infty}^+ e^{i \omega t}+ \langle \vec{T}_1 \rangle_{\infty}^- e^{-i \omega t},
\label{eq:RFsolution}
\end{equation}
where,
\begin{equation}
\langle \vec{T}_1 \rangle_{\infty}^{\pm} = \left( -\mathcal{A}_1 \pm i \omega \right)^{-1} \left( i \cos(b) -\sin(b) e^{\pm i \varphi} \right) \mathcal{B} /2. 
\end{equation}
Similar equations hold for the dynamical evolution of $\langle \vec{T}_{M=-1}\rangle$; in this case, the replacement $b \rightarrow -b$ should be applied.

We note that in situations where the longitudinal magnetic field significantly surpasses all other rates influencing the dynamics, the importance of one of the terms in Eq.\eqref{eq:RFsolution} -- either proportional to $\langle \vec{T}_M \rangle_{\infty}^+$ or $\langle \vec{T}_M \rangle_{\infty}^-$ -- becomes pronounced, with one outweighing the other. Nonetheless, when the longitudinal magnetic field does not satisfy the aforementioned condition (as in the SERF regime for example), the two terms become comparable in magnitude and should both be retained.

Taking into account that probe light detects the $g_\alpha$-weighted difference in hyperfine spin-components along the transverse $x$-direction, the measured response to a sinusoidal excitation at frequency $\nu$ can be expressed as:
\begin{equation}
\langle S_2^{\text{out}}(t)  \rangle = \frac{\Phi}{2} N_{\text{at}} \gamma_F B_{0,\perp} G A_c(\nu) \cos(2 \pi \nu t+\chi), \label{eq:MeansResponseToSine}
\end{equation}
where $\chi$ expresses a ($\nu$-dependent) phase lag of the weighted spin response with respect to the phase of the driving field, and the amplitude factor $A_c(\nu)$ is given by:
\begin{widetext}
\begin{equation}
A_c(\nu) = 2 \sqrt{ \left \{ \begin{pmatrix}
    D_a & -D_b
\end{pmatrix} \left[ \Re \left( \mathscr{A} \right) + \Re \left( \mathscr{B} \right) \right] 
\right \}^2 +  \left \{ \begin{pmatrix}
    D_a & -D_b
\end{pmatrix} \left[ \Im \left( \mathscr{A} \right) + \Im \left( \mathscr{B} \right) \right]
\right \}^2}\label{eq:Ac_omega}
\end{equation}
\end{widetext}
In the above equation, $\Re$ and $\Im$ denote respectively the real and imaginary part, and the matrices $\mathscr{A}$ and $\mathscr{B}$ are defined as:
\begin{align}
    \mathscr{A} & = \tilde{\mathfrak{M}} (\mathcal{A}_1 -i 2 \pi \nu ) ^{-1} \mathcal{B}\ \left( i \cos(b)-\sin(b) e^{i \phi} \right) \\
    \mathscr{B} & = \tilde{\mathfrak{M}} (\mathcal{A}_1^* -i 2 \pi \nu ) ^{-1}\mathcal{B} \left( -i \cos(b)-\sin(b) e^{i \phi} \right),
\end{align}
where $\tilde{\mathfrak{M}}$ is $2\times 4I$ matrix given by:
\begin{equation}
 \mathfrak{M} = \begin{pmatrix}
  \frac{\sqrt{(I+1) (2 I+1) (2 I+3)}}{2 \sqrt{3}} & 0 \\
 0 & \frac{\sqrt{I (2 I-1) (2 I+1)}}{2 \sqrt{3}}
 \end{pmatrix} \bigoplus \left[ 0 \right]_{2 \times (4I-2)}
\end{equation}

Typically, the magnetometer signals are obtained with a lock-in (phase-sensitive) amplifier, and the magnetic field amplitude $B_{0,\perp}$ is evaluated (assuming the relative strength and phase of the two transverse fields are known) from the amplitude of the sinusoidal response, see Eq.~\ref{eq:MeansResponseToSine}. We note that with a lock-in amplifier the phase angle $\chi$ can be identified. In principle, any two parameters from the triad $B_{0,\perp}, b, \phi$ can be determined from the two quadratures of a phase-sensitive detection of the light-modulation signal. In the following in the subsequent analysis, we will focus exclusively on magnetometers that solely measure $B_{0,\perp}$. 

\subsection{Demodulated signal}

We now proceed to characterize the magnetometer sensitivity, limited by spin projection noise and photon shot noise. Measurement-induced variations in spin-noise, such as dissipative spin-squeezing or measurement back-action noise (MBN) \cite{MitchellNatureCom,PhysRevLett.85.1594,PhysRevLett.106.143601} will not be considered here and will be investigated in a follow up work. Although MBN can be an important noise source in spin-polarized ensembles, it can be avoided, for example by using stroboscopic quantum-non-demolition measurements \cite{PhysRevLett.106.143601, Vasilakis2015, MartinPRL2017}.

The sensitivity is quantified by the signal to noise ratio (SNR). Magnetic field estimation is a parameter estimation problem where information about the field is extracted by leveraging the detected record over some finite measurement time $T$. Elaborate models for field estimation utilize maximum likelihood estimation functions  and Kalman filters \cite{PhysRevLett.120.040503,PhysRevX.5.031010}. Here, we use the simplest possible approach where the field is estimated from the time-average of the demodulated signal. To be concrete, we assume a simple model of lock-in amplifier where its phase is adjusted such that the response to a sinusoidal transverse excitation is fully represented in only one quadrature of the phase-sensitive detector. Magnetometry relies on the frequency-dependent quantity:
\begin{equation}
\mathfrak{K} = \frac{1}{T }\int_0^{T} dt \frac{1}{\TBW}\int_{t_0}^t e^{-\frac{t-t'}{\TBW}} \cos \left( 2 \pi \nu t' +\theta \right)  \mathcal{S}_{2}^{\text{out}} (t')  dt' \label{eq:VariableForSNR}
\end{equation}
where $\TBW$ is the time-constant of the lock-in filter. The phase $\theta$ of the lock-in is adjusted so that the response to the sinusoidal transverse magnetic field considered above is \cite{SuppInfo}:
\begin{equation}
\langle \mathfrak{K}  \rangle = \frac{\Phi}{4} N_{\text{at}} \gamma_F B_{0,\perp} G A_c(\nu).
\label{eq:SignalResponse}
\end{equation}
The noise in SNR quantifies the uncertainty in the magnetic field estimation and is given by the standard deviation of the $\mathfrak{K}$. This can be found from the integral over all frequencies of the measured noise spectrum filtered with a kernel function $ \mathscr{F}$ that depends on the frequency, the measurement time and lock-in filter: $\text{Var} \left[ \mathfrak{K} \right] =  \int_{-\infty}^{\infty} d \nu'  \mathscr{F}(\nu,\nu') S(\nu')$.  For spectra that exhibit minimal variation around the measurement frequency within the bandwidth defined by the measurement time $\text{BW}=1/(2T)$, the noise (variance) scales approximately as $1/T$, i.e., : $\text{Var} \left[ \mathfrak{K} \right] \approx S'(\nu)/(4 T)$, where $S'(\nu)=2S(\nu)$ is the single-sided spectrum (defined only for positive frequencies).

\subsection{Signal to noise ratio and magnetic sensitivity}

Then, the SNR of a magnetometer limited by spin projection noise and photon noise takes the form:
\begin{equation}
\text{SNR} = \frac{\langle \mathfrak{K}  \rangle}{\sqrt{\text{Var} \left[ \mathfrak{K} \right]}} \approx \frac{\gamma_F B_{0,\perp} A_c(\nu)}{\sqrt{\frac{2}{T}} \sqrt{\frac{2}{\Phi G^2 N_{\text{at}}^2}+\frac{\mathrm{S}'(\nu)}{N_{\text{at}}}}}, \label{eq:SNR0}
\end{equation}
where $\mathrm{S}'(\nu) = 2\mathrm{S}(\nu)$ is the single-sided spectrum. 
The magnetometer sensitivity, expressed in rms magnetic field units per square root bandwidth, is found from the rms value of the transverse magnetic field that gives a magnetometer response equal to noise (expressed as standard deviation) after a measurement time of $T=1/(2 \text{BW})$:
\begin{equation}
\frac{\delta B_{\text{rms}}}{\sqrt{\text{BW}}} = \frac{2 \sqrt{S'(\nu)}}{\Phi N_{\text{at}} \gamma_F G A_c(\nu)} = \frac{4 \sqrt{\frac{1}{\Phi G^2 N_{\text{at}}^2}+\frac{\mathrm{S}'(\nu)}{2 N_{\text{at}}}}}{\gamma_F A_c(\nu)}.
\end{equation}

\begin{figure}[t]
	\centering
\includegraphics[width=1\columnwidth]{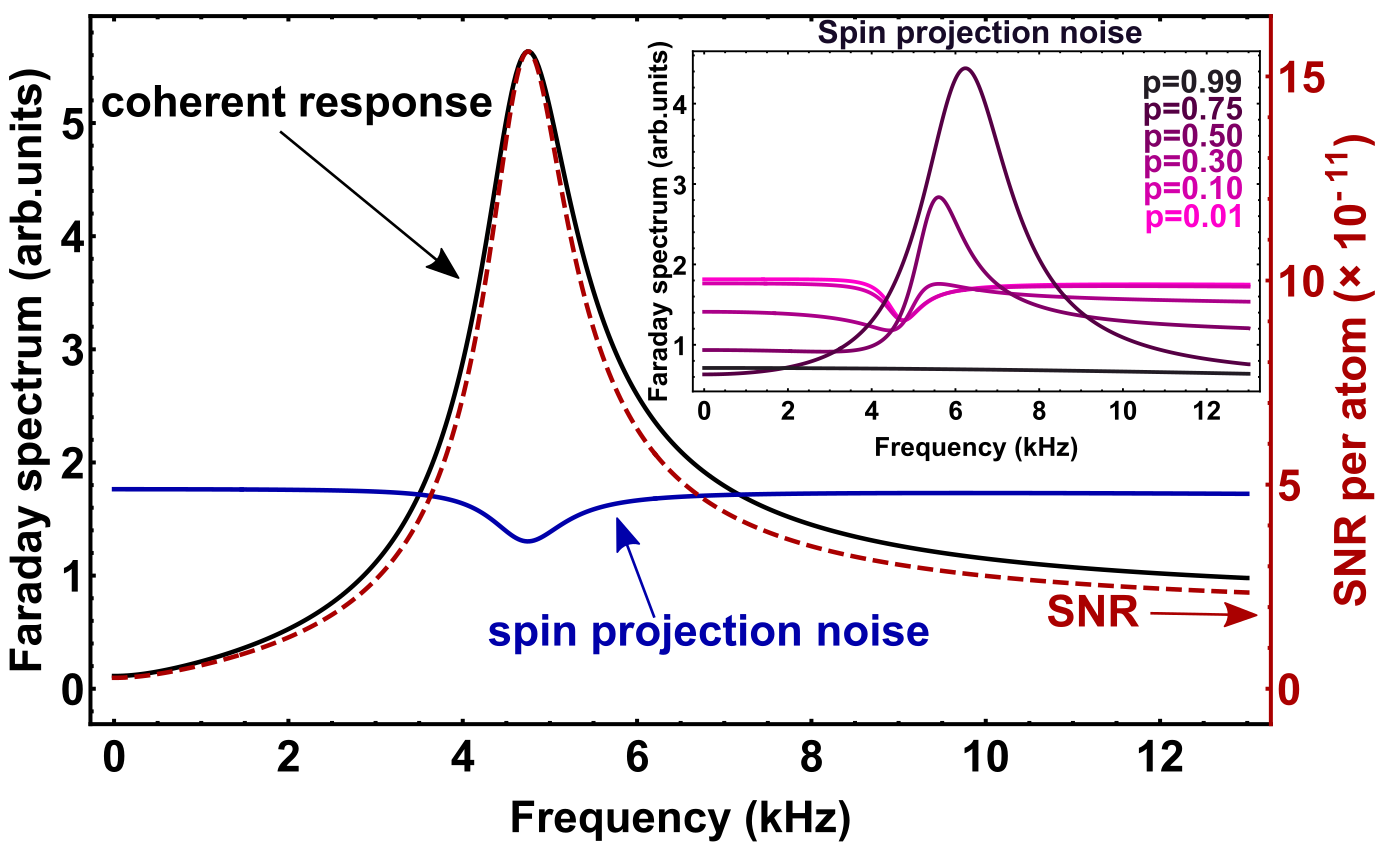}	
\caption{Calculation of the coherent response $A_c (\nu)$ (black solid line), the measured spin projection noise $\sqrt{S_{\mathcal{S}} (\nu)}$ (blue solid line) and the SNR spectra as given by Eq.\eqref{eq:SNR0} (dashed red line) for a \textsuperscript{87}Rb vapor at a state described by the spin-temperature distribution. The PSN is not taken into account. The spin polarization is $10 \%$ in a DC field of $10$ mG, and a transverse sinusoidal field with $\phi =0 $ and an amplitude $10^{12}$ times smaller than the DC. We assume a vapor at temperature $200$ \si{\celsius} corresponding to a \textsuperscript{87}Rb number density of $9.21 \times 10^{14}$ \si{\per\cubic\centi\meter}, a spin-exchange rate $\Rse= 8.4 \times 10^5$ \si{\per\second} and a 100-times smaller spin destruction rate. The detuning of $-5.7$ \si{\giga\hertz} from the $\rm{D}_1$ line and the optical linewidth (FWHM) of $1$ \si{\giga\hertz} are such that the noise dips are prominent \cite{PhysRevA.106.023112}. Tensor polarizability effects are less than $2\%$ of the vector polarizability and therefore negligible. The pumping rate is $\Rop= p\Rsd/(1-p)$ resulting in significant broadening of the magnetic linewidth at $p=0.99$ (see inset).  Inset shows the behavior of SPN spectra as a function of  spin polarization.} 
\label{fig:SNR}
\end{figure}

In Fig.\ref{fig:SNR} we plot the amplitude response $A_c (\nu)$, the noise  $\sqrt{S'(\nu)}$ and the $ \text{SNR}(\nu)$ as given by Eq.\eqref{eq:SNR0} for common experimental parameters (see caption). It is interesting to note that the measured spin-projection noise drops at the resonance frequency, where the response to a sinusoidal field is maximum. This feature appears only in the SERF regime, and vanishes at large magnetic fields where the Larmor precession occurs much faster than the spin-exchange collisions. In the SERF regime, frequent spin-exchange collisions force the two hyperfine spins to hybridize and precess with the same frequency. As a result, strong correlations are developed between the $a$ and $b$ spin manifolds which are positive at the common precession frequency. If the probe wavelength is tuned to measure strongly both manifolds and $D_a (\upnu) D_b (\upnu)>0$ the (large) cross-hyperfine terms in the spectrum are subtracted from the intra-hyperfine spin noise terms in Eq.~\ref{Eq:MeasuredSpectrum}, resulting in a reduction of the measured noise. At larger fields, the two hyperfine spins precess with opposite frequencies and their correlation is very weak.

In the inset of Fig.~\ref{fig:SNR}, we show how the reduction of SPN at frequencies close to the magnetic resonance depends on the ensemble polarization. At high spin polarization the cross-correlation features between the two hyperfine manifolds vanish due to the depletion of population in the lower hyperfine manifold. At the same time power broadening becomes detrimental for the magnetic resonance signal. 

In contrast to the noise spectrum, the coherent response of the magnetometer shows a Lorentzian-like behavior with a maximum response at the resonance frequency (see Fig.~\ref{fig:SNR}). It is interesting to ask whether the strong correlations between the two hyperfine states can be harnessed in order to enhance the magnetometer sensitivity. It turns out that reduction in noise is always associated with a corresponding reduction in the response of the measured signal, so that the SNR remains practically independent of the detuning if only spin-projection noise is considered.

\section{Conclusion}
\label{sec:conclusion}

In conclusion we have developed a theoretical framework for calculating noise spectra of spin-polarized atomic ensembles and for the first time determine analytically the SNR of an optically pumped magnetometer from first principles. Our model is based on the well developed mean-field theory of alkali-metal spins and addresses a broad range of experimental conditions encountered in sensitive magnetometry. A new SERF feature has been predicted resulting from noise redistribution due to strong hyperfine correlations in the ground electronic state. A new formula for the ultimate magnetic sensitivity is obtained. These findings have the possibility to improve quantum devices at the most fundamental level through SNR optimization and noise shaping.

\section{Acknowledgements}

GV acknowledges funding from EU QuantERA Project PACE-IN (GSRT Grant No. T11EPA4-00015) and from the Hellenic Foundation for Research and Innovation (HFRI) under the HFRI agreement  no. HFRI-00768 (project QCAT). KM acknowledges support from Grant FJC2021-047840-I funded by MCIN/AEI/ 10.13039/501100011033 and by the European Union ``NextGenerationEU/PRTR.''. MWM and KM acknowledge the European Commission project OPMMEG (101099379), Spanish Ministry of Science MCIN with funding from European Union NextGenerationEU (PRTR-C17.I1) and by Generalitat de Catalunya ``Severo Ochoa'' Center of Excellence CEX2019-000910-S; projects SAPONARIA (PID2021-123813NB-I00) and MARICHAS (PID2021-126059OA-I00) funded by MCIN/ AEI /10.13039/501100011033/ FEDER, EU; Generalitat de Catalunya through the CERCA program;  Ag\`{e}ncia de Gesti\'{o} d'Ajuts Universitaris i de Recerca Grant 2021-SGR-01453; Fundaci\'{o} Privada Cellex; Fundaci\'{o} Mir-Puig; I.K.K. acknowledges the cofinancing of this research by the European Union and Greek national funds through the Operational Program Crete 2020-2024, under the call ``Partnerships of Companies with Institutions for Research and Transfer of Knowledge in the Thematic Priorities of RIS3Crete", with project title ``Analyzing urban dynamics through monitoring the city magnetic environment'' (project KPHP1 - 0029067)

\bibliography{refs}

\begin{thebibliography}{49}%
\makeatletter
\providecommand \@ifxundefined [1]{%
 \@ifx{#1\undefined}
}%
\providecommand \@ifnum [1]{%
 \ifnum #1\expandafter \@firstoftwo
 \else \expandafter \@secondoftwo
 \fi
}%
\providecommand \@ifx [1]{%
 \ifx #1\expandafter \@firstoftwo
 \else \expandafter \@secondoftwo
 \fi
}%
\providecommand \natexlab [1]{#1}%
\providecommand \enquote  [1]{``#1''}%
\providecommand \bibnamefont  [1]{#1}%
\providecommand \bibfnamefont [1]{#1}%
\providecommand \citenamefont [1]{#1}%
\providecommand \href@noop [0]{\@secondoftwo}%
\providecommand \href [0]{\begingroup \@sanitize@url \@href}%
\providecommand \@href[1]{\@@startlink{#1}\@@href}%
\providecommand \@@href[1]{\endgroup#1\@@endlink}%
\providecommand \@sanitize@url [0]{\catcode `\\12\catcode `\$12\catcode `\&12\catcode `\#12\catcode `\^12\catcode `\_12\catcode `\%12\relax}%
\providecommand \@@startlink[1]{}%
\providecommand \@@endlink[0]{}%
\providecommand \url  [0]{\begingroup\@sanitize@url \@url }%
\providecommand \@url [1]{\endgroup\@href {#1}{\urlprefix }}%
\providecommand \urlprefix  [0]{URL }%
\providecommand \Eprint [0]{\href }%
\providecommand \doibase [0]{http://dx.doi.org/}%
\providecommand \selectlanguage [0]{\@gobble}%
\providecommand \bibinfo  [0]{\@secondoftwo}%
\providecommand \bibfield  [0]{\@secondoftwo}%
\providecommand \translation [1]{[#1]}%
\providecommand \BibitemOpen [0]{}%
\providecommand \bibitemStop [0]{}%
\providecommand \bibitemNoStop [0]{.\EOS\space}%
\providecommand \EOS [0]{\spacefactor3000\relax}%
\providecommand \BibitemShut  [1]{\csname bibitem#1\endcsname}%
\let\auto@bib@innerbib\@empty
\bibitem [{\citenamefont {Helstrom}(1976)}]{helstrom1976quantum}%
  \BibitemOpen
  \bibfield  {author} {\bibinfo {author} {\bibfnamefont {C.}~\bibnamefont {Helstrom}},\ }\href {https://books.google.es/books?id=fv9SAAAAMAAJ} {\emph {\bibinfo {title} {Quantum Detection and Estimation Theory}}},\ Mathematics in Science and Engineering : a series of monographs and textbooks\ (\bibinfo  {publisher} {Academic Press},\ \bibinfo {year} {1976})\BibitemShut {NoStop}%
\bibitem [{\citenamefont {Colangelo}\ \emph {et~al.}(2017)\citenamefont {Colangelo}, \citenamefont {Ciurana}, \citenamefont {Bianchet}, \citenamefont {Sewell},\ and\ \citenamefont {Mitchell}}]{Colangelo2017}%
  \BibitemOpen
  \bibfield  {author} {\bibinfo {author} {\bibfnamefont {G.}~\bibnamefont {Colangelo}}, \bibinfo {author} {\bibfnamefont {F.~M.}\ \bibnamefont {Ciurana}}, \bibinfo {author} {\bibfnamefont {L.~C.}\ \bibnamefont {Bianchet}}, \bibinfo {author} {\bibfnamefont {R.~J.}\ \bibnamefont {Sewell}}, \ and\ \bibinfo {author} {\bibfnamefont {M.~W.}\ \bibnamefont {Mitchell}},\ }\href {\doibase 10.1038/nature21434} {\bibfield  {journal} {\bibinfo  {journal} {Nature}\ }\textbf {\bibinfo {volume} {543}},\ \bibinfo {pages} {525} (\bibinfo {year} {2017})}\BibitemShut {NoStop}%
\bibitem [{\citenamefont {Savukov}\ \emph {et~al.}(2005)\citenamefont {Savukov}, \citenamefont {Seltzer}, \citenamefont {Romalis},\ and\ \citenamefont {Sauer}}]{PhysRevLett.95.063004}%
  \BibitemOpen
  \bibfield  {author} {\bibinfo {author} {\bibfnamefont {I.~M.}\ \bibnamefont {Savukov}}, \bibinfo {author} {\bibfnamefont {S.~J.}\ \bibnamefont {Seltzer}}, \bibinfo {author} {\bibfnamefont {M.~V.}\ \bibnamefont {Romalis}}, \ and\ \bibinfo {author} {\bibfnamefont {K.~L.}\ \bibnamefont {Sauer}},\ }\href {\doibase 10.1103/PhysRevLett.95.063004} {\bibfield  {journal} {\bibinfo  {journal} {Phys. Rev. Lett.}\ }\textbf {\bibinfo {volume} {95}},\ \bibinfo {pages} {063004} (\bibinfo {year} {2005})}\BibitemShut {NoStop}%
\bibitem [{\citenamefont {Ledbetter}\ \emph {et~al.}(2008)\citenamefont {Ledbetter}, \citenamefont {Savukov}, \citenamefont {Acosta}, \citenamefont {Budker},\ and\ \citenamefont {Romalis}}]{PhysRevA.77.033408}%
  \BibitemOpen
  \bibfield  {author} {\bibinfo {author} {\bibfnamefont {M.~P.}\ \bibnamefont {Ledbetter}}, \bibinfo {author} {\bibfnamefont {I.~M.}\ \bibnamefont {Savukov}}, \bibinfo {author} {\bibfnamefont {V.~M.}\ \bibnamefont {Acosta}}, \bibinfo {author} {\bibfnamefont {D.}~\bibnamefont {Budker}}, \ and\ \bibinfo {author} {\bibfnamefont {M.~V.}\ \bibnamefont {Romalis}},\ }\href {\doibase 10.1103/PhysRevA.77.033408} {\bibfield  {journal} {\bibinfo  {journal} {Phys. Rev. A}\ }\textbf {\bibinfo {volume} {77}},\ \bibinfo {pages} {033408} (\bibinfo {year} {2008})}\BibitemShut {NoStop}%
\bibitem [{\citenamefont {Deutsch}\ and\ \citenamefont {Jessen}(2010)}]{Deutsch2010}%
  \BibitemOpen
  \bibfield  {author} {\bibinfo {author} {\bibfnamefont {I.~H.}\ \bibnamefont {Deutsch}}\ and\ \bibinfo {author} {\bibfnamefont {P.~S.}\ \bibnamefont {Jessen}},\ }\href {\doibase 10.1016/j.optcom.2009.10.059} {\bibfield  {journal} {\bibinfo  {journal} {Optics Communications}\ }\textbf {\bibinfo {volume} {283}},\ \bibinfo {pages} {681} (\bibinfo {year} {2010})}\BibitemShut {NoStop}%
\bibitem [{\citenamefont {Tsang}(2023)}]{PhysRevA.107.012611}%
  \BibitemOpen
  \bibfield  {author} {\bibinfo {author} {\bibfnamefont {M.}~\bibnamefont {Tsang}},\ }\href {\doibase 10.1103/PhysRevA.107.012611} {\bibfield  {journal} {\bibinfo  {journal} {Phys. Rev. A}\ }\textbf {\bibinfo {volume} {107}},\ \bibinfo {pages} {012611} (\bibinfo {year} {2023})}\BibitemShut {NoStop}%
\bibitem [{\citenamefont {Mitchell}\ and\ \citenamefont {Palacios~Alvarez}(2020)}]{RevModPhys.92.021001}%
  \BibitemOpen
  \bibfield  {author} {\bibinfo {author} {\bibfnamefont {M.~W.}\ \bibnamefont {Mitchell}}\ and\ \bibinfo {author} {\bibfnamefont {S.}~\bibnamefont {Palacios~Alvarez}},\ }\href {\doibase 10.1103/RevModPhys.92.021001} {\bibfield  {journal} {\bibinfo  {journal} {Rev. Mod. Phys.}\ }\textbf {\bibinfo {volume} {92}},\ \bibinfo {pages} {021001} (\bibinfo {year} {2020})}\BibitemShut {NoStop}%
\bibitem [{\citenamefont {Degen}\ \emph {et~al.}(2017)\citenamefont {Degen}, \citenamefont {Reinhard},\ and\ \citenamefont {Cappellaro}}]{RevModPhys.89.035002}%
  \BibitemOpen
  \bibfield  {author} {\bibinfo {author} {\bibfnamefont {C.~L.}\ \bibnamefont {Degen}}, \bibinfo {author} {\bibfnamefont {F.}~\bibnamefont {Reinhard}}, \ and\ \bibinfo {author} {\bibfnamefont {P.}~\bibnamefont {Cappellaro}},\ }\href {\doibase 10.1103/RevModPhys.89.035002} {\bibfield  {journal} {\bibinfo  {journal} {Rev. Mod. Phys.}\ }\textbf {\bibinfo {volume} {89}},\ \bibinfo {pages} {035002} (\bibinfo {year} {2017})}\BibitemShut {NoStop}%
\bibitem [{\citenamefont {Troullinou}\ \emph {et~al.}(2021)\citenamefont {Troullinou}, \citenamefont {Jim\'enez-Mart\'{\i}nez}, \citenamefont {Kong}, \citenamefont {Lucivero},\ and\ \citenamefont {Mitchell}}]{PhysRevLett.127.193601}%
  \BibitemOpen
  \bibfield  {author} {\bibinfo {author} {\bibfnamefont {C.}~\bibnamefont {Troullinou}}, \bibinfo {author} {\bibfnamefont {R.}~\bibnamefont {Jim\'enez-Mart\'{\i}nez}}, \bibinfo {author} {\bibfnamefont {J.}~\bibnamefont {Kong}}, \bibinfo {author} {\bibfnamefont {V.~G.}\ \bibnamefont {Lucivero}}, \ and\ \bibinfo {author} {\bibfnamefont {M.~W.}\ \bibnamefont {Mitchell}},\ }\href {\doibase 10.1103/PhysRevLett.127.193601} {\bibfield  {journal} {\bibinfo  {journal} {Phys. Rev. Lett.}\ }\textbf {\bibinfo {volume} {127}},\ \bibinfo {pages} {193601} (\bibinfo {year} {2021})}\BibitemShut {NoStop}%
\bibitem [{\citenamefont {Troullinou}\ \emph {et~al.}(2023)\citenamefont {Troullinou}, \citenamefont {Lucivero},\ and\ \citenamefont {Mitchell}}]{PhysRevLett.131.133602}%
  \BibitemOpen
  \bibfield  {author} {\bibinfo {author} {\bibfnamefont {C.}~\bibnamefont {Troullinou}}, \bibinfo {author} {\bibfnamefont {V.~G.}\ \bibnamefont {Lucivero}}, \ and\ \bibinfo {author} {\bibfnamefont {M.~W.}\ \bibnamefont {Mitchell}},\ }\href {\doibase 10.1103/PhysRevLett.131.133602} {\bibfield  {journal} {\bibinfo  {journal} {Phys. Rev. Lett.}\ }\textbf {\bibinfo {volume} {131}},\ \bibinfo {pages} {133602} (\bibinfo {year} {2023})}\BibitemShut {NoStop}%
\bibitem [{\citenamefont {Jia}\ \emph {et~al.}(2023)\citenamefont {Jia}, \citenamefont {Novikov}, \citenamefont {Brasil}, \citenamefont {Zeuthen}, \citenamefont {M\"{u}ller},\ and\ \citenamefont {Polzik}}]{Jia2023}%
  \BibitemOpen
  \bibfield  {author} {\bibinfo {author} {\bibfnamefont {J.}~\bibnamefont {Jia}}, \bibinfo {author} {\bibfnamefont {V.}~\bibnamefont {Novikov}}, \bibinfo {author} {\bibfnamefont {T.~B.}\ \bibnamefont {Brasil}}, \bibinfo {author} {\bibfnamefont {E.}~\bibnamefont {Zeuthen}}, \bibinfo {author} {\bibfnamefont {J.~H.}\ \bibnamefont {M\"{u}ller}}, \ and\ \bibinfo {author} {\bibfnamefont {E.~S.}\ \bibnamefont {Polzik}},\ }\href {\doibase 10.1038/s41467-023-42059-y} {\bibfield  {journal} {\bibinfo  {journal} {Nature Communications}\ }\textbf {\bibinfo {volume} {14}} (\bibinfo {year} {2023}),\ 10.1038/s41467-023-42059-y}\BibitemShut {NoStop}%
\bibitem [{\citenamefont {Behbood}\ \emph {et~al.}(2014)\citenamefont {Behbood}, \citenamefont {Martin~Ciurana}, \citenamefont {Colangelo}, \citenamefont {Napolitano}, \citenamefont {T\'oth}, \citenamefont {Sewell},\ and\ \citenamefont {Mitchell}}]{PhysRevLett.113.093601}%
  \BibitemOpen
  \bibfield  {author} {\bibinfo {author} {\bibfnamefont {N.}~\bibnamefont {Behbood}}, \bibinfo {author} {\bibfnamefont {F.}~\bibnamefont {Martin~Ciurana}}, \bibinfo {author} {\bibfnamefont {G.}~\bibnamefont {Colangelo}}, \bibinfo {author} {\bibfnamefont {M.}~\bibnamefont {Napolitano}}, \bibinfo {author} {\bibfnamefont {G.}~\bibnamefont {T\'oth}}, \bibinfo {author} {\bibfnamefont {R.~J.}\ \bibnamefont {Sewell}}, \ and\ \bibinfo {author} {\bibfnamefont {M.~W.}\ \bibnamefont {Mitchell}},\ }\href {\doibase 10.1103/PhysRevLett.113.093601} {\bibfield  {journal} {\bibinfo  {journal} {Phys. Rev. Lett.}\ }\textbf {\bibinfo {volume} {113}},\ \bibinfo {pages} {093601} (\bibinfo {year} {2014})}\BibitemShut {NoStop}%
\bibitem [{\citenamefont {Behbood}\ \emph {et~al.}(2013)\citenamefont {Behbood}, \citenamefont {Colangelo}, \citenamefont {Martin~Ciurana}, \citenamefont {Napolitano}, \citenamefont {Sewell},\ and\ \citenamefont {Mitchell}}]{PhysRevLett.111.103601}%
  \BibitemOpen
  \bibfield  {author} {\bibinfo {author} {\bibfnamefont {N.}~\bibnamefont {Behbood}}, \bibinfo {author} {\bibfnamefont {G.}~\bibnamefont {Colangelo}}, \bibinfo {author} {\bibfnamefont {F.}~\bibnamefont {Martin~Ciurana}}, \bibinfo {author} {\bibfnamefont {M.}~\bibnamefont {Napolitano}}, \bibinfo {author} {\bibfnamefont {R.~J.}\ \bibnamefont {Sewell}}, \ and\ \bibinfo {author} {\bibfnamefont {M.~W.}\ \bibnamefont {Mitchell}},\ }\href {\doibase 10.1103/PhysRevLett.111.103601} {\bibfield  {journal} {\bibinfo  {journal} {Phys. Rev. Lett.}\ }\textbf {\bibinfo {volume} {111}},\ \bibinfo {pages} {103601} (\bibinfo {year} {2013})}\BibitemShut {NoStop}%
\bibitem [{\citenamefont {Vasilakis}\ \emph {et~al.}(2011)\citenamefont {Vasilakis}, \citenamefont {Shah},\ and\ \citenamefont {Romalis}}]{PhysRevLett.106.143601}%
  \BibitemOpen
  \bibfield  {author} {\bibinfo {author} {\bibfnamefont {G.}~\bibnamefont {Vasilakis}}, \bibinfo {author} {\bibfnamefont {V.}~\bibnamefont {Shah}}, \ and\ \bibinfo {author} {\bibfnamefont {M.~V.}\ \bibnamefont {Romalis}},\ }\href {\doibase 10.1103/PhysRevLett.106.143601} {\bibfield  {journal} {\bibinfo  {journal} {Phys. Rev. Lett.}\ }\textbf {\bibinfo {volume} {106}},\ \bibinfo {pages} {143601} (\bibinfo {year} {2011})}\BibitemShut {NoStop}%
\bibitem [{\citenamefont {Vasilakis}\ \emph {et~al.}(2015)\citenamefont {Vasilakis}, \citenamefont {Shen}, \citenamefont {Jensen}, \citenamefont {Balabas}, \citenamefont {Salart}, \citenamefont {Chen},\ and\ \citenamefont {Polzik}}]{Vasilakis2015}%
  \BibitemOpen
  \bibfield  {author} {\bibinfo {author} {\bibfnamefont {G.}~\bibnamefont {Vasilakis}}, \bibinfo {author} {\bibfnamefont {H.}~\bibnamefont {Shen}}, \bibinfo {author} {\bibfnamefont {K.}~\bibnamefont {Jensen}}, \bibinfo {author} {\bibfnamefont {M.}~\bibnamefont {Balabas}}, \bibinfo {author} {\bibfnamefont {D.}~\bibnamefont {Salart}}, \bibinfo {author} {\bibfnamefont {B.}~\bibnamefont {Chen}}, \ and\ \bibinfo {author} {\bibfnamefont {E.~S.}\ \bibnamefont {Polzik}},\ }\href {\doibase 10.1038/nphys3280} {\bibfield  {journal} {\bibinfo  {journal} {Nature Physics}\ }\textbf {\bibinfo {volume} {11}},\ \bibinfo {pages} {389} (\bibinfo {year} {2015})}\BibitemShut {NoStop}%
\bibitem [{\citenamefont {Kong}\ \emph {et~al.}(2020)\citenamefont {Kong}, \citenamefont {Jim{\'{e}}nez-Mart{\'{\i}}nez}, \citenamefont {Troullinou}, \citenamefont {Lucivero}, \citenamefont {T{\'{o}}th},\ and\ \citenamefont {Mitchell}}]{MitchellNatureCom}%
  \BibitemOpen
  \bibfield  {author} {\bibinfo {author} {\bibfnamefont {J.}~\bibnamefont {Kong}}, \bibinfo {author} {\bibfnamefont {R.}~\bibnamefont {Jim{\'{e}}nez-Mart{\'{\i}}nez}}, \bibinfo {author} {\bibfnamefont {C.}~\bibnamefont {Troullinou}}, \bibinfo {author} {\bibfnamefont {V.~G.}\ \bibnamefont {Lucivero}}, \bibinfo {author} {\bibfnamefont {G.}~\bibnamefont {T{\'{o}}th}}, \ and\ \bibinfo {author} {\bibfnamefont {M.~W.}\ \bibnamefont {Mitchell}},\ }\href {\doibase 10.1038/s41467-020-15899-1} {\bibfield  {journal} {\bibinfo  {journal} {Nature Communications}\ }\textbf {\bibinfo {volume} {\textbf{11}}} (\bibinfo {year} {2020}),\ 10.1038/s41467-020-15899-1}\BibitemShut {NoStop}%
\bibitem [{\citenamefont {Tsang}\ and\ \citenamefont {Caves}(2012)}]{PhysRevX.2.031016}%
  \BibitemOpen
  \bibfield  {author} {\bibinfo {author} {\bibfnamefont {M.}~\bibnamefont {Tsang}}\ and\ \bibinfo {author} {\bibfnamefont {C.~M.}\ \bibnamefont {Caves}},\ }\href {\doibase 10.1103/PhysRevX.2.031016} {\bibfield  {journal} {\bibinfo  {journal} {Phys. Rev. X}\ }\textbf {\bibinfo {volume} {2}},\ \bibinfo {pages} {031016} (\bibinfo {year} {2012})}\BibitemShut {NoStop}%
\bibitem [{\citenamefont {Budker}\ and\ \citenamefont {Romalis}(2007)}]{BudkerRomalis2007}%
  \BibitemOpen
  \bibfield  {author} {\bibinfo {author} {\bibfnamefont {D.}~\bibnamefont {Budker}}\ and\ \bibinfo {author} {\bibfnamefont {M.}~\bibnamefont {Romalis}},\ }\href {\doibase 10.1038/nphys566} {\bibfield  {journal} {\bibinfo  {journal} {Nature Physics}\ }\textbf {\bibinfo {volume} {3}},\ \bibinfo {pages} {227} (\bibinfo {year} {2007})}\BibitemShut {NoStop}%
\bibitem [{\citenamefont {Kuzmich}\ \emph {et~al.}(2000)\citenamefont {Kuzmich}, \citenamefont {Mandel},\ and\ \citenamefont {Bigelow}}]{PhysRevLett.85.1594}%
  \BibitemOpen
  \bibfield  {author} {\bibinfo {author} {\bibfnamefont {A.}~\bibnamefont {Kuzmich}}, \bibinfo {author} {\bibfnamefont {L.}~\bibnamefont {Mandel}}, \ and\ \bibinfo {author} {\bibfnamefont {N.~P.}\ \bibnamefont {Bigelow}},\ }\href {\doibase 10.1103/PhysRevLett.85.1594} {\bibfield  {journal} {\bibinfo  {journal} {Phys. Rev. Lett.}\ }\textbf {\bibinfo {volume} {85}},\ \bibinfo {pages} {1594} (\bibinfo {year} {2000})}\BibitemShut {NoStop}%
\bibitem [{\citenamefont {Napolitano}\ \emph {et~al.}(2011)\citenamefont {Napolitano}, \citenamefont {Koschorreck}, \citenamefont {Dubost}, \citenamefont {Behbood}, \citenamefont {Sewell},\ and\ \citenamefont {Mitchell}}]{Napolitano2011}%
  \BibitemOpen
  \bibfield  {author} {\bibinfo {author} {\bibfnamefont {M.}~\bibnamefont {Napolitano}}, \bibinfo {author} {\bibfnamefont {M.}~\bibnamefont {Koschorreck}}, \bibinfo {author} {\bibfnamefont {B.}~\bibnamefont {Dubost}}, \bibinfo {author} {\bibfnamefont {N.}~\bibnamefont {Behbood}}, \bibinfo {author} {\bibfnamefont {R.~J.}\ \bibnamefont {Sewell}}, \ and\ \bibinfo {author} {\bibfnamefont {M.~W.}\ \bibnamefont {Mitchell}},\ }\href {\doibase 10.1038/nature09778} {\bibfield  {journal} {\bibinfo  {journal} {Nature}\ }\textbf {\bibinfo {volume} {471}},\ \bibinfo {pages} {486–489} (\bibinfo {year} {2011})}\BibitemShut {NoStop}%
\bibitem [{\citenamefont {Brask}\ \emph {et~al.}(2015)\citenamefont {Brask}, \citenamefont {Chaves},\ and\ \citenamefont {Ko\l{}ody\ifmmode~\acute{n}\else \'{n}\fi{}ski}}]{PhysRevX.5.031010}%
  \BibitemOpen
  \bibfield  {author} {\bibinfo {author} {\bibfnamefont {J.~B.}\ \bibnamefont {Brask}}, \bibinfo {author} {\bibfnamefont {R.}~\bibnamefont {Chaves}}, \ and\ \bibinfo {author} {\bibfnamefont {J.}~\bibnamefont {Ko\l{}ody\ifmmode~\acute{n}\else \'{n}\fi{}ski}},\ }\href {\doibase 10.1103/PhysRevX.5.031010} {\bibfield  {journal} {\bibinfo  {journal} {Phys. Rev. X}\ }\textbf {\bibinfo {volume} {5}},\ \bibinfo {pages} {031010} (\bibinfo {year} {2015})}\BibitemShut {NoStop}%
\bibitem [{\citenamefont {Sewell}\ \emph {et~al.}(2012)\citenamefont {Sewell}, \citenamefont {Koschorreck}, \citenamefont {Napolitano}, \citenamefont {Dubost}, \citenamefont {Behbood},\ and\ \citenamefont {Mitchell}}]{PhysRevLett.109.253605}%
  \BibitemOpen
  \bibfield  {author} {\bibinfo {author} {\bibfnamefont {R.~J.}\ \bibnamefont {Sewell}}, \bibinfo {author} {\bibfnamefont {M.}~\bibnamefont {Koschorreck}}, \bibinfo {author} {\bibfnamefont {M.}~\bibnamefont {Napolitano}}, \bibinfo {author} {\bibfnamefont {B.}~\bibnamefont {Dubost}}, \bibinfo {author} {\bibfnamefont {N.}~\bibnamefont {Behbood}}, \ and\ \bibinfo {author} {\bibfnamefont {M.~W.}\ \bibnamefont {Mitchell}},\ }\href {\doibase 10.1103/PhysRevLett.109.253605} {\bibfield  {journal} {\bibinfo  {journal} {Phys. Rev. Lett.}\ }\textbf {\bibinfo {volume} {109}},\ \bibinfo {pages} {253605} (\bibinfo {year} {2012})}\BibitemShut {NoStop}%
\bibitem [{\citenamefont {Itano}\ \emph {et~al.}(1993)\citenamefont {Itano}, \citenamefont {Bergquist}, \citenamefont {Bollinger}, \citenamefont {Gilligan}, \citenamefont {Heinzen}, \citenamefont {Moore}, \citenamefont {Raizen},\ and\ \citenamefont {Wineland}}]{PhysRevA.47.3554}%
  \BibitemOpen
  \bibfield  {author} {\bibinfo {author} {\bibfnamefont {W.~M.}\ \bibnamefont {Itano}}, \bibinfo {author} {\bibfnamefont {J.~C.}\ \bibnamefont {Bergquist}}, \bibinfo {author} {\bibfnamefont {J.~J.}\ \bibnamefont {Bollinger}}, \bibinfo {author} {\bibfnamefont {J.~M.}\ \bibnamefont {Gilligan}}, \bibinfo {author} {\bibfnamefont {D.~J.}\ \bibnamefont {Heinzen}}, \bibinfo {author} {\bibfnamefont {F.~L.}\ \bibnamefont {Moore}}, \bibinfo {author} {\bibfnamefont {M.~G.}\ \bibnamefont {Raizen}}, \ and\ \bibinfo {author} {\bibfnamefont {D.~J.}\ \bibnamefont {Wineland}},\ }\href {\doibase 10.1103/PhysRevA.47.3554} {\bibfield  {journal} {\bibinfo  {journal} {Phys. Rev. A}\ }\textbf {\bibinfo {volume} {47}},\ \bibinfo {pages} {3554} (\bibinfo {year} {1993})}\BibitemShut {NoStop}%
\bibitem [{\citenamefont {Romalis}(2013)}]{RomalisQuantumNoiseChapterInBudkerBook2013}%
  \BibitemOpen
  \bibfield  {author} {\bibinfo {author} {\bibfnamefont {M.~V.}\ \bibnamefont {Romalis}},\ }\enquote {\bibinfo {title} {Quantum noise in atomic magnetometers},}\ in\ \href {\doibase 10.1017/CBO9780511846380.003} {\emph {\bibinfo {booktitle} {Optical Magnetometry}}},\ \bibinfo {editor} {edited by\ \bibinfo {editor} {\bibfnamefont {D.}~\bibnamefont {Budker}}\ and\ \bibinfo {editor} {\bibfnamefont {D.~F.}\ \bibnamefont {Jackson~Kimball}}}\ (\bibinfo  {publisher} {Cambridge University Press},\ \bibinfo {year} {2013})\ pp.\ \bibinfo {pages} {25--39}\BibitemShut {NoStop}%
\bibitem [{\citenamefont {Budker}\ and\ \citenamefont {Kozlov}(2020)}]{budker2020sensing}%
  \BibitemOpen
  \bibfield  {author} {\bibinfo {author} {\bibfnamefont {D.}~\bibnamefont {Budker}}\ and\ \bibinfo {author} {\bibfnamefont {M.~G.}\ \bibnamefont {Kozlov}},\ }\href@noop {} {\enquote {\bibinfo {title} {Sensing: Equation one},}\ } (\bibinfo {year} {2020}),\ \Eprint {http://arxiv.org/abs/2011.11043} {arXiv:2011.11043 [quant-ph]} \BibitemShut {NoStop}%
\bibitem [{\citenamefont {Mouloudakis}\ \emph {et~al.}(2022)\citenamefont {Mouloudakis}, \citenamefont {Vasilakis}, \citenamefont {Lucivero}, \citenamefont {Kong}, \citenamefont {Kominis},\ and\ \citenamefont {Mitchell}}]{PhysRevA.106.023112}%
  \BibitemOpen
  \bibfield  {author} {\bibinfo {author} {\bibfnamefont {K.}~\bibnamefont {Mouloudakis}}, \bibinfo {author} {\bibfnamefont {G.}~\bibnamefont {Vasilakis}}, \bibinfo {author} {\bibfnamefont {V.~G.}\ \bibnamefont {Lucivero}}, \bibinfo {author} {\bibfnamefont {J.}~\bibnamefont {Kong}}, \bibinfo {author} {\bibfnamefont {I.~K.}\ \bibnamefont {Kominis}}, \ and\ \bibinfo {author} {\bibfnamefont {M.~W.}\ \bibnamefont {Mitchell}},\ }\href {\doibase 10.1103/PhysRevA.106.023112} {\bibfield  {journal} {\bibinfo  {journal} {Phys. Rev. A}\ }\textbf {\bibinfo {volume} {106}},\ \bibinfo {pages} {023112} (\bibinfo {year} {2022})}\BibitemShut {NoStop}%
\bibitem [{\citenamefont {Happer}\ and\ \citenamefont {Tang}(1973)}]{Happer-Tang}%
  \BibitemOpen
  \bibfield  {author} {\bibinfo {author} {\bibfnamefont {W.}~\bibnamefont {Happer}}\ and\ \bibinfo {author} {\bibfnamefont {H.}~\bibnamefont {Tang}},\ }\href {https://link.aps.org/doi/10.1103/PhysRevLett.31.273} {\bibfield  {journal} {\bibinfo  {journal} {Phys. Rev. Lett.}\ }\textbf {\bibinfo {volume} {\textbf{31}}} (\bibinfo {year} {1973})}\BibitemShut {NoStop}%
\bibitem [{\citenamefont {Happer}\ and\ \citenamefont {Tam}(1977)}]{Happer-Tam}%
  \BibitemOpen
  \bibfield  {author} {\bibinfo {author} {\bibfnamefont {W.}~\bibnamefont {Happer}}\ and\ \bibinfo {author} {\bibfnamefont {A.~C.}\ \bibnamefont {Tam}},\ }\href@noop {} {\bibfield  {journal} {\bibinfo  {journal} {Phys. Rev. A}\ }\textbf {\bibinfo {volume} {\textbf{16}}} (\bibinfo {year} {1977})}\BibitemShut {NoStop}%
\bibitem [{\citenamefont {Savukov}\ and\ \citenamefont {Romalis}(2005)}]{SavukovRomalis}%
  \BibitemOpen
  \bibfield  {author} {\bibinfo {author} {\bibfnamefont {I.~M.}\ \bibnamefont {Savukov}}\ and\ \bibinfo {author} {\bibfnamefont {M.~V.}\ \bibnamefont {Romalis}},\ }\href {https://link.aps.org/doi/10.1103/PhysRevA.71.023405} {\bibfield  {journal} {\bibinfo  {journal} {Phys. Rev. A}\ }\textbf {\bibinfo {volume} {71}},\ \bibinfo {pages} {023405} (\bibinfo {year} {2005})}\BibitemShut {NoStop}%
\bibitem [{\citenamefont {Allred}\ \emph {et~al.}(2002)\citenamefont {Allred}, \citenamefont {Lyman}, \citenamefont {Kornack},\ and\ \citenamefont {Romalis}}]{Allred}%
  \BibitemOpen
  \bibfield  {author} {\bibinfo {author} {\bibfnamefont {J.~C.}\ \bibnamefont {Allred}}, \bibinfo {author} {\bibfnamefont {R.~N.}\ \bibnamefont {Lyman}}, \bibinfo {author} {\bibfnamefont {T.~W.}\ \bibnamefont {Kornack}}, \ and\ \bibinfo {author} {\bibfnamefont {M.~V.}\ \bibnamefont {Romalis}},\ }\href {https://link.aps.org/doi/10.1103/PhysRevLett.89.130801} {\bibfield  {journal} {\bibinfo  {journal} {Phys. Rev. Lett.}\ }\textbf {\bibinfo {volume} {\textbf{89}}},\ \bibinfo {pages} {130801} (\bibinfo {year} {2002})}\BibitemShut {NoStop}%
\bibitem [{\citenamefont {Kominis}\ \emph {et~al.}(2003)\citenamefont {Kominis}, \citenamefont {Kornack}, \citenamefont {Allred},\ and\ \citenamefont {Romalis}}]{Kominis2003}%
  \BibitemOpen
  \bibfield  {author} {\bibinfo {author} {\bibfnamefont {I.~K.}\ \bibnamefont {Kominis}}, \bibinfo {author} {\bibfnamefont {T.~W.}\ \bibnamefont {Kornack}}, \bibinfo {author} {\bibfnamefont {J.~C.}\ \bibnamefont {Allred}}, \ and\ \bibinfo {author} {\bibfnamefont {M.~V.}\ \bibnamefont {Romalis}},\ }\href {https://doi.org/10.1038/nature01484} {\bibfield  {journal} {\bibinfo  {journal} {Nature}\ }\textbf {\bibinfo {volume} {\textbf{422}}} (\bibinfo {year} {2003})}\BibitemShut {NoStop}%
\bibitem [{\citenamefont {Boto}\ \emph {et~al.}(2018)\citenamefont {Boto}, \citenamefont {Holmes}, \citenamefont {Leggett}, \citenamefont {Roberts}, \citenamefont {Shah}, \citenamefont {Meyer}, \citenamefont {Mu{\~n}oz}, \citenamefont {Mullinger}, \citenamefont {Tierney}, \citenamefont {Bestmann} \emph {et~al.}}]{boto}%
  \BibitemOpen
  \bibfield  {author} {\bibinfo {author} {\bibfnamefont {E.}~\bibnamefont {Boto}}, \bibinfo {author} {\bibfnamefont {N.}~\bibnamefont {Holmes}}, \bibinfo {author} {\bibfnamefont {J.}~\bibnamefont {Leggett}}, \bibinfo {author} {\bibfnamefont {G.}~\bibnamefont {Roberts}}, \bibinfo {author} {\bibfnamefont {V.}~\bibnamefont {Shah}}, \bibinfo {author} {\bibfnamefont {S.~S.}\ \bibnamefont {Meyer}}, \bibinfo {author} {\bibfnamefont {L.~D.}\ \bibnamefont {Mu{\~n}oz}}, \bibinfo {author} {\bibfnamefont {K.~J.}\ \bibnamefont {Mullinger}}, \bibinfo {author} {\bibfnamefont {T.~M.}\ \bibnamefont {Tierney}}, \bibinfo {author} {\bibfnamefont {S.}~\bibnamefont {Bestmann}},  \emph {et~al.},\ }\href@noop {} {\bibfield  {journal} {\bibinfo  {journal} {Nature}\ }\textbf {\bibinfo {volume} {\textbf{555}}},\ \bibinfo {pages} {657} (\bibinfo {year} {2018})}\BibitemShut {NoStop}%
\bibitem [{\citenamefont {Limes}\ \emph {et~al.}(2020)\citenamefont {Limes}, \citenamefont {Foley}, \citenamefont {Kornack}, \citenamefont {Caliga}, \citenamefont {McBride}, \citenamefont {Braun}, \citenamefont {Lee}, \citenamefont {Lucivero},\ and\ \citenamefont {Romalis}}]{PhysRevApplied.14.011002}%
  \BibitemOpen
  \bibfield  {author} {\bibinfo {author} {\bibfnamefont {M.}~\bibnamefont {Limes}}, \bibinfo {author} {\bibfnamefont {E.}~\bibnamefont {Foley}}, \bibinfo {author} {\bibfnamefont {T.}~\bibnamefont {Kornack}}, \bibinfo {author} {\bibfnamefont {S.}~\bibnamefont {Caliga}}, \bibinfo {author} {\bibfnamefont {S.}~\bibnamefont {McBride}}, \bibinfo {author} {\bibfnamefont {A.}~\bibnamefont {Braun}}, \bibinfo {author} {\bibfnamefont {W.}~\bibnamefont {Lee}}, \bibinfo {author} {\bibfnamefont {V.}~\bibnamefont {Lucivero}}, \ and\ \bibinfo {author} {\bibfnamefont {M.}~\bibnamefont {Romalis}},\ }\href {\doibase 10.1103/PhysRevApplied.14.011002} {\bibfield  {journal} {\bibinfo  {journal} {Phys. Rev. Appl.}\ }\textbf {\bibinfo {volume} {14}},\ \bibinfo {pages} {011002} (\bibinfo {year} {2020})}\BibitemShut {NoStop}%
\bibitem [{\citenamefont {Zhang}\ \emph {et~al.}(2020)\citenamefont {Zhang}, \citenamefont {Xiao}, \citenamefont {Ding}, \citenamefont {Feng}, \citenamefont {Peng}, \citenamefont {Shen}, \citenamefont {Sun}, \citenamefont {Wu}, \citenamefont {Wu}, \citenamefont {Yang}, \citenamefont {Zheng}, \citenamefont {Zhang}, \citenamefont {Chen},\ and\ \citenamefont {Guo}}]{Zhang2020}%
  \BibitemOpen
  \bibfield  {author} {\bibinfo {author} {\bibfnamefont {R.}~\bibnamefont {Zhang}}, \bibinfo {author} {\bibfnamefont {W.}~\bibnamefont {Xiao}}, \bibinfo {author} {\bibfnamefont {Y.}~\bibnamefont {Ding}}, \bibinfo {author} {\bibfnamefont {Y.}~\bibnamefont {Feng}}, \bibinfo {author} {\bibfnamefont {X.}~\bibnamefont {Peng}}, \bibinfo {author} {\bibfnamefont {L.}~\bibnamefont {Shen}}, \bibinfo {author} {\bibfnamefont {C.}~\bibnamefont {Sun}}, \bibinfo {author} {\bibfnamefont {T.}~\bibnamefont {Wu}}, \bibinfo {author} {\bibfnamefont {Y.}~\bibnamefont {Wu}}, \bibinfo {author} {\bibfnamefont {Y.}~\bibnamefont {Yang}}, \bibinfo {author} {\bibfnamefont {Z.}~\bibnamefont {Zheng}}, \bibinfo {author} {\bibfnamefont {X.}~\bibnamefont {Zhang}}, \bibinfo {author} {\bibfnamefont {J.}~\bibnamefont {Chen}}, \ and\ \bibinfo {author} {\bibfnamefont {H.}~\bibnamefont {Guo}},\ }\href {\doibase 10.1126/sciadv.aba8792} {\bibfield  {journal} {\bibinfo  {journal} {Science Advances}\ }\textbf {\bibinfo {volume} {6}} (\bibinfo {year}
  {2020}),\ 10.1126/sciadv.aba8792}\BibitemShut {NoStop}%
\bibitem [{\citenamefont {Bloch}\ \emph {et~al.}(2023)\citenamefont {Bloch}, \citenamefont {Shaham}, \citenamefont {Hochberg}, \citenamefont {Kuflik}, \citenamefont {Volansky},\ and\ \citenamefont {Katz}}]{Bloch2023}%
  \BibitemOpen
  \bibfield  {author} {\bibinfo {author} {\bibfnamefont {I.~M.}\ \bibnamefont {Bloch}}, \bibinfo {author} {\bibfnamefont {R.}~\bibnamefont {Shaham}}, \bibinfo {author} {\bibfnamefont {Y.}~\bibnamefont {Hochberg}}, \bibinfo {author} {\bibfnamefont {E.}~\bibnamefont {Kuflik}}, \bibinfo {author} {\bibfnamefont {T.}~\bibnamefont {Volansky}}, \ and\ \bibinfo {author} {\bibfnamefont {O.}~\bibnamefont {Katz}},\ }\href {\doibase 10.1038/s41467-023-41162-4} {\bibfield  {journal} {\bibinfo  {journal} {Nature Communications}\ }\textbf {\bibinfo {volume} {14}} (\bibinfo {year} {2023}),\ 10.1038/s41467-023-41162-4}\BibitemShut {NoStop}%
\bibitem [{\citenamefont {Vasilakis}\ \emph {et~al.}(2009)\citenamefont {Vasilakis}, \citenamefont {Brown}, \citenamefont {Kornack},\ and\ \citenamefont {Romalis}}]{VasilakisPRL2009}%
  \BibitemOpen
  \bibfield  {author} {\bibinfo {author} {\bibfnamefont {G.}~\bibnamefont {Vasilakis}}, \bibinfo {author} {\bibfnamefont {J.~M.}\ \bibnamefont {Brown}}, \bibinfo {author} {\bibfnamefont {T.~W.}\ \bibnamefont {Kornack}}, \ and\ \bibinfo {author} {\bibfnamefont {M.~V.}\ \bibnamefont {Romalis}},\ }\href {\doibase 10.1103/PhysRevLett.103.261801} {\bibfield  {journal} {\bibinfo  {journal} {Phys. Rev. Lett.}\ }\textbf {\bibinfo {volume} {103}},\ \bibinfo {pages} {261801} (\bibinfo {year} {2009})}\BibitemShut {NoStop}%
\bibitem [{\citenamefont {Mouloudakis}\ \emph {et~al.}(2023{\natexlab{a}})\citenamefont {Mouloudakis}, \citenamefont {Kong}, \citenamefont {Sierant}, \citenamefont {Arkin}, \citenamefont {Ruiz}, \citenamefont {Jiménez-Martínez},\ and\ \citenamefont {Mitchell}}]{mouloudakis2023anomalous}%
  \BibitemOpen
  \bibfield  {author} {\bibinfo {author} {\bibfnamefont {K.}~\bibnamefont {Mouloudakis}}, \bibinfo {author} {\bibfnamefont {J.}~\bibnamefont {Kong}}, \bibinfo {author} {\bibfnamefont {A.}~\bibnamefont {Sierant}}, \bibinfo {author} {\bibfnamefont {E.}~\bibnamefont {Arkin}}, \bibinfo {author} {\bibfnamefont {M.~H.}\ \bibnamefont {Ruiz}}, \bibinfo {author} {\bibfnamefont {R.}~\bibnamefont {Jiménez-Martínez}}, \ and\ \bibinfo {author} {\bibfnamefont {M.~W.}\ \bibnamefont {Mitchell}},\ }\href@noop {} {\enquote {\bibinfo {title} {Anomalous spin projection noise in a spin-exchange-relaxation-free alkali-metal vapor},}\ } (\bibinfo {year} {2023}{\natexlab{a}}),\ \Eprint {http://arxiv.org/abs/2307.16869} {arXiv:2307.16869 [physics.atom-ph]} \BibitemShut {NoStop}%
\bibitem [{\citenamefont {Mouloudakis}\ \emph {et~al.}(2023{\natexlab{b}})\citenamefont {Mouloudakis}, \citenamefont {Vouzinas}, \citenamefont {Margaritakis}, \citenamefont {Koutsimpela}, \citenamefont {Mouloudakis}, \citenamefont {Koutrouli}, \citenamefont {Skotiniotis}, \citenamefont {Tsironis}, \citenamefont {Loulakis}, \citenamefont {Mitchell}, \citenamefont {Vasilakis},\ and\ \citenamefont {Kominis}}]{PhysRevA.108.052822}%
  \BibitemOpen
  \bibfield  {author} {\bibinfo {author} {\bibfnamefont {K.}~\bibnamefont {Mouloudakis}}, \bibinfo {author} {\bibfnamefont {F.}~\bibnamefont {Vouzinas}}, \bibinfo {author} {\bibfnamefont {A.}~\bibnamefont {Margaritakis}}, \bibinfo {author} {\bibfnamefont {A.}~\bibnamefont {Koutsimpela}}, \bibinfo {author} {\bibfnamefont {G.}~\bibnamefont {Mouloudakis}}, \bibinfo {author} {\bibfnamefont {V.}~\bibnamefont {Koutrouli}}, \bibinfo {author} {\bibfnamefont {M.}~\bibnamefont {Skotiniotis}}, \bibinfo {author} {\bibfnamefont {G.~P.}\ \bibnamefont {Tsironis}}, \bibinfo {author} {\bibfnamefont {M.}~\bibnamefont {Loulakis}}, \bibinfo {author} {\bibfnamefont {M.~W.}\ \bibnamefont {Mitchell}}, \bibinfo {author} {\bibfnamefont {G.}~\bibnamefont {Vasilakis}}, \ and\ \bibinfo {author} {\bibfnamefont {I.~K.}\ \bibnamefont {Kominis}},\ }\href {\doibase 10.1103/PhysRevA.108.052822} {\bibfield  {journal} {\bibinfo  {journal} {Phys. Rev. A}\ }\textbf {\bibinfo {volume} {108}},\ \bibinfo {pages} {052822} (\bibinfo {year}
  {2023}{\natexlab{b}})}\BibitemShut {NoStop}%
\bibitem [{\citenamefont {Shah}\ \emph {et~al.}(2010)\citenamefont {Shah}, \citenamefont {Vasilakis},\ and\ \citenamefont {Romalis}}]{PhysRevLett.104.013601}%
  \BibitemOpen
  \bibfield  {author} {\bibinfo {author} {\bibfnamefont {V.}~\bibnamefont {Shah}}, \bibinfo {author} {\bibfnamefont {G.}~\bibnamefont {Vasilakis}}, \ and\ \bibinfo {author} {\bibfnamefont {M.~V.}\ \bibnamefont {Romalis}},\ }\href {\doibase 10.1103/PhysRevLett.104.013601} {\bibfield  {journal} {\bibinfo  {journal} {Phys. Rev. Lett.}\ }\textbf {\bibinfo {volume} {104}},\ \bibinfo {pages} {013601} (\bibinfo {year} {2010})}\BibitemShut {NoStop}%
\bibitem [{\citenamefont {Budker}\ \emph {et~al.}(2002)\citenamefont {Budker}, \citenamefont {Gawlik}, \citenamefont {Kimball}, \citenamefont {Rochester}, \citenamefont {Yashchuk},\ and\ \citenamefont {Weis}}]{RevModPhys.74.1153}%
  \BibitemOpen
  \bibfield  {author} {\bibinfo {author} {\bibfnamefont {D.}~\bibnamefont {Budker}}, \bibinfo {author} {\bibfnamefont {W.}~\bibnamefont {Gawlik}}, \bibinfo {author} {\bibfnamefont {D.~F.}\ \bibnamefont {Kimball}}, \bibinfo {author} {\bibfnamefont {S.~M.}\ \bibnamefont {Rochester}}, \bibinfo {author} {\bibfnamefont {V.~V.}\ \bibnamefont {Yashchuk}}, \ and\ \bibinfo {author} {\bibfnamefont {A.}~\bibnamefont {Weis}},\ }\href {\doibase 10.1103/RevModPhys.74.1153} {\bibfield  {journal} {\bibinfo  {journal} {Rev. Mod. Phys.}\ }\textbf {\bibinfo {volume} {74}},\ \bibinfo {pages} {1153} (\bibinfo {year} {2002})}\BibitemShut {NoStop}%
\bibitem [{\citenamefont {Happer}\ \emph {et~al.}(2010)\citenamefont {Happer}, \citenamefont {Jau},\ and\ \citenamefont {Walker}}]{HapperBook2010}%
  \BibitemOpen
  \bibfield  {author} {\bibinfo {author} {\bibfnamefont {W.}~\bibnamefont {Happer}}, \bibinfo {author} {\bibfnamefont {Y.}~\bibnamefont {Jau}}, \ and\ \bibinfo {author} {\bibfnamefont {T.}~\bibnamefont {Walker}},\ }\href@noop {} {\emph {\bibinfo {title} {Optically Pumped Atoms}}}\ (\bibinfo  {publisher} {Wiley},\ \bibinfo {year} {2010})\BibitemShut {NoStop}%
\bibitem [{\citenamefont {Happer}(1972)}]{RevModPhys.44.169}%
  \BibitemOpen
  \bibfield  {author} {\bibinfo {author} {\bibfnamefont {W.}~\bibnamefont {Happer}},\ }\href {\doibase 10.1103/RevModPhys.44.169} {\bibfield  {journal} {\bibinfo  {journal} {Rev. Mod. Phys.}\ }\textbf {\bibinfo {volume} {44}},\ \bibinfo {pages} {169} (\bibinfo {year} {1972})}\BibitemShut {NoStop}%
\bibitem [{\citenamefont {Gardiner}(2009)}]{gardiner2009stochastic}%
  \BibitemOpen
  \bibfield  {author} {\bibinfo {author} {\bibfnamefont {C.}~\bibnamefont {Gardiner}},\ }\href@noop {} {\emph {\bibinfo {title} {\textit{Stochastic methods}}}},\ Vol.\ \bibinfo {volume} {\textbf{4}}\ (\bibinfo  {publisher} {Springer Berlin},\ \bibinfo {year} {2009})\BibitemShut {NoStop}%
\bibitem [{Sup()}]{SuppInfo}%
  \BibitemOpen
  \href@noop {} {}\bibinfo {note} {See Supplemental Online Material at [URL will be inserted by publisher] for more details.}\BibitemShut {Stop}%
\bibitem [{\citenamefont {Happer}\ and\ \citenamefont {Mathur}(1967)}]{HapperMathur}%
  \BibitemOpen
  \bibfield  {author} {\bibinfo {author} {\bibfnamefont {W.}~\bibnamefont {Happer}}\ and\ \bibinfo {author} {\bibfnamefont {B.~S.}\ \bibnamefont {Mathur}},\ }\href@noop {} {\bibfield  {journal} {\bibinfo  {journal} {Phys. Rev.}\ }\textbf {\bibinfo {volume} {\textbf{163}}} (\bibinfo {year} {1967})}\BibitemShut {NoStop}%
\bibitem [{\citenamefont {Kozbial}\ \emph {et~al.}(2023)\citenamefont {Kozbial}, \citenamefont {Elson}, \citenamefont {Rushton}, \citenamefont {Akbar}, \citenamefont {Meraki}, \citenamefont {Jensen},\ and\ \citenamefont {Kolodynski}}]{kozbial2023spin}%
  \BibitemOpen
  \bibfield  {author} {\bibinfo {author} {\bibfnamefont {M.}~\bibnamefont {Kozbial}}, \bibinfo {author} {\bibfnamefont {L.}~\bibnamefont {Elson}}, \bibinfo {author} {\bibfnamefont {L.~M.}\ \bibnamefont {Rushton}}, \bibinfo {author} {\bibfnamefont {A.}~\bibnamefont {Akbar}}, \bibinfo {author} {\bibfnamefont {A.}~\bibnamefont {Meraki}}, \bibinfo {author} {\bibfnamefont {K.}~\bibnamefont {Jensen}}, \ and\ \bibinfo {author} {\bibfnamefont {J.}~\bibnamefont {Kolodynski}},\ }\href@noop {} {\enquote {\bibinfo {title} {Spin noise spectroscopy of an alignment-based atomic magnetometer},}\ } (\bibinfo {year} {2023}),\ \Eprint {http://arxiv.org/abs/2312.05577} {arXiv:2312.05577 [physics.atom-ph]} \BibitemShut {NoStop}%
\bibitem [{\citenamefont {Appelt}\ \emph {et~al.}(1998)\citenamefont {Appelt}, \citenamefont {Baranga}, \citenamefont {Erickson}, \citenamefont {Romalis}, \citenamefont {Young},\ and\ \citenamefont {Happer}}]{appelt}%
  \BibitemOpen
  \bibfield  {author} {\bibinfo {author} {\bibfnamefont {S.}~\bibnamefont {Appelt}}, \bibinfo {author} {\bibfnamefont {A.~B.~A.}\ \bibnamefont {Baranga}}, \bibinfo {author} {\bibfnamefont {C.~J.}\ \bibnamefont {Erickson}}, \bibinfo {author} {\bibfnamefont {M.~V.}\ \bibnamefont {Romalis}}, \bibinfo {author} {\bibfnamefont {A.~R.}\ \bibnamefont {Young}}, \ and\ \bibinfo {author} {\bibfnamefont {W.}~\bibnamefont {Happer}},\ }\href {\doibase 10.1103/PhysRevA.58.1412} {\bibfield  {journal} {\bibinfo  {journal} {Phys. Rev. A}\ }\textbf {\bibinfo {volume} {\textbf{58}}},\ \bibinfo {pages} {1412} (\bibinfo {year} {1998})}\BibitemShut {NoStop}%
\bibitem [{\citenamefont {Martin~Ciurana}\ \emph {et~al.}(2017)\citenamefont {Martin~Ciurana}, \citenamefont {Colangelo}, \citenamefont {Slodi\ifmmode~\check{c}\else \v{c}\fi{}ka}, \citenamefont {Sewell},\ and\ \citenamefont {Mitchell}}]{MartinPRL2017}%
  \BibitemOpen
  \bibfield  {author} {\bibinfo {author} {\bibfnamefont {F.}~\bibnamefont {Martin~Ciurana}}, \bibinfo {author} {\bibfnamefont {G.}~\bibnamefont {Colangelo}}, \bibinfo {author} {\bibfnamefont {L.}~\bibnamefont {Slodi\ifmmode~\check{c}\else \v{c}\fi{}ka}}, \bibinfo {author} {\bibfnamefont {R.~J.}\ \bibnamefont {Sewell}}, \ and\ \bibinfo {author} {\bibfnamefont {M.~W.}\ \bibnamefont {Mitchell}},\ }\href {\doibase 10.1103/PhysRevLett.119.043603} {\bibfield  {journal} {\bibinfo  {journal} {Physical Review Letters}\ }\textbf {\bibinfo {volume} {119}},\ \bibinfo {pages} {043603} (\bibinfo {year} {2017})}\BibitemShut {NoStop}%
\bibitem [{\citenamefont {Jim\'enez-Mart\'{\i}nez}\ \emph {et~al.}(2018)\citenamefont {Jim\'enez-Mart\'{\i}nez}, \citenamefont {Ko\l{}ody\ifmmode~\acute{n}\else \'{n}\fi{}ski}, \citenamefont {Troullinou}, \citenamefont {Lucivero}, \citenamefont {Kong},\ and\ \citenamefont {Mitchell}}]{PhysRevLett.120.040503}%
  \BibitemOpen
  \bibfield  {author} {\bibinfo {author} {\bibfnamefont {R.}~\bibnamefont {Jim\'enez-Mart\'{\i}nez}}, \bibinfo {author} {\bibfnamefont {J.}~\bibnamefont {Ko\l{}ody\ifmmode~\acute{n}\else \'{n}\fi{}ski}}, \bibinfo {author} {\bibfnamefont {C.}~\bibnamefont {Troullinou}}, \bibinfo {author} {\bibfnamefont {V.~G.}\ \bibnamefont {Lucivero}}, \bibinfo {author} {\bibfnamefont {J.}~\bibnamefont {Kong}}, \ and\ \bibinfo {author} {\bibfnamefont {M.~W.}\ \bibnamefont {Mitchell}},\ }\href {\doibase 10.1103/PhysRevLett.120.040503} {\bibfield  {journal} {\bibinfo  {journal} {Phys. Rev. Lett.}\ }\textbf {\bibinfo {volume} {120}},\ \bibinfo {pages} {040503} (\bibinfo {year} {2018})}\BibitemShut {NoStop}%
\end{thebibliography}%


\begin{thebibliography}{6}%
\makeatletter
\providecommand \@ifxundefined [1]{%
 \@ifx{#1\undefined}
}%
\providecommand \@ifnum [1]{%
 \ifnum #1\expandafter \@firstoftwo
 \else \expandafter \@secondoftwo
 \fi
}%
\providecommand \@ifx [1]{%
 \ifx #1\expandafter \@firstoftwo
 \else \expandafter \@secondoftwo
 \fi
}%
\providecommand \natexlab [1]{#1}%
\providecommand \enquote  [1]{``#1''}%
\providecommand \bibnamefont  [1]{#1}%
\providecommand \bibfnamefont [1]{#1}%
\providecommand \citenamefont [1]{#1}%
\providecommand \href@noop [0]{\@secondoftwo}%
\providecommand \href [0]{\begingroup \@sanitize@url \@href}%
\providecommand \@href[1]{\@@startlink{#1}\@@href}%
\providecommand \@@href[1]{\endgroup#1\@@endlink}%
\providecommand \@sanitize@url [0]{\catcode `\\12\catcode `\$12\catcode
  `\&12\catcode `\#12\catcode `\^12\catcode `\_12\catcode `\%12\relax}%
\providecommand \@@startlink[1]{}%
\providecommand \@@endlink[0]{}%
\providecommand \url  [0]{\begingroup\@sanitize@url \@url }%
\providecommand \@url [1]{\endgroup\@href {#1}{\urlprefix }}%
\providecommand \urlprefix  [0]{URL }%
\providecommand \Eprint [0]{\href }%
\providecommand \doibase [0]{http://dx.doi.org/}%
\providecommand \selectlanguage [0]{\@gobble}%
\providecommand \bibinfo  [0]{\@secondoftwo}%
\providecommand \bibfield  [0]{\@secondoftwo}%
\providecommand \translation [1]{[#1]}%
\providecommand \BibitemOpen [0]{}%
\providecommand \bibitemStop [0]{}%
\providecommand \bibitemNoStop [0]{.\EOS\space}%
\providecommand \EOS [0]{\spacefactor3000\relax}%
\providecommand \BibitemShut  [1]{\csname bibitem#1\endcsname}%
\let\auto@bib@innerbib\@empty
\bibitem [{\citenamefont {Happer}\ and\ \citenamefont {Tam}(1977)}]{HapperTam}%
  \BibitemOpen
  \bibfield  {author} {\bibinfo {author} {\bibfnamefont {W.}~\bibnamefont
  {Happer}}\ and\ \bibinfo {author} {\bibfnamefont {A.~C.}\ \bibnamefont
  {Tam}},\ }\href@noop {} {\bibfield  {journal} {\bibinfo  {journal} {Phys.
  Rev. A}\ }\textbf {\bibinfo {volume} {\textbf{16}}} (\bibinfo {year}
  {1977})}\BibitemShut {NoStop}%
\bibitem [{Note1()}]{Note1}%
  \BibitemOpen
  \bibinfo {note} {$|00SS\rangle = \protect \mathbb {I}/\protect \sqrt {[S]}$,
  $\protect \text {Tr}\left [ T_{\Lambda \mu }^{\protect \dag } (I I) e^{\beta
  I_z} \right ] = (-1)^{\mu } \sum _{m=-I}^I \langle I m| T_{\Lambda -\mu }(II)
  e^{\beta m} |Im \rangle $.}\BibitemShut {Stop}%
\bibitem [{Note2()}]{Note2}%
  \BibitemOpen
  \bibinfo {note} {We use the symmetry properties of the 9j symbols: \begin
  {equation} \begin {pmatrix} I & 1/2 & \protect \tilde {F} \\ I & 1/2 &
  \protect \tilde {F} \\ \Lambda & 1 & L \end {pmatrix} = \begin {pmatrix} I &
  I & \Lambda \\ 1/2 & 1/2 & 1 \\ \protect \tilde {F} & \protect \tilde {F} & L
  \end {pmatrix}= \begin {pmatrix} \protect \tilde {F} & \protect \tilde {F} &
  L \\ I & I & \Lambda \\ 1/2 & 1/2 & 1 \end {pmatrix}. \end
  {equation}}\BibitemShut {NoStop}%
\bibitem [{\citenamefont {Appelt}\ \emph {et~al.}(1998)\citenamefont {Appelt},
  \citenamefont {Baranga}, \citenamefont {Erickson}, \citenamefont {Romalis},
  \citenamefont {Young},\ and\ \citenamefont {Happer}}]{HapperAppelt}%
  \BibitemOpen
  \bibfield  {author} {\bibinfo {author} {\bibfnamefont {S.}~\bibnamefont
  {Appelt}}, \bibinfo {author} {\bibfnamefont {A.~B.~A.}\ \bibnamefont
  {Baranga}}, \bibinfo {author} {\bibfnamefont {C.~J.}\ \bibnamefont
  {Erickson}}, \bibinfo {author} {\bibfnamefont {M.~V.}\ \bibnamefont
  {Romalis}}, \bibinfo {author} {\bibfnamefont {A.~R.}\ \bibnamefont {Young}},
  \ and\ \bibinfo {author} {\bibfnamefont {W.}~\bibnamefont {Happer}},\ }\href
  {\doibase 10.1103/PhysRevA.58.1412} {\bibfield  {journal} {\bibinfo
  {journal} {Phys. Rev. A}\ }\textbf {\bibinfo {volume} {\textbf{58}}},\
  \bibinfo {pages} {1412} (\bibinfo {year} {1998})}\BibitemShut {NoStop}%
\bibitem [{Note3()}]{Note3}%
  \BibitemOpen
  \bibinfo {note} {The situation changes for buffer-gas-free atomic ensembles;
  however, the methodology presented here can be seamlessly applied even in
  this case.}\BibitemShut {Stop}%
\bibitem [{Note4()}]{Note4}%
  \BibitemOpen
  \bibinfo {note} {This approximation eliminates the non-linear Zeeman
  splitting from the analysis.}\BibitemShut {Stop}%
\end{thebibliography}%

\end{document}


\author{K. Mouloudakis}
\email[Corresponding author: ]{kostas.mouloudakis@icfo.eu}
\ICFO

\author{V. Koutrouli}
\UOC
\ITE

\author{I. K. Kominis}
\UOC

\author{M. W. Mitchell}
\ICFO
\ICREA

\author{G. Vasilakis}
\email[Corresponding author: ]{gvasilak@iesl.forth.gr}
\ITE

{\onecolumngrid
\centering
\noindent
{\Large {Supplementary Information for: \textit{Spin projection noise and the magnetic sensitivity of optically pumped magnetometers}}}
\maketitle
\date{\today}
}
{\onecolumngrid
\tableofcontents
}

\section{Preliminaries}
We will use the notation of \cite{HapperTam}. Equation * in \cite{HapperTam} will be refered to as HT*. The only change will be in the notation of the $X(FF')$ and $Y(FF')$ defined in HT65 and HT66; to avoid confusion we will adopt the notation $X_L(FF')$ and $Y_L(FF')$.  The operator $T_{LM}$ will always refer to the coupled basis. The multiplicity of any angular momentum operator $Q$ will be denoted as $[Q]=2Q+1$.

We will frequently encounter the quantity: $\langle 001 m| \rho \rangle$. We use HT64 and find:
\begin{equation}
\langle 001 m| \rho \rangle = \sum_{FF'} Y_1(FF') \langle 1 m F F' | \rho \rangle \approx Y_1(aa) \langle T_{1m}^{\dag}(aa) \rangle+Y_1(bb) \langle T_{1m}^{\dag}(bb) \rangle, \label{eq:rhoElectron}
\end{equation}
where the approximation refers to the fact that we neglect the hyperfine coherences so that for $F \neq F':  \langle L m F F' | \rho \rangle =0$.

Similarly, from HT63 and neglecting hyperfine coherences:
\begin{equation}
\langle 1 m 00 | \rho \rangle = \sum_{FF'} X_1(FF') \langle 1 m F F' | \rho \rangle \approx X_1(aa) \langle T_{1m}^{\dag}(aa) \rangle+X_1(bb) \langle T_{1m}^{\dag}(bb) \rangle. \label{eq:rhoNuclear}
\end{equation}

 We define the spherical tensor operators as in \cite{HapperTam}:
\begin{equation}
T_{LM}(KK')=  \sum_{m} \ket{Km} \bra{K' m-M} (-1)^{m-M-K'}
\times C^{LM}_{Km;K' M-m}.
\end{equation}

\subsection{Useful equations}
Here, we derive equations that will be used in the following derivations.

Inverting HT62 we find:
\begin{align}
&|\Lambda \mu j m \rangle =|\Lambda \mu II\rangle | j m SS \rangle = \sum_{L} \left[ |\Lambda II \rangle |j SS \rangle \right]_{L(\mu+m)} C_{\Lambda \mu; j m}^{L (\mu+m)} \label{eq:ConvertOperatorUncoupledToCoupled0} \\
&\overset{\text{HT61}}{=} \sqrt{[\Lambda] [j]} \sum_{L} \sum_{F F'} \sqrt{[F] [F']} C_{\Lambda \mu ;j m}^{L (\mu+m)}  \begin{pmatrix}
I & S & F \\ I & S & F' \\ \Lambda & j & L 
\end{pmatrix} |L (\mu+m) F F' \rangle. \label{eq:ConvertOperatorUncoupledToCoupled1}
\end{align}
Then:
\begin{equation}
\langle T^{\dag}_{1M}(\tilde{F} \tilde{F} )  |\Lambda \mu 1  m \rangle = \sqrt{3[\Lambda]} \sqrt{[\tilde{F}] [\tilde{F}]} C_{\Lambda \mu; 1 m}^{1 M} \begin{pmatrix}
I & 1/2 & \tilde{F} \\ I & 1/2 & \tilde{F} \\ \Lambda & 1 & 1 
\end{pmatrix}, \label{eq:PolarizationT1MGeneralAverage}
\end{equation}
and more generally:
\begin{equation}
\langle T^{\dag}_{LM}(\tilde{F} \tilde{F} )  |\Lambda \mu 1  m \rangle = \sqrt{3[\Lambda]} \sqrt{[\tilde{F}] [\tilde{F}]} C_{\Lambda \mu; 1 m}^{L M} \begin{pmatrix}
I & 1/2 & \tilde{F} \\ I & 1/2 & \tilde{F} \\ \Lambda & 1 & L 
\end{pmatrix}.
\end{equation}
From the 9J symbol properties (permutation of the first and second row) we find:
\begin{equation}
\begin{pmatrix}
I & 1/2 & I\pm 1/2 \\ I & 1/2 & I\pm 1/2 \\ \Lambda & 1 & L 
\end{pmatrix} = (-1)^{4I+2+L+\Lambda \pm 1} \begin{pmatrix}  I & 1/2 & I\pm 1/2 \\ I & 1/2 & I\pm 1/2 \\ \Lambda & 1 & L 
\end{pmatrix} = (-1)^{L+\Lambda+1} \begin{pmatrix}
I & 1/2 & I\pm 1/2 \\ I & 1/2 & I\pm 1/2 \\ L & 1 & L 
\end{pmatrix}  . \label{eq:Polarization9JSymmetry}
\end{equation}
Also, operator averages in the spin-temperature density matrix state give\footnote{$|00SS\rangle = \mathbb{I}/\sqrt{[S]}$, $\Tr \left[ T_{\Lambda \mu}^{\dag} (I I) e^{\beta I_z} \right] = (-1)^{\mu} \sum_{m=-I}^I \langle I m| T_{\Lambda -\mu}(II) e^{\beta m} |Im \rangle$.}:
\begin{equation}
\langle \Lambda \mu 0 0 | \rho_{\text{ST}} \rangle = \Tr \left[ T_{\Lambda \mu}^{\dag} (I I) \frac{e^{\beta I_z}}{Z(I,\beta)}\right] \Tr \left[ T_{0 0}^{\dag} (S S) \frac{e^{\beta S_z}}{Z(S,\beta)}\right] = \frac{1}{\sqrt{[S]}} \sum_{m=-I}^{I} (-1)^{m-I} C^{\Lambda 0}_{I m;I-m} \frac{e^{\beta m}}{Z(I,\beta)} \delta_{\mu 0},
\end{equation}
\begin{equation}
\langle 001m|\rho_{\text{ST}} \rangle = \frac{1}{\sqrt{[I]}} \frac{\tanh \left[ \beta/2\right]}{\sqrt{2}} \delta_{m 0}, \label{eq:Polarization1mSE}
\end{equation}
\begin{align}
& \Tr \left[ T_{LM}^{\dag}(FF) \rho_{\text{ST}} \right] = \sum_{m_s m_i} \langle m_i m_s | T_{LM}^{\dag}(FF) \rho_{\text{ST}} |m_i m_s \rangle \\
&=  \sum_{m_s m_i \mu} \langle m_i m_s  |F \mu-M \rangle \langle F \mu|  \rho_{\text{ST}} |m_i m_s \rangle (-1)^{\mu-M-F} C_{F\mu;F (M-\mu)}^{L M} \\
& =  \sum_{\substack{ m_s m_i \mu \\ \mu_i \mu_s \\ \mu_i' \mu_s'} } \langle m_i m_s  |\mu_i \mu_s \rangle   \langle \mu_i' \mu_s'|  \frac{e^{\beta (\hat{I}_z+\hat{S}_z)}}{Z} |m_i m_s \rangle   C_{I \mu_i;S \mu_s }^{F \mu-M} C_{I \mu_i';S \mu_s' }^{F \mu} (-1)^{\mu-M-F} C_{F\mu;F (M-\mu)}^{L M} \\
&=\sum_{m_s m_i  }  \frac{e^{\beta (m_i+m_s)}}{Z}  \left[ C_{I m_i;S m_s }^{F (m_i+m_s)} \right]^2   C_{F(m_i+m_s);F (-m_i-m_s)}^{L 0} (-1)^{m_i+m_s-F} \delta_{M0}
\end{align}

\section{Polarized atoms}
Here, we examine the equations of motion for the mean values (and through the regression formula the quantum correlations) when the atoms are polarized. We will consider the experimentally relevant case, where the density matrix deviates only a little from the spin-temperature density matrix $\rho_{\text{ST}}$ (we assume polarized spin ensemble along the $z$ direction):
\begin{equation}
\rho_{\text{ST}} = \frac{e^{\beta S_z}}{Z(1/2,\beta)} \frac{e^{\beta I_z}}{Z(I,\beta)}, \phantom{aa} Z(K,\beta) = \frac{\sinh \left[ \beta (K+1/2)\right]}{\sinh\left[ \beta/2\right]},
\end{equation}
where the spin-temperature parameter $\beta$ is related to the mean spin:
\begin{equation}
\langle K_z \rangle = \frac{(2K+1) \coth (\beta/2) \coth \left[ \beta (K+1/2) \right] -\coth^2(\beta/2) }{2} \tanh \frac{\beta}{2}
\end{equation}
We define the degree of the electron spin polarization as $p=2 | \langle K_z\rangle |$ for $K=1/2$. In this case, the density matrix can be written in the form:
\begin{equation}
\rho = \rho_{\text{ST}}+\tilde{\rho},
\end{equation}
where $\tilde{\rho}$ has the part that the density matrix that describes the perturbation from the spin-temperature state. Physically, the polarization of the ensemble is captured mainly in the spin-temperature part and the polarization in $\tilde{\rho}$ is small (see below for a quantitative statement).

Since $\Tr [ \rho_{\text{ST}}]=1$, the part of the density matrix $\tilde{\rho}$ is necessarily traceless, and cannot be considered as a density matrix. In this respect, $\Tr[\hat{A} \tilde{\rho} ]$ does not correspond to a mean value of an operator $\hat{A}$. In the following, we will use $\langle\tilde{ \hat{A}} \rangle$ and $\langle\hat{A} \rangle_{\text{ST}}$ to denote $\Tr [ \hat{A} \tilde{\rho}]$ and $\Tr [ \hat{A} \rho_{\text{ST}}]$ respectively. We note that for any operator: $\langle \hat{A} \rangle = \langle\tilde{ \hat{A}} \rangle+ \langle\hat{A} \rangle_{\text{ST}}$.

\subsection{Spin dynamics}

As described in the main text, the overall evolution of the single-atom density matrix due to spin-exchange, spin-destruction, optical pumping and Hamiltonian effects is given by:
\begin{equation}
 \frac{d  \rho}{d t} =  A_{\rm{hfs}}\frac{[ \mathbf{I} \cdot \mathbf{S},\rho]}{i\hbar}  +g_s \mu_{_B} \frac{[\mathbf{S} \cdot \mathbf{B} (t), \rho]}{i \hbar} + R_{\rm{sd}} (\rho_{\mathbf{0}} -\rho)+ R_{\mathrm{se}} (\rho_{\langle \mathbf{S} \rangle} - \rho) + R_{\text{op}} (\rho_{\mathbf{s}} - \rho). \label{eq:TotalDensityMatrixEquation}
 \end{equation} 
In the following, we express the r.h.s of each of the above terms in the spherical tensor representation in order to keep track of the different velocity contributions and distinguish between linear and non-linear terms. To facilitate clarity in the presentation, we break down the various contributions on the right-hand side of Eq.~\ref{eq:TotalDensityMatrixEquation} and analyze their individual effects. We write:
\begin{equation}
    \frac{d  \rho}{d t} = V_{\text{HF}}+V_{\text{MG}}+V_{\text{SD}}+V_{\text{SE}}+V_{\text{OP}},
\end{equation}
where:
\begin{equation}
V_{\text{HF}} = A_{\rm{hfs}}\frac{[ \mathbf{I} \cdot \mathbf{S},\rho]}{i\hbar}, \phantom{a} V_{\text{MG}} = g_s \mu_{_B} \frac{[\mathbf{S} \cdot \mathbf{B} (t), \rho]}{i \hbar}, \phantom{a} V_{\text{SD}} =R_{\rm{sd}} (\rho_{\mathbf{0}} -\rho), \phantom{a}  V_{\text{SE}} = R_{\mathrm{se}} (\rho_{\langle \mathbf{S} \rangle} - \rho), \phantom{a} V_{\text{OP}} = R_{\text{op}} (\rho_{\mathbf{s}} - \rho).
\end{equation}

\subsection{Spin-exchange}

Spin-exchange dynamics for a single species are described by the non-linear density matrix equation $\frac{d  \rho}{d t}=R_{\mathrm{se}} (\rho_{\langle \mathbf{S} \rangle} - \rho)$ (HT77):

\begin{equation}
V_{\text{SE}} = \sum_{\substack{\Lambda \mu m} }  \sqrt{2 [I]} \langle \Lambda \mu 00|\rho\rangle \langle 001m|\rho \rangle  |\Lambda \mu 1 m \rangle - \sum_{\substack{\Lambda \mu m} }  \langle \Lambda \mu 1 m |\rho\rangle |\Lambda \mu 1 m \rangle. \label{eq:Happer78Nonliner}
\end{equation}

From HT72 we find that:
\begin{align}
\rho & = \sum_{\Lambda \mu} |\Lambda \mu 00 \rangle \langle \Lambda \mu 00 |\rho \rangle + \sum_{\Lambda \mu m} |\Lambda \mu I I  \rangle \langle \Lambda \mu 1 m |\rho \rangle S_m\\
& \stackrel{\text{HT71}}{=} \sum_{\Lambda \mu} |\Lambda \mu 00 \rangle \langle \Lambda \mu 00 |\rho \rangle + \sum_{\Lambda \mu m} |\Lambda \mu 1 m \rangle \langle \Lambda \mu 1 m |\rho \rangle \\
& \Rightarrow  \sum_{\Lambda \mu m} |\Lambda \mu 1 m  \rangle \langle \Lambda \mu 1 m |\rho \rangle = \rho-\sum_{\Lambda \mu} |\Lambda \mu 00 \rangle \langle \Lambda \mu 00 |\rho \rangle.
\end{align}
We use the above result to write:
\begin{equation}
V_{\text{SE}} = - (\rho-\sum_{\Lambda \mu} |\Lambda \mu 00 \rangle \langle \Lambda \mu 00 |\rho \rangle)+ \sum_{\substack{\Lambda \mu m} }  \sqrt{2 [I]} \langle \Lambda \mu 00|\rho\rangle \langle 001m|\rho \rangle  |\Lambda \mu 1 m \rangle. \label{eq:Happer78NonlinerNewForm}
\end{equation}

The above equation is formulated using operators in the uncoupled basis. Nevertheless, the observables in the experiment are represented in the coupled basis. For calculations, it is therefore convenient to reframe Eq.~\ref{eq:Happer78NonlinerNewForm} using only operators in the coupled basis. This transformation can be achieved by utilizing the HT63, HT64 and Eq.~\ref{eq:ConvertOperatorUncoupledToCoupled1}. The density matrix equation due to spin-exchange is:
\begin{align}
&V_{\text{SE}}  = -\rho+\sum_{\Lambda=0}^{2I} \sum_{\mu=-\Lambda}^{\Lambda} \sum_{\substack{FF'\\f f'}} X_{\Lambda } (FF') X_{\Lambda } (ff') \langle T_{\Lambda \mu}^{\dag}(f f')\rangle T_{\Lambda \mu}(F F')+\sqrt{6 [I]} \sum_{\Lambda=0}^{2I} \sum_{\mu=-\Lambda}^{\Lambda} \sum_{m=-1}^{1} \sum_{\substack{FF'\\f f'\\ \Phi \Phi'}}  \sum_{K=|\Lambda-1|}^{\Lambda+1} \Bigg \{  \nonumber \\
& \sqrt{ [\Lambda] [\Phi] [\Phi'] } X_{\Lambda } (FF') \langle T_{\Lambda \mu}^{\dag}(F F')\rangle Y_1 (ff') \langle T_{1 m}^{\dag}(f f')\rangle   C_{\Lambda \mu; 1 m}^{K \mu+m} \begin{pmatrix}
I & S & \Phi \\ I & S & \Phi' \\ \Lambda & 1 & K 
\end{pmatrix} T_{K (\mu+m)} ( \Phi \Phi') \Bigg \}. \label{eq:SEGeneralAlternativeForm1}
\end{align}

Multiplying both sides of Eq.~\ref{eq:SEGeneralAlternativeForm1} with $T_{LM}^{\dag}(\tilde{F} \tilde{F})$ and taking the trace we find:
\begin{align} 
&\frac{1}{\Rse}\frac{d}{dt}\langle T_{LM}^{\dag}(\tilde{F} \tilde{F}) \rangle   = -\langle T_{LM}^{\dag}(\tilde{F} \tilde{F}) \rangle+  X_{L} (\tilde{F} \tilde{F}) \sum_{\substack{f f'}} X_{L } (ff') \langle T_{LM}^{\dag}(f f')\rangle+[\tilde{F}]\sqrt{6 [I]} \sum_{\Lambda=0}^{2I} \sum_{\mu=-\Lambda}^{\Lambda} \sum_{\substack{FF'\\f f'}} \Bigg \{  \nonumber \\ 
& \sqrt{ [\Lambda]} X_{\Lambda } (FF') \langle T_{\Lambda \mu}^{\dag}(F F')\rangle Y_1 (ff') \langle T_{1 (M-\mu)}^{\dag}(f f')\rangle  C_{\Lambda \mu; 1 (M-\mu)}^{L M} \begin{pmatrix}
I & S & \tilde{F} \\ I & S & \tilde{F} \\ \Lambda & 1 & L 
\end{pmatrix} \Bigg \}.  \label{eq:SEMeanEvolutionVeryGeneral00}
\end{align}
To derive the above equation, we used the orthogonality property: $\langle T^{\dag}_{LM}(FF') T_{lm}(ff')  \rangle=\delta_{Ll} \delta_{Mm} \delta_{Ff} \delta_{Ff'}$.

In the following we neglect the effect of hyperfine coherences in the evolution of the Zeeman coherences. This is physically justifiable on the basis that the hyperfine coherences oscillate with the hyperfine frequency, much faster than any other time-scale in a typical experiment. Then Eq.~\ref{eq:SEMeanEvolutionVeryGeneral00} takes the simpler form:
\begin{align} 
&\frac{1}{\Rse}\frac{d}{dt}\langle T_{LM}^{\dag}(\tilde{F} \tilde{F}) \rangle   = -\langle T_{LM}^{\dag}(\tilde{F} \tilde{F}) \rangle\nonumber \\
& +  X_{L} (\tilde{F} \tilde{F}) \left[  X_{L } (aa) \langle T_{LM}^{\dag}(aa)\rangle + X_{L } (bb) \langle T_{LM}^{\dag}(bb)\rangle  \right] 
+Y_1(\tilde{F} \tilde{F}) \left[ Y_1(aa) \langle T^{\dag}_{1M}(aa) \rangle+ Y_1(bb) \langle T^{\dag}_{1M}(bb) \rangle \right] \delta_{L1}  \nonumber \\
& +[\tilde{F}]\sqrt{6 [I]} \sum_{\Lambda=1}^{2I} \sum_{\mu=-\Lambda}^{\Lambda} \sum_{F,f} \Bigg \{ \sqrt{ [\Lambda]} X_{\Lambda } (FF) \langle T_{\Lambda \mu}^{\dag}(F F)\rangle Y_1 (ff) \langle T_{1 (M-\mu)}^{\dag}(f f)\rangle  C_{\Lambda \mu; 1 (M-\mu)}^{L M} \begin{pmatrix}
I & S & \tilde{F} \\ I & S & \tilde{F} \\ \Lambda & 1 & L 
\end{pmatrix} \Bigg \},  \label{eq:SEMeanEvolutionVeryGeneral}
\end{align}
where we used the properties:
\begin{equation}
C_{00; 1 M}^{L M}  = 1, 
\end{equation}
\begin{equation}
\sqrt{[F]} C_{Fm; F -m}^{0 0} (-1)^{m-F}  = 1, 
\end{equation}
\begin{equation}
X_0(FF) = \frac{[F]}{\sqrt{[S]}} W(FS0I;I F) = \sqrt{ \frac{[F]}{[I] [S]}},
\end{equation}
\begin{equation}
\sqrt{[a]} \langle T_{00}^{\dag}(aa)\rangle + \sqrt{[b]} \langle T_{00}^{\dag}(bb)\rangle  = 1, 
\end{equation}
\begin{equation}
\begin{pmatrix}
I & S & \tilde{F} \\ I & S & \tilde{F} \\ 0 & 1 & L 
\end{pmatrix} =  \frac{W(S,I,1,\tilde{F};\tilde{F},S)}{\sqrt{3[I]}} \delta_{L1} = \frac{W(I,\tilde{F},S,1;S, \tilde{F})}{\sqrt{3[I]}} \delta_{L1} = \frac{Y_1(\tilde{F} \tilde{F})}{[\tilde{F}] \sqrt{3}}  \delta_{L1}.
\end{equation}

The 9j symbols appearing in the evolution of operators have two rows identical (within the approximation of neglecting hyperfine coherences). Then, from Eq.~\ref{eq:Polarization9JSymmetry} the only non-zero 9j symbol terms are those which:
\begin{equation}
\Lambda+1+L \text{ is even} \Rightarrow \Lambda+L \text{ odd}.
\end{equation}
In addition, from the triangular condition for the 9j symbol or the Clebsch-Gordan coefficient, the terms in the third line of Eq.~\ref{eq:SEMeanEvolutionVeryGeneral} are non-zero only when $L-1\leq \Lambda \leq L+1$. Combining this with the condition $\Lambda+L \text{ odd}$, we find that the summation over $\Lambda$ gives nonzero terms only when $\Lambda=L\pm 1$.

For calculations it is convenient to express the 9j symbols in terms of 6j symbols (which are more readily available in Mathematica). Using the formula:
\begin{equation}
\begin{Bmatrix}c & f & 1 \\ \frac{1}{2} & \frac{1}{2} & g \end{Bmatrix} \begin{pmatrix}
a & b & c \\ d & e & f \\ \frac{1}{2} & \frac{1}{2} & 1
\end{pmatrix} = \frac{(-1)^{2g}}{3} \begin{Bmatrix}
a & b & c \\ \frac{1}{2} & g & e 
\end{Bmatrix} 
\begin{Bmatrix}
e & d & f \\ \frac{1}{2} & g & a
\end{Bmatrix}
-\frac{(-1)^{b+d-g}}{6 [c]}
\begin{Bmatrix}
a & b & c \\ e & d & \frac{1}{2}
\end{Bmatrix} \delta_{fc},
\end{equation}
where $g=f+\frac{1}{2}$ if $f=c$ and $g=\frac{f+c}{2}$ if $f \neq c$ and taking into account that the 9j symbol is non-zero only for $\Lambda=L\pm1$ (which means that the 9j symbol is zero for $L=\Lambda$), we find\footnote{We use the symmetry properties of the 9j symbols:
\begin{equation}
\begin{pmatrix}
I & 1/2 & \tilde{F} \\ I & 1/2 & \tilde{F} \\ \Lambda & 1 & L
\end{pmatrix}  = \begin{pmatrix} I & I & \Lambda \\ 1/2 & 1/2 & 1 \\ \tilde{F} & \tilde{F} & L \end{pmatrix}= \begin{pmatrix}  \tilde{F} & \tilde{F} & L \\ I & I & \Lambda \\ 1/2 & 1/2 & 1  \end{pmatrix}.
\end{equation}}:
\begin{equation}
\begin{pmatrix}
I & \frac{1}{2} & \tilde{F} \\ I & \frac{1}{2} & \tilde{F} \\ L\pm1 & 1 & L
\end{pmatrix} = -\frac{ \begin{Bmatrix}
\tilde{F} & \tilde{F} & L \\ \frac{1}{2} & L\pm \frac{1}{2} & I 
\end{Bmatrix} 
\begin{Bmatrix}
I & I & L\pm1 \\ \frac{1}{2} &  L\pm\frac{1}{2} & \tilde{F}
\end{Bmatrix}}{3\begin{Bmatrix}L & L\pm1 & 1 \\ \frac{1}{2} & \frac{1}{2} &  L\pm\frac{1}{2} \end{Bmatrix}}.
\end{equation}


\subsubsection{Linearization}
The third line in Eq.~\ref{eq:SEMeanEvolutionVeryGeneral} has nonlinear terms (the products of mean values of operators), whereas the first two lines only include linear terms. In the case of unpolarized atoms, linearization of the equations of motion entails disregarding the third line,  given that it involves the multiplication of two small quantities. However, this approach changes when considering (longitudinally) polarized atoms, where the small factors pertain solely to mean values of operators linked to transverse spins ($T_{LM}(FF)$ for $M\neq0$).
To address this, we write the mean value of each operator as a sum of two terms (see discussion above):
\begin{equation}
\langle T_{LM}(FF) \rangle = \langle T_{LM}(FF) \rangle_{\text{ST}}+ \langle \tilde{T}_{LM}(FF) \rangle
\end{equation}
The first term describes mean values of operators in the state defined by the spin-temperature distribution and is nonzero solely for longitudinal operators ($M=0$). The second term delineates deviations from the spin-temperature distribution and is small for all operators (irrespective of the rank or component). As in the case for unpolarized atoms,  we will disregard terms which are second order in these small quantities and linearize the equation of motions.

Based on the above, we find:
\begin{align}
&\langle T_{\Lambda \mu}^{\dag}(F F)\rangle \langle T_{1 (M-\mu)}^{\dag}(f f)\rangle = \langle T_{\Lambda \mu}^{\dag}(F F)\rangle_{\text{ST}} \langle T_{1 (M-\mu)}^{\dag}(f f)\rangle_{\text{ST}}+ \langle T_{\Lambda \mu}^{\dag}(F F)\rangle_{\text{ST}} \langle \tilde{T}_{1 (M-\mu)}^{\dag}(f f)\rangle \nonumber \\
&+ \langle \tilde{T}_{\Lambda \mu}^{\dag}(F F)\rangle \langle T_{1 (M-\mu)}^{\dag}(f f)\rangle_{\text{ST}}+\langle \tilde{T}_{\Lambda \mu}^{\dag}(F F)\rangle\langle \tilde{T}_{1 (M-\mu)}^{\dag}(f f)\rangle \\
&\approx \langle T_{\Lambda \mu}^{\dag}(F F)\rangle_{\text{ST}} \langle T_{1 (M-\mu)}^{\dag}(f f)\rangle_{\text{ST}}+ \langle T_{\Lambda \mu}^{\dag}(F F)\rangle_{\text{ST}} \langle \tilde{T}_{1 (M-\mu)}^{\dag}(f f)\rangle +\langle \tilde{T}_{\Lambda \mu}^{\dag}(F F)\rangle \langle T_{1 (M-\mu)}^{\dag}(f f)\rangle_{\text{ST}} \\
& = \langle T_{\Lambda 0}^{\dag}(F F)\rangle_{\text{ST}} \langle T_{1 0}^{\dag}(f f)\rangle_{\text{ST}} \delta_{\mu 0} \delta_{M 0} + \langle T_{\Lambda 0}^{\dag}(F F)\rangle_{\text{ST}} \langle \tilde{T}_{1 M}^{\dag}(f f)\rangle \delta_{\mu 0} +\langle \tilde{T}_{\Lambda M}^{\dag}(F F)\rangle \langle T_{1 0}^{\dag}(f f)\rangle_{\text{ST}} \delta_{\mu M}
\end{align}

The \textit{linearized} equations of motion for $M\neq0$ are written as:
\begin{align} 
&\frac{1}{\Rse}\frac{d}{dt}\langle T_{LM}^{\dag}(\tilde{F} \tilde{F}) \rangle   = -\langle T_{LM}^{\dag}(\tilde{F} \tilde{F}) \rangle\nonumber \\
& +  X_{L} (\tilde{F} \tilde{F}) \left[  X_{L } (aa) \langle T_{LM}^{\dag}(aa)\rangle + X_{L } (bb) \langle T_{LM}^{\dag}(bb)\rangle  \right] 
+Y_1(\tilde{F} \tilde{F}) \left[ Y_1(aa) \langle T^{\dag}_{1M}(aa) \rangle+ Y_1(bb) \langle T^{\dag}_{1M}(bb) \rangle \right] \delta_{L1}  \nonumber \\
& +[\tilde{F}]\sqrt{6 [I]} \sum_{\substack{\Lambda=L-1 \\ \Lambda \geq 1}}^{L+1} \sum_{F,f} \Bigg \{ \sqrt{ [\Lambda]} X_{\Lambda } (FF)  \langle T_{\Lambda 0}^{\dag}(F F)\rangle_{\text{ST}}  Y_1 (ff) \langle T_{1 M}^{\dag}(f f)\rangle  C_{\Lambda 0; 1 M}^{L M} \begin{pmatrix}
I & S & \tilde{F} \\ I & S & \tilde{F} \\ \Lambda & 1 & L 
\end{pmatrix} \Bigg \} \nonumber \\
& +[\tilde{F}]\sqrt{6 [I]} \sum_{\substack{\Lambda=L-1 \\ \Lambda \geq 1}}^{L+1}  \sum_{F,f} \Bigg \{ \sqrt{ [\Lambda]} X_{\Lambda } (FF) \langle T_{\Lambda M}^{\dag}(F F)\rangle Y_1 (ff) \rangle \langle T_{1 0}^{\dag}(f f)\rangle_{\text{ST}}  C_{\Lambda M; 1 0}^{L M} \begin{pmatrix}
I & S & \tilde{F} \\ I & S & \tilde{F} \\ \Lambda & 1 & L 
\end{pmatrix} \Bigg \}. \label{eq:SEMeanEvolutionLinear0}
\end{align}
In writing the above Equation we used the fact that for $M\neq0$:
\begin{equation}
\langle T_{LM}(FF) \rangle_{\text{ST}} =0 \Rightarrow \langle T_{LM}(FF) \rangle  =\langle \tilde{T}_{LM}(FF) \rangle.
\end{equation}

We highlight that in the case of polarized atoms, the dynamics intermingle operators of different ranks, i.e. differing values of $L$. However,  these dynamics do not introduce any intermingling of operators associated with different $M$ values.

From the properties of the Hermitian conjugate: $T_{LM}^{\dag}(FF)=(-1)^M T_{L-M}(FF)$ and the Clebsch-Gordan coefficient: $C^{c\gamma}_{a\alpha,b\beta}=(-1)^{a+b-c}C^{c-\gamma}_{a-\alpha,b-\beta}$, the equations of motion for the mean values $\langle T_{LM}(FF)\rangle$ follow exactly the same dynamics as for $\langle T_{LM}^{\dag}(FF)\rangle$.

Overall, we can express the linear dynamics in matrix form:
\begin{equation}
\frac{d \langle \vec{T}_M \rangle}{dt}\Bigg|_{\text{se}}=\mathcal{A}_{\text{SE}} \langle \vec{T}_M \rangle,
\end{equation} \label{eq:SpinExchangeEvolMatriEqu}
where:
\begin{equation}
\vec{T}_M = \left[  T_{1M}(aa),  T_{1M}(bb), T_{2M}(aa), T_{2M}(bb), ... T_{LM}(aa), T_{LM}(bb),...   \right], \text{ } L=1,2,...2I,
\end{equation}
where $a=I+1/2$ and $b=I-1/2$ denote the hyperfine manifold.

The spin-exchange evolution matrix does not depend on $M$ (it is identical for $M=\pm1$) and can be written in the block form:
\setcounter{MaxMatrixCols}{20}
\begin{equation}
\mathcal{A}_{\text{SE}}= \Rse \begin{pmatrix}
\mathbb{C}_{1} + \mathbb{A}_{1}+ \mathbb{A}_{0}(1)& \mathbb{A}_{+}(1) & 0_{2\times2}  &0_{2\times2} & ... & 0_{2\times2} & 0_{2\times2} & 0_{2\times2} & ...&  0_{2\times2} & 0_{2\times2} \\
\mathbb{A}_{1}+ \mathbb{A}_{-}(2) & \mathbb{A}_{0}(2) & \mathbb{A}_{+}(2) & 0_{2\times2} & ...  & 0_{2\times2} & 0_{2\times2} & 0_{2\times2} &...& 0_{2\times2} & 0_{2\times2} \\
\mathbb{A}_{1}& \mathbb{A}_{-}(3) & \mathbb{A}_{0}(3) & \mathbb{A}_{+}(3) &... & 0_{2\times2} & 0_{2\times2} & 0_{2\times2} & ... & 0_{2\times2} & 0_{2\times2} \\
. & . & . & . & . & . & . & . & .& .& .&\\
\mathbb{A}_{1}  & 0_{2\times2} & 0_{2\times2} & 0_{2\times2} & ...& \mathbb{A}_{-}(L) & \mathbb{A}_{0}(L) & \mathbb{A}_{+}(L) & ... & 0_{2\times2} & 0_{2\times2} \\
. & . & . & . & . & . & . & . & .& .& .&\\
\mathbb{A}_{1}  & 0_{2\times2} & 0_{2\times2} & 0_{2\times2} & 0_{2\times2} &0_{2\times2} & 0_{2\times2} & 0_{2\times2} & ... &\mathbb{A}_{-}(2I) & \mathbb{A}_{0}(2I)
\end{pmatrix}, \label{eq:SpinExchangeMatrix}
\end{equation} 
where the $2\times2$ matrices are:
\begin{equation}
\mathbb{A}_{\pm}(L)=\begin{pmatrix}
Z_{L\pm1}(a)X_{L\pm1}(aa)  & Z_{L\pm1}(a)X_{L\pm1}(bb)  \\
Z_{L\pm1}(b)X_{L\pm1}(aa)  & Z_{L\pm1}(b)X_{L\pm1}(bb)
\end{pmatrix}
\Big [ Y_1(aa) \langle T_{10}(aa) \rangle+ Y_1(bb) \langle T_{10}(aa) \rangle  \Big], \label{eq:EvolutionSubMatrix1}
\end{equation}
\begin{equation}
\mathbb{A}_{0}(L)=\begin{pmatrix}
-1+X_L(aa) X_L(aa)  & X_L(aa) X_L(bb)   \\
X_L(aa) X_L(bb) & -1+X_L(aa) X_L(aa)
\end{pmatrix}, \label{eq:EvolutionSubMatrix2}
\end{equation}
\begin{equation}
\mathbb{C}_{1}=\begin{pmatrix}
Y_1(aa) Y_1(aa)  &Y_1(aa) Y_1(bb)   \\
Y_1(aa) Y_1(bb) & Y_1(bb) Y_1(bb)
\end{pmatrix}, \label{eq:EvolutionSubMatrix3}
\end{equation}
and
\begin{equation}
\mathbb{A}_{1}=\left[ \mathbb{D}_{+} + \mathbb{D}_{+} \right] \begin{pmatrix}
Y_1(aa) & Y_1(bb) \\
Y_1(aa) & Y_1(bb)
\end{pmatrix},\label{eq:EvolutionSubMatrix4}
\end{equation}
with
\begin{equation}
\mathbb{D}_{\pm} =\begin{pmatrix}
 Z'_{L\pm1}(a) & 0 \\
 0 & Z'_{L\pm1}(b)
\end{pmatrix} \left[ X_{L\pm1}(aa)\langle T_{L\pm1,0}(aa)\rangle+X_{L\pm1}(bb)\langle T_{L\pm1,0}(bb)\rangle \right], \label{eq:EvolutionSubMatrix5}
\end{equation}
and:
\begin{equation}
    Z_{\Lambda }(F) = [F]\sqrt{6 [I]}  \sqrt{ [\Lambda]}  C_{\Lambda 0; 1 M}^{L M} \begin{pmatrix}
I & S & F \\ I & S & F \\ \Lambda & 1 & L 
\end{pmatrix}
\end{equation}
\begin{equation}
    Z'_{\Lambda}(F) = [F]\sqrt{6 [I]}  \sqrt{ [\Lambda]}  C_{\Lambda M; 1 0}^{L M} \begin{pmatrix}
I & S & F \\ I & S & F \\ \Lambda & 1 & L 
\end{pmatrix}
\end{equation}

\subsection{Spin destruction and Optical pumping}

\subsubsection{Spin destruction}
We consider ``S-damping'' processes causing spin-relaxation, i.e. processes that part of the density matrix $\rho$ with electron polarization but does not affect the part with purely nuclear polarization \cite{HapperAppelt}:
\begin{equation}
V_{\text{SD}} = \left( \alpha - \rho \right) \overset{\text{HT74}}{\scalebox{1}{=}} -\rho+\sum_{\Lambda=0}^{2I} \sum_{\mu=-\Lambda}^{\Lambda}  \langle \Lambda \mu 00 |\rho \rangle |\lambda\mu 00 \rangle. \label{eq:SpinDestruction0}
\end{equation}
As above, we use HT63 and express $|\lambda\mu 00 \rangle$ in the form of a sum of operators in the coupled spherical basis. We ignore hyperfine coherences and find:
\begin{align}
\frac{1}{R_{\text{sd}}} \frac{d}{dt} \langle T_{LM}(FF) \rangle & = - \langle  T_{LM}(FF) \rangle + \sum_f X_L(ff) X_L(FF) \langle T_{LM}(ff) \rangle \\
&= - \langle  T_{LM}(FF) \rangle + \left[X_L(aa) \langle T_{LM}(aa) \rangle  +X_L(bb)  \langle T_{LM}(bb) \rangle \right]X_L(FF). \label{eq:SpinDestruction1}
\end{align}

Using the definition of Eq.~\ref{eq:EvolutionSubMatrix2} we write:
\begin{equation}
    \frac{d \langle \vec{T}_M \rangle}{dt}\Bigg|_{\text{sd}} = \mathcal{A}_{\text{SD},M} \langle \vec{T}_M \rangle,
\end{equation}
where:
\begin{equation}
\mathcal{A}_{\text{SD},M}= R_{\text{sd}} \begin{pmatrix}
 \mathbb{A}_{0}(1)&  0_{2\times2}  &0_{2\times2} & ... & 0_{2\times2} & ... & 0_{2\times2} & 0_{2\times2} \\
0_{2\times2} & \mathbb{A}_{0}(2) &  0_{2\times2} & ...  & 0_{2\times2} & ... & 0_{2\times2} & 0_{2\times2} \\
. & . & . & . & . & . & . & . \\
 0_{2\times2}  & 0_{2\times2} & 0_{2\times2}  & ...& \mathbb{A}_{0}(L) &  ... & 0_{2\times2} & 0_{2\times2} \\
. & . & . & . & . & . & . & . \\
 0_{2\times2}  & 0_{2\times2} & 0_{2\times2} & ... & 0_{2\times2} &... & 0_{2\times2} & \mathbb{A}_{0}(2I)
\end{pmatrix}. \label{eq:SpinDestrMatrix}
\end{equation}

\subsubsection{Optical pumping}
Here, we analyze the case where the electron spin of the excited atom (produced from optical pumping) undergoes complete depolarization, while the nuclear spin remains polarized before the atom reverts to its ground state \cite{HapperAppelt}.  This  situation is  encountered at the typical buffer gas pressures employed in experiments\footnote{The situation changes for buffer-gas-free atomic ensembles; however, the methodology presented here can be seamlessly applied even in this case.}.

The density matrix evolves according to the following equation \cite{HapperAppelt}:
\begin{equation}
V_{\text{OP}} = \left( \alpha - \rho \right)+2\alpha \mathbf{s} \mathbf{S}, \label{eq:OpticalPumping0}
\end{equation}
where $\mathbf{s}$ is the mean photon spin of the pump light and $\mathbf{S}$ is the electron spin operator. The first two in the right hand side of Eq.~\ref{eq:OpticalPumping0} are identical to those describing spin-destruction (see Eq.~\ref{eq:SpinDestruction0}). We concentrate on the third term and it as:
\begin{align}
& \alpha \mathbf{s} \mathbf{S}  = \alpha \sum_{m=-1}^{1}(-1)^m s_{-m} S_m = \sum_{m=-1}^{1} \sum_{\Lambda=0}^{2I} \sum_{\mu=-\Lambda}^{\Lambda}  s_{-m} \langle \Lambda \mu 00 |\rho \rangle |\Lambda\mu 00 \rangle S_m \\
 & =\sum_{m=-1}^{1}  \sum_{\Lambda=0}^{2I} \sum_{\mu=-\Lambda}^{\Lambda} (-1)^m s_{-m} \langle \Lambda \mu 00 |\rho \rangle \left( |\Lambda \mu II \rangle \otimes |00SS\rangle \right) (\mathbb{I} \otimes |1mSS\rangle/\sqrt{2} ) \\
&=  \sum_{m=-1}^{1} \sum_{\Lambda=0}^{2I} \sum_{\mu=-\Lambda}^{\Lambda}  (-1)^m s_{-m } \langle \Lambda \mu 00 |\rho \rangle \left( |\Lambda \mu II \rangle \otimes \mathbb{I}/\sqrt{2S+1} \right)(\mathbb{I} \otimes |1mSS\rangle/\sqrt{2} ) \\
&=  \sum_{m=-1}^{1} \sum_{\Lambda=0}^{2I} \sum_{\mu=-\Lambda}^{\Lambda}  (-1)^m \frac{s_{-m }}{\sqrt{2(2S+1)}} \langle \Lambda \mu 00 |\rho \rangle \left( |\Lambda \mu II \rangle \otimes |1mSS\rangle \right) \\
&=\sum_{m=-1}^{1} \sum_{\Lambda=0}^{2I} \sum_{\mu=-\Lambda}^{\Lambda}  (-1)^m \frac{s_{-m }}{2} \langle \Lambda \mu 00 |\rho \rangle  |\Lambda \mu 1m \rangle  \\
 &= \sum_{m=-1}^{1} \sum_{\Lambda=0}^{2I} \sum_{\mu=-\Lambda}^{\Lambda}  \sum_{K} \sum_{\substack{FF'\\f f'}} (-1)^m \frac{s_{-m }}{2} \sqrt{3[ \Lambda] } \sqrt{[F] [F']} X_{\Lambda}(ff') \langle T^{\dag}_{\Lambda \mu}(ff') \rangle C_{\Lambda \mu ;1 m}^{K (\mu+m)}  \begin{pmatrix}
I & S & F \\ I & S & F' \\ \Lambda & 1 & K 
\end{pmatrix} |K (\mu+m) F F' \rangle.
\end{align}
To derive the above equations we used:
\begin{equation}
T_{1M}(SS) = \frac{\sqrt{3}}{\sqrt{s(s+1)(2s+1)}} S_m, M=0,\pm1
\end{equation}
and
\begin{equation}
T_{00}(SS)=\frac{1}{\sqrt{2S+1}}\mathbb{I}.
\end{equation}

Neglecting hyperfine coherences, we find for the dynamics of optical pumping:
\begin{align}
&\frac{1}{R_{\text{op}}} \frac{d}{dt} \langle  T_{LM}(FF) \rangle  = - \langle  T_{LM}(FF) \rangle + \left[X_L(aa) \langle T_{LM}(aa) \rangle  +X_L(bb)  \langle T_{LM}(bb) \rangle \right]X_L(FF)\\
&+\sum_{m=-1}^{1} \sum_{\Lambda=L-1}^{L+1} (-1)^m  s_{-m } \sqrt{3[ \Lambda] } [F] \left[ X_{\Lambda}(aa) \langle T_{\Lambda (M-m)}(aa) \rangle+ X_{\Lambda}(bb) \langle T_{\Lambda (M-m)}(bb) \rangle \right] C_{\Lambda (M-m) ;1 m}^{L M}  \begin{pmatrix}
I & S & F \\ I & S & F \\ \Lambda & 1 & L \end{pmatrix}. \label{eq:OpticalPumpingF00}
\end{align}

In a typical experiment the pump beam lies along the longitudinal $z$ direction, so that $s_m = s_0 \delta_{m0}$. Then we find:
\begin{align}
&\frac{1}{R_{\text{op}}} \frac{d}{dt} \langle  T_{LM}(FF) \rangle  = - \langle  T_{LM}(FF) \rangle + \left[X_L(aa) \langle T_{LM}(aa) \rangle  +X_L(bb)  \langle T_{LM}(bb) \rangle \right]X_L(FF)\\
&+ s_0\sum_{\Lambda=L-1}^{L+1}  \sqrt{3[ \Lambda] } [F] \left[ X_{\Lambda}(aa) \langle T_{\Lambda M}(aa) \rangle+ X_{\Lambda}(bb) \langle T_{\Lambda M}(bb) \rangle \right] C_{\Lambda M ;1 0}^{L M}  \begin{pmatrix}
I & S & F \\ I & S & F \\ \Lambda & 1 & L \end{pmatrix}. \label{eq:OpticalPumpingF0}
\end{align}

In matrix form:
\begin{equation}
   \frac{d \langle \vec{T}_M \rangle}{dt}\Bigg|_{\text{op}} = \mathcal{A}_{\text{OP}} \langle \vec{T}_M \rangle, 
\end{equation}
where:
\setcounter{MaxMatrixCols}{20}
\begin{equation}
\mathcal{A}_{\text{OP}}= R_{\text{op}} \begin{pmatrix}
 \mathbb{A}_{0}(1)& \mathbb{A}'_{+}(1) & 0_{2\times2}  &0_{2\times2} & ... & 0_{2\times2} & 0_{2\times2} & 0_{2\times2} & ...&  0_{2\times2} & 0_{2\times2} \\
\mathbb{A}'_{-}(2) & \mathbb{A}_{0}(2) & \mathbb{A}'_{+}(2) & 0_{2\times2} & ...  & 0_{2\times2} & 0_{2\times2} & 0_{2\times2} &...& 0_{2\times2} & 0_{2\times2} \\
. & . & . & . & . & . & . & . & .& .& .&\\
0_{2\times2}  & 0_{2\times2} & 0_{2\times2} & 0_{2\times2} & ...& \mathbb{A}'_{-}(L) & \mathbb{A}_{0}(L) & \mathbb{A}'_{+}(L) & ... & 0_{2\times2} & 0_{2\times2} \\
. & . & . & . & . & . & . & . & .& .& .&\\
0_{2\times2}  & 0_{2\times2} & 0_{2\times2} & 0_{2\times2} & 0_{2\times2} &0_{2\times2} & 0_{2\times2} & 0_{2\times2} & ... &\mathbb{A}'_{-}(2I) & \mathbb{A}_{0}(2I)
\end{pmatrix}, \label{eq:OpticalPumpingMatrix}
\end{equation} 
and:
\begin{equation}
\mathbb{A}'_{\pm}(L)=\frac{s_0}{\sqrt{2[I]}}\begin{pmatrix}
Z_{L\pm1}(a)X_{L\pm1}(aa)  & Z_{L\pm1}(a)X_{L\pm1}(bb)  \\
Z_{L\pm1}(b)X_{L\pm1}(aa)  & Z_{L\pm1}(b)X_{L\pm1}(bb)
\end{pmatrix}. \label{eq:EvolutionSubMatrix10}
\end{equation}


\subsection{Magnetic field}
The evolution of the operators due to magnetic field is most easily found using the Heisenberg equation. 

\subsubsection{Longitudinal field}
We will neglect the coupling of the magnetic field to the nuclear spin and only consider the coupling to the electron spin. We have:
\begin{align}
S_z &= \mathbb{I} \otimes S_z = \sqrt{2I+1}T_{00}(II)\otimes T_{10}(SS)/\sqrt{2} = \frac{\sqrt{2I+1}}{\sqrt{2}}|0010 \rangle =  \frac{\sqrt{2I+1}}{\sqrt{2}} \left[ |0\rangle |1 \rangle \right]_{10} \\
& = \frac{\sqrt{3[I]}}{\sqrt{2}} \sum_{FF'} \sqrt{[F][F']} \begin{pmatrix}
I & 1/2 & F\\I & 1/2 & F' \\ 0 & 1 & 1
\end{pmatrix} T_{10}(FF').
\end{align}
Ignoring hyperfine coherences\footnote{This approximation eliminates the non-linear Zeeman splitting from the analysis.} we have:
\begin{align}
S_z & = \frac{\sqrt{3[I]}}{\sqrt{2}} \sum_{F} [F] \begin{pmatrix}
I & 1/2 & F\\I & 1/2 & F \\ 0 & 1 & 1
\end{pmatrix} T_{10}(FF) = \frac{\sqrt{3[I]}}{\sqrt{2}} \sum_{F} [F] \begin{pmatrix}
I & 1/2 & F\\I & 1/2 & F \\ 0 & 1 & 1
\end{pmatrix} \frac{\sqrt{3}}{\sqrt{F(F+1)(2F+1)}}\hat{F}_z \\
& = \frac{3\sqrt{2I+1}}{\sqrt{2}} \sum_F \frac{\sqrt{2F+1}}{\sqrt{F(F+1)}} \frac{(-1)^{1/2+F+I+1}}{\sqrt{2I+1}\sqrt{3}} \begin{Bmatrix} F & 1/2 & I \\ 1/2 & F & 1\end{Bmatrix} F_z= \frac{\hat{F}_z(a)-\hat{F}_z(b)}{2I+1} .
\end{align}
The evolution of a (spherical) operator $T_{LM}(FF)$ due to the magnetic field is:
\begin{align}
\frac{d \langle T_{LM}(FF) \rangle}{dt} & = \langle \imath \omega_e \left[ S_z , T_{LM}(FF) \right]\rangle = \langle \imath \frac{\omega_e}{2 I+1} (-1)^{F-(I+1/2)} \left[ \hat{F}_z , T_{LM}(FF) \right]\rangle \\
&= \imath  \omega_0  (-1)^{F-(I+1/2)} M \langle T_{LM} (FF) \rangle,
\end{align}
with $\omega_0=\frac{g_s \mu_{\text{B}}}{(2I+1) \hbar} B_z = \frac{\gamma_e}{2I+1} B_z = \gamma_F B_z$, where $g$ is the electron g-factor, $\mu_{\text{B}}$ is the Bohr magneton, and $\gamma_e$ and $\gamma_F$ are respectively the free electron and the atomic gyromagnetic ratio.

In matrix form:
\begin{equation}
    \frac{d \langle \vec{T}_M \rangle}{dt}\Bigg|_{\text{mg}} = \mathcal{A}_{\text{MG},M} \langle \vec{T}_M \rangle,
\end{equation}
where $\mathcal{A}_{\text{MG},M}$ is a diagonal matrix with identical absolute values for its diagonal elements:
\begin{equation}
     \mathcal{A}_{\text{MG},M} = M \imath \omega_0 \text{diag}\left[ 1,-1,1,-1,...,1,-1\right].
\end{equation}
Unlike the relaxation processes, the evolution matrices that describe the dynamics induced by the magnetic field are different for $M=\pm1$, and they are interconnected by a complex conjugation transformation.

\subsubsection{Transverse field}
As before, we will neglect the coupling to the nuclear spin. We assume transverse fields in the $x$ and $y$ direction, so that for the evolution of a spherical tensor in the $|F m_F \rangle$ basis we need to calculate the commutator:
\begin{equation}
\frac{d T_{LM}(FF')}{dt} = \imath g_s \mu_{\text{B}} B_x  \left[S_x,T_{LM}(FF') \right]+\imath g_s \mu_{\text{B}} B_y  \left[S_y,T_{LM}(FF') \right].
\end{equation}
We recall that:
\begin{equation}
S_x = \frac{1}{\sqrt{2}} \left( S_{-1}-S_{+1}\right), S_y = \frac{\imath}{\sqrt{2}} \left( S_{-1}+S_{+1} \right), S_z =S_0, 
\end{equation}
and that:
\begin{align}
S_m &= \mathbb{I} \otimes S_z = \sqrt{2I+1}T_{00}(II)\otimes T_{1m}(SS)/\sqrt{2} = \frac{\sqrt{2I+1}}{\sqrt{2}}|001m \rangle =  \frac{\sqrt{2I+1}}{\sqrt{2}} \left[ |0\rangle |1 \rangle \right]_{1m} \\
& = \frac{\sqrt{3[I]}}{\sqrt{2}} \sum_{FF'} \sqrt{[F][F']} \begin{pmatrix}
I & 1/2 & F\\I & 1/2 & F' \\ 0 & 1 & 1
\end{pmatrix} T_{1m}(FF').
\end{align}
Ignoring hyperfine coherences:
\begin{align}
S_m & = \frac{\sqrt{3[I]}}{\sqrt{2}} \sum_{F} [F] \begin{pmatrix}
I & 1/2 & F\\I & 1/2 & F \\ 0 & 1 & 1
\end{pmatrix} T_{1m}(FF) = \frac{\sqrt{3[I]}}{\sqrt{2}} \sum_{F} [F] \begin{pmatrix}
I & 1/2 & F\\I & 1/2 & F \\ 0 & 1 & 1
\end{pmatrix} \frac{\sqrt{3}}{\sqrt{F(F+1)(2F+1)}}\hat{F}_m \\
& = \frac{3\sqrt{2I+1}}{\sqrt{2}} \sum_F \frac{\sqrt{2F+1}}{\sqrt{F(F+1)}} \frac{(-1)^{1/2+F+I+1}}{\sqrt{2I+1}\sqrt{3}} \begin{Bmatrix} F & 1/2 & I \\ 1/2 & F & 1\end{Bmatrix} \hat{F}_m= \frac{\hat{F}_m(a)-\hat{F}_m(b)}{2I+1}.
\end{align}
Overall:
\begin{align}
S_x & =\frac{\hat{F}_{-1}(a)-\hat{F}_{1}(a)-\hat{F}_{-1}(b)+\hat{F}_{1}(b)}{\sqrt{2}[I]}, \\
S_y & =\imath \frac{\hat{F}_{-1}(a)+\hat{F}_{1}(a)-\hat{F}_{-1}(b)-\hat{F}_{1}(b)}{\sqrt{2}[I]}
\end{align}
For $\hat{F}_\mu(F) = T_{1\mu}(FF) \frac{\sqrt{F(F+1)(2F+1)}}{\sqrt{3}}$:
\begin{equation}
\left[ \hat{F}_{\mu}(F), T_{LM}(F'F') \right] = \sqrt{L(L+1)}C_{L M;1\mu}^{L(M+\mu)} T_{L(M+\mu)}(FF) \delta_{FF'}. 
\end{equation}

The evolution of a (spherical) operator $T_{LM}(FF)$ due to the $B_x$ transverse magnetic field is:
\begin{align}
\frac{d T_{LM}(FF) }{dt} = \imath \Omega_{\text{R}x} \frac{\sqrt{L(L+1)}}{\sqrt{2}} &\Bigg\{  \left[ C_{L M;1-1}^{L(M-1)} T_{L(M-1)}(aa) - C_{L M;11}^{L(M+1)} T_{L(M+1)}(aa) \right] \delta_{Fa} \nonumber \\
& -\left[ C_{L M;1-1}^{L(M-1)} T_{L(M-1)}(bb) - C_{L M;11}^{L(M+1)} T_{L(M+1)}(bb) \right] \delta_{Fb} \Bigg\} \nonumber \\
&-\Omega_{\text{R}y} \frac{\sqrt{L(L+1)}}{\sqrt{2}} \Bigg\{  \left[ C_{L M;1-1}^{L(M-1)} T_{L(M-1)}(aa) + C_{L M;11}^{L(M+1)} T_{L(M+1)}(aa) \right] \delta_{Fa} ,  \nonumber \\
& -\left[ C_{L M;1-1}^{L(M-1)} T_{L(M-1)}(bb) + C_{L M;11}^{L(M+1)} T_{L(M+1)}(bb) \right] \delta_{Fb} \Bigg\}
\end{align}
where $\Omega_{\text{R}x}=\gamma_F B_x$ and $\Omega_{\text{R}y}=\gamma_F B_y$, which can be time varying.

\subsection{Hyperfine interaction}
In the Heisenberg representation, the evolution of a spherical tensor due to the hyperfine interaction is:
\begin{equation}
    \frac{d T_{LM}(FF') }{dt} = \frac{\imath}{\hbar} A_{\text{hf}}\left[\mathbf{I} \cdot \mathbf{S},T_{LM}(FF') \right] = \frac{\imath}{\hbar}\frac{1}{2} \left[ F(F+1)-F'(F'+1)\right]A_{\text{hf}}.
\end{equation}
Therefore, neglecting hyperfine coherences, we find that the hyperfine interaction does not affect the evolution of the Zeeman coherences, i.e.:
\begin{equation}
     \frac{d \langle \vec{T}_M \rangle}{dt}\Bigg|_{\text{hf}} = 0
\end{equation}

\subsection{Overall evolution}
Overall the expectation values of the transverse spin operators follow:

\begin{equation}
\frac{d \langle \vec{T}_M \rangle}{dt}= \frac{d \langle \vec{T}_M \rangle}{dt}\Bigg|_{\text{hf}}+\frac{d \langle \vec{T}_M \rangle}{dt}\Bigg|_{\text{mg}}+\frac{d \langle \vec{T}_M \rangle}{dt}\Bigg|_{\text{se}}+\frac{d \langle \vec{T}_M \rangle}{dt}\Bigg|_{\text{sd}}+\frac{d \langle \vec{T}_M \rangle}{dt}\Bigg|_{\text{op}}=( \mathcal{A}_{\text{MG},M}+\mathcal{A}_{\text{SE}} +\mathcal{A}_{\text{SD},M} +\mathcal{A}_{\text{OP}} ) \langle \vec{T}_M \rangle \equiv \mathcal{A}_M \langle \vec{T}_M \rangle,
\end{equation} \label{eq:SpinExchangeEvolMatriEqu22}
where:
\begin{equation}
\vec{T}_M = \left[  T_{1M}(aa),  T_{1M}(bb), T_{2M}(aa), T_{2M}(bb), ... T_{LM}(aa), T_{LM}(bb),...   \right], \text{ } L=1,2,...2I,
\end{equation}

\section{Measured signal and spectrum}

Though the spherical tensors are more convenient for calculations, for a direct comparison with the experiment we need to find the evolution of the transverse components $\langle F_{x,y} \rangle$. For instance, assuming probe propagation along the $x$ axis, the measured quantity in a single species experiment can be written in the form:
\begin{equation}
\mathcal{S} = D_a F_x(aa)-D_b F_x(bb),
\end{equation}
where $D_a=D(\nu - \nu_{0a})$, $D_b=D(\nu - \nu_{0b})$, with $D(\nu)$ being the dispersion function, $\nu$ the probe frequency, and $\nu_{0a}$, $\nu_{0b}$ the resonance frequencies for transitions from the $a$ and $b$ hyperfine manifolds, respectively .

\subsection{Change of basis}
We need to transform from the spherical basis to the Cartesian-experimentally relevant basis. For this we use the relationships:
\begin{equation}
\hat{T}_{1M} (\tilde{F}\tilde{F}) = \frac{\sqrt{3}}{\sqrt{F (F+1)(2F+1)}} \hat{F}_M (\tilde{F}\tilde{F}) ,  \label{eq:SphericalToFM}
\end{equation}
\begin{equation}
\hat{F}_1(\tilde{F}\tilde{F}) = -\frac{\hat{F}_x(\tilde{F}\tilde{F})+\imath \hat{F}_y(\tilde{F}\tilde{F})}{\sqrt{2}}, \phantom{aa} \hat{F}_{-1}(\tilde{F}\tilde{F}) = \frac{\hat{F}_x(\tilde{F}\tilde{F})-\imath \hat{F}_y(\tilde{F}\tilde{F})}{\sqrt{2}}. \label{eq:FMToFxFy}
\end{equation}
Converting to the Cartesian basis requires both $T_{11}(\tilde{F}\tilde{F})$ and $T_{1-1}(\tilde{F}\tilde{F})$. However, it doesn't necessitate tensors of a rank higher than one.

We define the vectors:
\begin{equation}
\vec{T}_{|1|}=\left[\vec{T}_{1}, \vec{T}_{-1} \right]^{\top},
\end{equation}
and
\begin{equation}
\mathbf{F} = \left[ F_{x}(aa), F_{x}(bb), F_{y}(aa), F_{y}(bb) \right]^{\top},
\end{equation}
where the $x$, $y$ indices denote Cartesian components and $a$ and $b$ denote the hyperfine manifold. The two vectors are related through the equation (change of basis transformation):
\begin{equation}
\mathbf{F} = \mathcal{M} \vec{T}_{|1|}, \label{eq:change2Cartesian1}
\end{equation}
where:
\begin{equation}
 \mathcal{M} = \begin{pmatrix}
 -\mathfrak{M} & 0_{2 \times (4I-2)} & \mathfrak{M} & 0_{2 \times (4I-2)} \\
 \imath \mathfrak{M} & 0_{2\times (4I-2)}  & \imath \mathfrak{M} & 0_{2\times (4I-2)} 
 \end{pmatrix} \label{eq:change2Cartesian2}
\end{equation}
and:
\begin{equation}
 \mathfrak{M} = \begin{pmatrix}
  \frac{\sqrt{(I+1) (2 I+1) (2 I+3)}}{2 \sqrt{3}} & 0 \\
 0 & \frac{\sqrt{I (2 I-1) (2 I+1)}}{2 \sqrt{3}}
 \end{pmatrix}.
\end{equation}

The measured signal can be written succinctly in the form:
\begin{equation}
\mathcal{S} = \mathcal{V} \mathcal{M} \vec{T}_{|1|},
\end{equation}
where:
\begin{equation}
\mathcal{V} = \begin{pmatrix}
 D_a & -D_b & 0 & 0
\end{pmatrix}.
\end{equation}

\subsection{Spectrum}
Assuming optical pumping in the $z$ direction, the linear dynamics do not mix components of different $M$ and can be written in the form:
\begin{equation}
\frac{d}{dt} \langle \vec{T}_{|1|} \rangle = \mathfrak{A} \langle \vec{T}_{|1|} \rangle  = \begin{pmatrix}
\mathcal{A} & 0_{4I\times4I} \\
0_{4I\times4I} & \mathcal{A}^*
\end{pmatrix} \langle \vec{T}_{|1|} \rangle ,
\end{equation}
where $*$ denotes complex conjugation. The matrix $\mathcal{A}$ includes the contribution from all the processes that affect the spin dynamics:
\begin{equation}
    \mathcal{A} = \mathcal{A}_{\text{SE}}+\mathcal{A}_{\text{SD}}+\mathcal{A}_{\text{OP}}+\mathcal{A}_{\text{MG},M=1}.
\end{equation} 

The spectrum in the spherical basis is:
\begin{equation}
S_{\text{sp}} (\omega) = -\frac{1}{2\pi} \left( -\mathfrak{A}+\imath \omega \right)^{-1} \left( \mathfrak{A} \Sigma_{\text{sp}}+\Sigma_{\text{sp}} \mathfrak{A}^{\top}  \right)    \left( -\mathfrak{A}^{\top}-\imath \omega \right)^{-1}, \label{eq:SpectrumSpherical0}
\end{equation}
where $\Sigma_{\text{sp}}$ is the equal-time ($\tau=0$) covariance matrix, which is calculated at the equilibrium state described by the spin-temperature distribution.

For this state, we have:
\begin{align}
& \langle T_{LM}(FF) (t) T_{L'M}(F'F') (t) \rangle = \nonumber \\
&\Tr \Big [ \sum_{m m'} |F m \rangle \langle F m-M | (-1)^{m+m'-2M-F-F'} C_{Fm;FM-m}^{LM} C_{F'm';F'M-m'}^{L'M}  |F'm'\rangle \langle F' m'-M| \frac{1}{Z}e^{\beta F_z}  \Big ] =\\
& \sum_{m m'\tilde{F} k  m_i m_s} \Big \{ (-1)^{m+m'-2M-F-F'} C_{Fm;FM-m}^{LM} C_{F'm';F'M-m'}^{L'M} C^{\tilde{F} k}_{I m_i;S m_s }  \times \nonumber \\
& \phantom{aaaaaaaaaaaaaa} \langle \tilde{F} k |F m \rangle \langle F m-M  |F'm'\rangle \langle F' m'-M|  \frac{1}{Z}e^{\beta (\hat{I}_z+\hat{S}_z ) } |m_i m_s \rangle \Big \} = \\
& \sum_{m m'\tilde{F} k  m_i m_s} \Big \{ (-1)^{m+m'-2M-F-F'} C_{Fm;FM-m}^{LM} C_{F'm';F'M-m'}^{L'M} C^{\tilde{F} k}_{I m_i;S m_s } C^{\tilde{F}' k}_{I m_i;S m_s } \frac{1}{Z}e^{\beta k }  \times \nonumber \\
& \phantom{aaaaaaaaaaaaaa} \delta_{\tilde{F} F} \delta_{km} \delta_{FF'} \delta_{(m-M) m'} \delta_{F'\tilde{F}'} \delta_{(m'-M) k} \Big \}.
\end{align}
The result has  terms proportional to $\delta_{(m'-M)m} \delta_{(m-M)m'}$, which are nonzero only for $M=0$. Therefore, for $M=\pm 1$:
\begin{equation}
\langle T_{LM}(FF) (t) T_{L'M}(F'F') (t) \rangle=0. \label{eq:CovSameMSameTime}
\end{equation}

Similarly, we find:
\begin{align}
& \langle T_{L1}(FF) (t) T_{L'(-1)}(F'F') (t) \rangle = \nonumber \\
& \sum_{\substack{m m'\tilde{F} \tilde{F}' \\ k m_i m_s}}  (-1)^{m+m'-F-F'} C_{Fm;F(1-m)}^{L1} C_{F'm';F'(-1-m')}^{L'-1} C^{\tilde{F} k}_{I m_i;S m_s } C^{\tilde{F}' k}_{I m_i;S m_s } \delta_{\tilde{F} F} \delta_{km} \delta_{FF'} \delta_{(m-1) m'} \delta_{F'\tilde{F}'} \delta_{(m'+1) k}  \frac{1}{Z}e^{\beta k} \\
& = -\frac{ \delta_{FF'}}{Z}\sum_{m m_i m_s}  C_{Fm;F(1-m)}^{L1} C_{F(m-1);F(-m)}^{L'-1} \left[ C_{Im_i;S m_s}^{Fm} \right]^2 e^{\beta m}, \label{eq:CovMatrixElements1}
\end{align}
and:
\begin{align}
\langle T_{L(-1)}(FF) (t) T_{L'1}(F'F') (t) \rangle = -\frac{ \delta_{FF'}}{Z} \sum_{m m_i m_s}  C_{Fm;F(-1-m)}^{L-1} C_{F(m+1);F(-m)}^{L'1} \left[ C_{Im_i;S m_s}^{Fm} \right]^2 e^{\beta m}. \label{eq:CovMatrixElements2}
\end{align}

Given Eq.~\ref{eq:CovSameMSameTime} and the symmetry of the covariance matrix (expressed in the symmetrized form: $\hat{A}\hat{B}+\hat{B}\hat{A}$), we write:
\begin{equation}
\Sigma_{\text{sp}} = \begin{pmatrix}
0_{4I\times4I} & \Sigma \\
\Sigma^{\top} & 0_{4I\times4I}
\end{pmatrix}=\begin{pmatrix}
0_{4I\times4I} & \Sigma \\
\Sigma & 0_{4I\times4I}
\end{pmatrix},
\end{equation}
where $\Sigma$ is a $4I\times4I$ matrix defined by Eq.~\ref{eq:CovMatrixElements1} (or Eq.~\ref{eq:CovMatrixElements2}).
In writing the last equation, we employed the property that the elements of $\Sigma$ are strictly real and that $\Sigma$ is symmetric as can be seen from taking into account the Clebsch-Gordan property: $C_{a \alpha; b \beta}^{c \gamma} = (-1)^{a+b-c}C_{a -\alpha; b -\beta}^{c -\gamma} =(-1)^{a+b-c}C_{b \beta; a \alpha}^{c \gamma} $.

The covariance matrix in Cartesian coordinates is then $\Sigma_{\text{C}}=\mathcal{M} \Sigma_{\text{sp}} \mathcal{M}^{\top}$. 

Eq.~\ref{eq:SpectrumSpherical0} gives:
\begin{align}
S_{sp}(\omega)& = -\frac{1}{2\pi}
\begin{pmatrix}
( \mathcal{A}-\imath \omega )^{-1} & 0_{4I\times4I} \\
0_{4I\times4I} & (\mathcal{A}^*-\imath \omega )^{-1}
\end{pmatrix} 
\begin{pmatrix}
0_{4I\times4I} & \mathcal{A} \Sigma+\Sigma \mathcal{A}^{\dag} \\
\mathcal{A}^* \Sigma^{\dag}+\Sigma^{\dag} \mathcal{A}^\top  & 0_{4I\times4I}
\end{pmatrix} 
\begin{pmatrix}
( \mathcal{A}^{\top}+\imath \omega )^{-1} & 0_{4I\times4I} \\
0_{4I\times4I} & (\mathcal{A}^\dag+\imath \omega )^{-1}
\end{pmatrix}\\
&=-\frac{1}{2\pi}\begin{pmatrix}
0_{4I\times4I} & ( \mathcal{A}-\imath \omega )^{-1} (\mathcal{A} \Sigma+\Sigma \mathcal{A}^{\dag}) \left[ ( \mathcal{A}-\imath \omega )^{-1} \right]^{\dag} \\
(\mathcal{A}^*-\imath \omega )^{-1} (\mathcal{A}^* \Sigma+\Sigma \left[ \mathcal{A}^*\right]^{\dag}) \left[ (\mathcal{A}^*-\imath \omega )^{-1} \right]^{\dag} & 0_{4I\times4I} 
\end{pmatrix}.
\end{align} 

The spectrum in the Cartesian coordinates is:
\begin{equation}
S_{\text{C}} = \mathcal{M} S_{\text{sp}} \mathcal{M}^{\top} = \frac{1}{2\pi}
\begin{pmatrix}
   \tilde{\mathfrak{M}} \left[ \mathbb{A}+\mathbb{B} \right]  \tilde{\mathfrak{M}}^\top & \imath \tilde{\mathfrak{M}} \left[ \mathbb{A}-\mathbb{B} \right]  \tilde{\mathfrak{M}}^\top \\
   -\imath \tilde{\mathfrak{M}} \left[ \mathbb{A}-\mathbb{B} \right]  \tilde{\mathfrak{M}}^\top & \tilde{\mathfrak{M}} \left[ \mathbb{A}+\mathbb{B} \right]  \tilde{\mathfrak{M}}^\top 
\end{pmatrix},
\end{equation}
where:
\begin{equation}
\tilde{\mathfrak{M}} = \begin{pmatrix}
\mathfrak{M} & 0_{2 \times (4I-2)}
\end{pmatrix},
\end{equation} 
and:
\begin{equation}
    \mathbb{A} = ( \mathcal{A}-\imath \omega )^{-1} (\mathcal{A} \Sigma+\Sigma \mathcal{A}^{\dag}) \left[ ( \mathcal{A}-\imath \omega )^{-1} \right]^{\dag},
\end{equation}
\begin{equation}
    \mathbb{B} = (\mathcal{A}^*-\imath \omega )^{-1} (\mathcal{A}^* \Sigma+\Sigma \left[ \mathcal{A}^*\right]^{\dag}) \left[ (\mathcal{A}^*-\imath \omega )^{-1} \right]^{\dag}.
\end{equation}
The measured spectrum is:
\begin{equation}
S_{\mathcal{S} } = \Re \left \{ \mathcal{V}\mathcal{M} S_{\text{sp}} \mathcal{M}^{\top} \mathcal{V}^{\top} \right \} = \Re \left\{ 
\begin{pmatrix}
    D_a & -D_b
\end{pmatrix} \tilde{\mathfrak{M}} \left[ \mathbb{A}+\mathbb{B} \right]  \tilde{\mathfrak{M}}^\top
\begin{pmatrix}
    D_a \\ -D_b
\end{pmatrix} \right \},
\end{equation}
where the real part comes from the fact that in the actual measurement the measured correlation is the symmetric form: $\frac{1}{2} \left[ \hat{F}_x(t+\tau) \hat{F}_x(t)+\hat{F}_x(t) \hat{F}_x(t+\tau) \right]$.

\section{Response to a coherent signal}
Here, we consider the response of spins to a coherent, sinusoidally-time-varying, transverse magnetic field:
\begin{equation}
B=B_{x0} \cos(\omega t) \mathbf{x} +B_{y0} \cos(\omega t +\phi) \mathbf{y} =B_{0\perp} \left [ \cos(b) \frac{e^{\imath \omega t}+e^{-\imath \omega t} }{2} \mathbf{x} +\sin(b) \frac{e^{\imath \omega t+\imath \phi}+e^{-\imath \omega t-\imath \phi} }{2} \mathbf{y} \right] = \frac{B_{0\perp}}{2} \mathbf{p}e^{\imath \omega t}+\text{c.c},
\end{equation}
where $\mathbf{p} = \cos(b) \mathbf{x}+e^{\imath \phi} \sin({b}) \mathbf{y}$ is the unit vector describing the polarization of the transverse magnetic field.
We assume a small magnetic field excitation (Rabi frequency $\Omega_{R0}=\gamma_F B_{0\perp}$ much smaller than the Larmor frequency and the slowest spin-relaxation rate) oscillating at a frequency $\omega$ close to the Larmor frequency. We neglect coherences in the harmonics of $\omega$, setting (for the coherent response): $\langle T_{LM}(FF) \rangle =0$ for $|M|>1$. We also take that $\langle T_{L0}(FF) \rangle \approx \langle T_{L0}(FF) \rangle_{\text{ST}}$.

In this case, we can write:
\begin{equation}
\frac{d}{dt} \langle \vec{T}_{|M|} \rangle = \mathfrak{A} \langle \vec{T}_{|M|} \rangle+\mathfrak{B}_x \left( e^{\imath \omega t}+e^{-\imath \omega t} \right)/2+\mathfrak{B}_y \left( e^{\imath \omega t+\imath \phi}+e^{-\imath \omega t -\imath \phi} \right)/2 , \label{eq:MatrixEquationForCoherentEvolution}
\end{equation}
where $\mathfrak{B}_x$, $\mathfrak{B}_y$ are column vectors given by:
\begin{equation}
    \mathfrak{B}_{x,y} = \begin{pmatrix}
    \mathcal{B}_{x,y} \\
        -\mathcal{B}_{x,y}^*
        \end{pmatrix},
\end{equation}
with $\mathcal{B}_{x}$ being a column vector found from the evolution:
\begin{equation}
\frac{d \langle T_{L1}(FF) \rangle }{dt} \Bigg |_{\substack{\textrm{transverse} \\ \textrm{$x$ magnetic}}} = \imath \Omega_{Rx0}\frac{\sqrt{L(L+1)}}{\sqrt{2}}\left[ \langle T_{L0}(aa) \rangle\delta_{Fa}-\langle T_{L0}(bb)\rangle \delta_{Fb} \right],
\end{equation}
and $\mathcal{B}_y$ found from the evolution:
\begin{equation}
\frac{d \langle T_{L1}(FF) \rangle }{dt} \Bigg |_{\substack{\textrm{transverse} \\ \textrm{$y$ magnetic}}} = - \Omega_{Ry0}\frac{\sqrt{L(L+1)}}{\sqrt{2}}\left[ \langle T_{L0}(aa) \rangle\delta_{Fa}-\langle T_{L0}(bb)\rangle \delta_{Fb} \right].
\end{equation}
In the above equations, we used the Rabi frequencies:
$\Omega_{Rx0}=\gamma_F  B_{x0}$ and $\Omega_{Ry0}=\gamma_F  B_{y0}$

Notice that $\mathcal{B}_y = \imath \mathcal{B}_x \Omega_{Ry0}/\Omega_{Rx0}$ and we can write:
\begin{equation}
    \mathcal{B}_x= \Omega_{Rx0}e^{\imath \pi/2} \mathcal{B}, \text{ } \mathcal{B}_y=\Omega_{Ry0} e^{\imath \pi} \mathcal{B},  
\end{equation}
\begin{equation}
    \mathcal{B} = \frac{\sqrt{L(L+1)}}{\sqrt{2} }\left[ \langle T_{L0}(aa) \rangle\delta_{Fa}-\langle T_{L0}(bb)\rangle \delta_{Fb} \right].
\end{equation}

The solution of Eq.~\ref{eq:MatrixEquationForCoherentEvolution} in the time domain is:
\begin{equation}
\langle \vec{T}_M \rangle(t) = e^{\mathfrak{A} t} \langle \vec{T}_M \rangle(0)+\frac{1}{2}\int_{0}^{t} dt' e^{\mathfrak{A}(t-t')} \left( \mathfrak{B}_x + \mathfrak{B}_y e^{\imath \phi} \right)  e^{\imath \omega t'}+ \frac{1}{2}\int_{0}^{t} dt' e^{\mathfrak{A}(t-t')} \left( \mathfrak{B}_x + \mathfrak{B}_y e^{-\imath \phi} \right)  e^{-\imath \omega t'}.
\end{equation}
The (steady state) long-time limit —$t$ much larger than the slowest relaxation time scale—  can be found either from the above equation or more directly by assuming a steady state solution of the form $ \langle \vec{T}_M \rangle_{\infty}^+ e^{\imath \omega t}$ (or $ \langle \vec{T}_M \rangle_{\infty}^- e^{\imath \omega t}$) and equating the terms proportional to $e^{\imath \omega t}$ (or $e^{-\imath \omega t}$) in Eq.~\ref{eq:MatrixEquationForCoherentEvolution}. The steady state solution is:
\begin{equation}
\langle \vec{T}_M \rangle(t) = \langle \vec{T}_M \rangle_{\infty}^+ e^{\imath \omega t}+ \langle \vec{T}_M \rangle_{\infty}^- e^{-\imath \omega t},
\end{equation}
where:
\begin{equation}
\langle \vec{T}_M \rangle_{\infty}^+ = \left( -\mathfrak{A} + \imath \omega \right)^{-1} \left( \mathfrak{B}_x + \mathfrak{B}_y e^{\imath \phi} \right) /2, \text{ } \langle \vec{T}_M \rangle_{\infty}^- = \left( -\mathfrak{A} -\imath \omega \right)^{-1} \left( \mathfrak{B}_x + \mathfrak{B}_y e^{-\imath \phi} \right)/2.
\end{equation}

We note that in situations where the longitudinal magnetic field significantly surpasses all other rates influencing the dynamics, the importance of one of the terms—either proportional to $\langle \vec{T}_M \rangle_{\infty}^+$ or $\langle \vec{T}_M \rangle_{\infty}^-$—becomes pronounced, with one vastly outweighing the other. Nonetheless, when the longitudinal magnetic field does not satisfy the aforementioned condition (as in the SERF regime), the two terms become comparable in magnitude and should both be retained.

The steady state solution for the vector $\langle \vec{T} \rangle$ is:
\begin{equation}
\langle \vec{T} \rangle_{\infty} = \frac{1}{2}
\begin{pmatrix}
\left( -\mathcal{A} + \imath \omega \right)^{-1} \left( \mathcal{B}_x+ \mathcal{B}_y e^{\imath \phi} \right) e^{\imath \omega t}+ \left( -\mathcal{A} -\imath \omega \right)^{-1} \left (\mathcal{B}_x+ \mathcal{B}_y e^{-\imath \phi} \right) e^{-\imath \omega t} \\
-\left( -\mathcal{A}^* + \imath \omega \right)^{-1}\left( \mathcal{B}_x^*+ \mathcal{B}_y^* e^{\imath \phi} \right) e^{\imath \omega t}-\left( -\mathcal{A}^* -\imath \omega \right)^{-1} \left( \mathcal{B}_x^*+ \mathcal{B}_y^* e^{-\imath \phi} \right)  e^{-\imath \omega t} 
\end{pmatrix}.
\end{equation}

In Cartesian components:
\begin{equation}
\langle \vec{F} \rangle_{\infty} = \mathcal{M} \langle \vec{T} \rangle_{\infty},
\end{equation}
and the steady state measured signal is:
\begin{equation}
\mathcal{S}_{\infty} = \mathcal{V} \mathcal{M} \langle \vec{T} \rangle_{\infty}.
\label{eq:sin_signal}
\end{equation}

After some straightforward matrix algebra, taking into account Eqs.~\ref{eq:change2Cartesian1} and \ref{eq:change2Cartesian2}  we find:
\begin{equation}
\langle \vec{F} \rangle_{\infty} = \begin{pmatrix}
\tilde{\mathfrak{M}} \left[ (\mathcal{A} -\imath \omega ) ^{-1} \mathcal{B}\ \left( \imath \Omega_{Rx0}-\Omega_{Ry0} e^{\imath \phi} \right) e^{\imath \omega t}+\text{c.c} \right]+\tilde{\mathfrak{M}} \left[ (\mathcal{A}^* -\imath \omega ) ^{-1} \mathcal{B} \left( -\imath \Omega_{Rx0}-\Omega_{Ry0} e^{\imath \phi} \right) e^{\imath \omega t}+\text{c.c} \right] \\
-\imath \tilde{\mathfrak{M}} \left[ (\mathcal{A} -\imath \omega ) ^{-1}\ \mathcal{B} \left( \imath \Omega_{Rx0}-\Omega_{Ry0} e^{\imath \phi} \right) e^{\imath \omega t}-\text{c.c} \right]+\imath \tilde{\mathfrak{M}} \left[ (\mathcal{A}^* -\imath \omega ) ^{-1}\mathcal{B} \left( -\imath \Omega_{Rx0}-\Omega_{Ry0} e^{\imath \phi} \right) e^{\imath \omega t}-\text{c.c} \right]  \\
\end{pmatrix}.
\label{eq:SteadyStateResponseXY2}
\end{equation}

The coherent signal consists of a combination of two sinusoidal terms originating from the two terms present in each row of the vector in Eq.~\ref{eq:SteadyStateResponseXY2}. These sinusoidal terms are proportional to $\Omega_{R0}$ (and thus proportional to $B_{0\perp}$), and in general they have different phases and amplitudes: $B_{0\perp} \mathcal{G}_1 (\omega) \cos(\omega t + \phi_1) + B_{0\perp} \mathcal{G}_2 (\omega) \cos(\omega t + \phi_2)$. In order to compare with noise, this sum should be expressed as a single cosine term with an amplitude of: $B_{0\perp} A_c (\omega)$, $A_c(\omega)= \sqrt{\mathcal{G}_1^2 (\omega) + \mathcal{G}_2^2 (\omega) + 2 \mathcal{G}_1 (\omega)\mathcal{G}_2 (\omega)\cos(\phi_2 - \phi_1)}$. This amplitude is to be compared with the noise level at the corresponding frequency to find the signal to noise ratio. It's important to note that in an actual experiment, the demodulation phase (corresponding to the phase of the lock-in amplifier) is adjusted such that the amplitude $A_c$ can be directly determined.

For analytical calculations, it is useful to write Eq.~\ref{eq:SteadyStateResponseXY2} in the block form:

\begin{equation}
\langle \vec{F} \rangle_{\infty} = 2 \gamma_F B_{0\perp} \begin{pmatrix}
\left[ \Re \left( \mathscr{A} \right) + \Re \left( \mathscr{B} \right) \right] \cos (\omega t) - \left[ \Im \left( \mathscr{A} \right) + \Im \left( \mathscr{B} \right) \right] \sin (\omega t)  \\
 \left[ \Im \left( \mathscr{A} \right) - \Im \left( \mathscr{B} \right) \right] \cos (\omega t) + \left[ \Re \left( \mathscr{A} \right) - \Re \left( \mathscr{B} \right) \right] \sin (\omega t)
\end{pmatrix},
\end{equation}\label{eq:SteadyStateResponseXY3}
where:
\begin{align}
    \mathscr{A} & = \tilde{\mathfrak{M}} (\mathcal{A} -\imath \omega ) ^{-1} \mathcal{B}\ \left( \imath \cos(b)-\sin(b) e^{\imath \phi} \right) \\
    \mathscr{B} & = \tilde{\mathfrak{M}} (\mathcal{A}^* -\imath \omega ) ^{-1}\mathcal{B} \left( -\imath cos(b)-\sin(b) e^{\imath \phi} \right).
\end{align}
The measured signal (amplitude) is $B_{0\perp} A_c(\omega)$, where:
\begin{equation}
 A_c (\omega)= 2 \gamma_F \sqrt{ \left \{ \begin{pmatrix}
    D_a & -D_b
\end{pmatrix} \left[ \Re \left( \mathscr{A} \right) + \Re \left( \mathscr{B} \right) \right] 
 \right \}^2 +  \left \{ \begin{pmatrix}
    D_a & -D_b
\end{pmatrix} \left[ \Im \left( \mathscr{A} \right) + \Im \left( \mathscr{B} \right) \right]
 \right \}^2}
\label{eq:Ac_omega}
\end{equation}

\section{Signal to noise ratio}
Here, we derive the signal to noise ratio (SNR) in the type of atomic-optical magnetometer described above.

The magnetometer signal at a frequency $\omega$ is taken to be the polarimetry output demodulated at the same frequency, with phase adjusted so that the signal is maximized. 

For concreteness, we take a 3 dB/oct lock-in filter, and model the demodulated signal at the output of a lock-in amplifier to be at time $t$:
\begin{equation}
\mathcal{K}(t) = \frac{1}{\TBW}\int_{t_0}^{t} e^{-\frac{t-t'}{\TBW}}\cos \left(\omega t' \right) S_{\text{out}}(t') dt', \label{eq:LockinSignal}
\end{equation}
where $\TBW$ is the time-constant of the lock-in filter determining the bandwidth of the measurement, and $S_{\text{out}}$ is the polarimetry output. Since we are interested in the steady state SNR, we extended the integration to start from a distant point in the past at $t_0$. In the following, the limit  $t_0 \rightarrow -\infty$ is implied.
In a typical measurement, the following condition is satisfied:
\begin{equation}
\omega \TBW \gg 1. \label{eq:LockinConditions}
\end{equation}
In the context of the magnetometer under study, magnetic field detection is essentially a parameter estimation problem, wherein the magnetic field estimation after measurement time $T$ relies on the quantity:
\begin{equation}
\mathfrak{K}=\frac{1}{T}\int_{0}^{T} \mathcal{K}(t) dt.
\end{equation} 

When considering the SNR, the signal refers to the response to a sinusoidal excitation, neglecting noise. In this case, the polarimetry output is (see Eqs.~\eqref{eq:sin_signal} and \eqref{eq:Ac_omega}):
\begin{equation}
S_{\text{out}} = B_{0\perp} A_c(\omega) \cos \left( \omega t \right),
\end{equation}
where $A_c$ represents the measured response to a transverse AC magnetic field of amplitude $B_{0\perp}$ and frequency $\omega$. 
Then:
\begin{equation}
\bar{\mathfrak{K}} = \frac{1}{T \TBW}\int_0^{T} dt \int_{t_0}^t dt' e^{-\frac{t-t'}{\TBW}}\cos \left(\omega t' \right) B_{0\perp} A_{c}(\omega)\cos \left( \omega t' \right) dt \approx\frac{B_{0\perp} A_c(\omega)}{2}, \label{eq:LockinSignalResponse}
\end{equation}
where the approximation holds for the typical condition stated in Eq.\eqref{eq:LockinConditions}.

The noise in SNR quantifies the uncertainty in the magnetic field estimation and is given by the standard deviation of the quantity $\mathfrak{K}$:
\begin{align}
\text{Var}\left[ \mathfrak{K} \right] & = \frac{1}{T^2\TBW^2}\int_0^{T} dt \int_0^{T} dt' \int_{t_0}^t dx \int_{t_0}^{t'} dx' e^{-\frac{t-x}{\TBW}}e^{-\frac{t'-x'}{\TBW}} \cos \left(\omega x \right) \cos \left(\omega x' \right) \langle S_{\text{out}}(x) S_{\text{out}}(x') \rangle \label{eq:Noise1} \\
& =  \frac{1}{T^2\TBW^2}\int_0^{T} dt \int_0^{T} dt'\int_{t_0}^{t} dx \int_{x-t'}^{x-t_0} d\tau e^{-\frac{t-x}{\TBW}}e^{-\frac{t'-(x-\tau)}{\TBW}} \cos \left(\omega x \right) \cos \left[\omega (x-\tau) \right] \langle S_{\text{out}}(x) S_{\text{out}}(x-\tau) \rangle   \label{eq:Noise2} \\
& =  \frac{1}{T^2\TBW^2}\int_0^{T} dt \int_0^{T} dt'\int_{t_0}^{t} dx \int_{x-t'}^{x-t_0} d\tau e^{-\frac{t-x}{\TBW}}e^{-\frac{t'-(x-\tau)}{\TBW}} \cos \left(\omega x \right) \cos \left[\omega (x-\tau) \right] R(\tau)   \label{eq:Noise3} \\
& =  \frac{1}{T^2\TBW^2}\int_{-\infty}^{\infty} d \omega' \int_0^{T} dt \int_0^{T} dt'\int_{t_0}^{t} dx \int_{x-t'}^{x-t_0} d\tau e^{-\frac{t-x}{\TBW}}e^{-\frac{t'-(x-\tau)}{\TBW}} \cos \left(\omega x \right) \cos \left[\omega (x-\tau) \right] e^{\imath \omega' t} S(\omega')   \label{eq:Noise4} \\
& =  \int_{-\infty}^{\infty} d \omega'  \mathscr{F}(\omega,\omega') S(\omega') =  \int_{0}^{\infty} W(\omega') \mathscr{F}(\omega,\omega')  d\omega' \label{eq:Noise5},
\end{align}
where $\langle \cdot \rangle$ denotes statistical averaging, $S(\omega)$ and $W(\omega)=2S(\omega)$ are respectively the two-sided and one-sided spectrum, $R(\tau) = \langle S_{\text{out}}(t) S_{\text{out}}(t-\tau) \rangle $ is the correlation function of the polarimetry output, and the filter function is:
\begin{dmath}
\mathscr{F} = \Bigg \{-2 {\omega'}  \omega \left[2 \TBW^2 \left({\omega'}
   ^2+\omega^2\right)+1\right] \cos [ ({\omega'}
   -\omega) T]+2 {\omega'}  \omega \left[ 2
   \TBW^2 \left({\omega'} ^2+\omega^2\right)+1\right] \cos [ ({\omega'} +\omega)T]+4
   \cos ({\omega'} T ) \left[\TBW \omega
   (\omega^2-{\omega'}^2 )  \sin
   (\omega T)-\cos ( \omega T)
   \left ( \TBW^2 {\omega'} ^4+2 \TBW^2 {\omega'} ^2
   \omega^2+\TBW^2 \omega^4+{\omega'}
   ^2\right)\right]+({\omega'}^2 -\omega^2)  \left[\cos (2  \omega T)
   \left(\TBW^2 ({\omega'}^2 -\omega^2)+1\right)+2 \TBW \omega
   \sin (2  \omega T)\right]+\left(3 {\omega'}
   ^2+\omega^2\right) \left(\TBW^2
   \left({\omega'} ^2+3 \omega^2\right)+1\right) \Bigg \} \Bigg \{ 2 T^2 ({\omega'}^2 -\omega^2)^2  \left(\TBW^4 \left({\omega'} ^2-\omega^2\right)^2+2 \TBW^2 \left({\omega'} ^2+\omega^2\right)+1\right)  \Bigg \}^{-1}. \label{eq:filterFunction}
\end{dmath}
In writing the last equation in \ref{eq:Noise5}, we used the fact that both the noise spectrum and the filter $\mathscr{F}$ are even functions.  
For the experimentally relevant case where $\omega' / \omega \sim 1$ and $\omega \TBW \gg1$ the filter function takes (for positive frequencies) the simple form:
\begin{equation}
    \mathscr{F} \approx \frac{1}{4}\left ( \frac{\sin\left[ \frac{(\omega-\omega')T}{2}\right]}{\frac{(\omega-\omega')T}{2}}\right)^2,
\end{equation}
i.e. a sinc-squared function where the effective width (quantified for instance as the FWHM) drops linearly with the measurement time $T$.

Overall, in the steady state of the magnetometer considered in this work, the signal (response) remains independent of the measurement time $T$, while the noise decreases as the measurement time increases. The precise relationship between noise and $T$ depends on the spectrum. For spectra that exhibit minimal variation around the measurement frequency within the bandwidth defined by the measurement time ($\text{BW}=1/(2T)$), the noise (variance) scales approximately as $1/T$.

The magnetometer sensitivity, typically expressed in magnetic field units per square root bandwidth, is found from the rms value of the transverse magnetic field that gives a magnetometer response equal to noise (expressed as standard deviation) after a measurement time of $T=1/(2 \text{BW})$.

We consider the experimentally relevant case of $W$ varying very little in the region where the filter function is non-negligible. Then $W$ can be taken out of the integral in Eq.~\ref{eq:Noise5} and the variance takes the approximate form:
\begin{equation}
    \text{Var}\left[ \mathfrak{K} \right] \approx W(\omega) \int_0^{\infty} d\omega' \mathscr{F}(\omega,\omega') \approx W(\omega) \int_{-\infty}^{\infty} d\omega' \mathscr{F}(\omega,\omega') \approx W(\omega) \frac{\pi}{2 T} = \frac{\Tilde{W}(f)}{4T},
\end{equation}
where $\Tilde{W}(f) = 2 \pi W(\omega)$ indicates the one-sided spectral density in Hertz, and $f=\omega/(2 \pi)$. The extension of the integration to $-\infty$ is justified on the basis that the filter function is appreciable only over a region on the order of the filter width.
Unity SNR gives:
\begin{equation}
    \frac{B_{0\perp} \tilde{A}_c(f)/2}{\sqrt{\Tilde{W}(f)/(4 T)}} = 1 \Rightarrow \frac{B_{\text{rms}}}{\sqrt{\text{BW}}} = \text{Sensitivity = }\frac{\sqrt{\tilde{W}(f)}}{\tilde{A}_c(f)},
\end{equation}
where $B_{\text{rms}} = B_{0\perp}/\sqrt{2}$ is the rms value, and  $\tilde{A}_c(f)=A_c(2 \pi f)$ is the response expressed as a function of the frequency $f$.

Operationally, sensitivity can be determined through the following procedure: apply a known calibration transverse sinusoidal magnetic field and measure the one-sided spectral density in Hz, utilizing a frequency bin of 1 Hz (i.e., a measurement time per repetition of 1 second). To avoid distortion stemming from Fourier windowing, ensure that the AC frequency matches one of the frequencies measured in the sampled Fourier spectrum. Under these conditions, the measured spectral density yields the $\text{rms}$-squared value of a sinusoidal signal. Calculate the ratio between the height of the coherent response and the height of the noise at the same frequency. The sensitivity, expressed in magnetic field units per square root Hz, is simply the rms value of the calibration magnetic field divided by the square root of the measured ratio.

\bibliography{supp}